\documentclass[preprint,11pt,pre]{revtex4}

\usepackage{dcolumn}
\usepackage{bm}
\usepackage{graphicx}
\usepackage{amssymb,amsfonts,amsmath}
\usepackage{rotating}
\usepackage{subfigure}
\usepackage{url}
\usepackage{hyperref}
\usepackage[usenames]{color}
\graphicspath{{./Figures/}}

\begin{document}

\preprint{PRE/Self-Assembly}

\title{Fast Generation of Potentials for Self Assembly of Particles}

\author{Philip du Toit}
 \affiliation{Control and Dynamical Systems, California Institute of Technology, MC 107-81, Pasadena, CA 91125}
 \email{pdutoit@cds.caltech.edu}
 \homepage{https://www.cds.caltech.edu/~pdutoit}
\author{Katalin Grubits\footnote{Current affiliation: Department of Mathematics, Marymount Manhattan College, 221 East 71st St, New York, NY 10021}}
 \affiliation{Control and Dynamical Systems, California Institute of Technology, MC 107-81, Pasadena, CA 91125}
\author{Sorin Costiner}
\affiliation{United Technologies Research Center, East Hartford, CT 06108}%
\author{Jerrold Marsden}
 \affiliation{Control and Dynamical Systems, California Institute of Technology, MC 107-81, Pasadena, CA 91125}

\date{\today}

\begin{abstract}
We address the inverse problem of designing isotropic pairwise particle interaction potentials that lead to the formation of a desired lattice when a system of particles is cooled.  The design problem is motivated by the desire to produce materials with pre-specified structure and properties.  We present a heuristic computation-free geometric method, as well as a fast and robust trend optimization method that lead to the formation of high quality honeycomb lattices.  The trend optimization method is particularly successful since it is well-suited to efficient optimization of the noisy and expensive objective functions encountered in the self-assembly design problem.  We also present anisotropic potentials that robustly lead to the formation of the kagome lattice, a lattice that has not previously been obtained with isotropic potentials.
\end{abstract}

\maketitle

\section{Introduction\label{sec:Introduction}}

During the process of self-assembly, randomly distributed components acting under the influence of short-range mutual interactions, arrange themselves into a highly ordered final configuration.  An important feature of self-assembly is that the components arrange \emph{themselves} into this more ordered state without the influence of external factors.  The ordered arrangement observed at the level of the resulting superstructure is predicated on the design of the individual components and their local interactions.  Understanding the self-assembly process and how local properties of the components may be manipulated to influence the resulting global ordering, is an active area of research spanning a broad range of disciplines including materials science,  chemical engineering, bioengineering, and nanotechnology.  

In a seminal article, Whitesides and Grzybowski \cite{whitesides2002a} provided a broad definition of self-assembly, and addressed promising applications in a wide range of disciplines.  They observed that self-assembly of cells to form tissue, organs, and ultimately organisms, is fundamental to life, and motivates a determined study of the self-assembly process.  Subsequent studies and applications of self-assembly are as fascinating as they are plentiful. Murr \textit{et al}  \cite{murr2005a} identified self-assembly as the mechanism by which silicatein monomers combine to form protein fibers in certain marine sponges.  Zheng \textit{et al} used shape recognition and selective binding to induce self-assembly and packaging of integrated semiconductor microsystem devices by agitating the components in an aqueous environment \cite{zheng2004a}, and Stauth \textit{et al} \cite{stauth2006a} have manufactured field-effect transistors via self-assembly of micrometer scale components using similar techniques.  In \cite{gross2008a}, self-assembly is shown to enhance synergistic group transport of autonomous robots.  Jakab \textit{et al} recognized that although biological systems are genetically controlled, the formation of biological superstructures is ultimately governed by physical interactions, and demonstrated the formation of prescribed shapes using self-assembling multicellular systems \cite{jakab2004a}.

Typically, studies of self-assembly examine the ordered superstructures that arise from a given fixed interaction potential.  For example, Manoharan \textit{et al} used optical and electron microscopy to identify the range of structures produced by self-assembly of colloidal microspheres as fluid is removed from the emulsion droplets containing the spheres \cite{manoharan2003a}.  Maksymovych \textit{et al} investigated the formation and reactivity of linear chains of dimethyldisulfide molecules on a gold surface  \cite{maksymovych2008a}.   Engel \textit{et al} (\cite{engel2007a}, see also \cite{hoang2008a,liu2008a,glaser2007a,quandt1999a}) have observed the formation of complex crystals and quasicrystals arising from a simple double-well interaction potential.

Interest in the fabrication of nanomaterials and photonic crystals with desired material properties \cite{ho1990a, sigmund1996a, mary1996a,xu1999a, hyun2001a,ferey1999a,chen2001a,greer2000a} motivates the inverse problem: \emph{design} the shortrange interactions in order to induce self-assembly of the components into a desired lattice structure.  Laboratory techniques now available allow for modification and tuning of particle interaction potentials \cite{murray1996a}, and hence experimentalists are achieving ever-increasing control over the local interactions that influence the global properties of the material formed via self-assembly.  These methods typically use colloidal suspensions and optical forcing to alter the chemical environment and screening properties of the solution in which the assembly occurs.   

In this paper, we specifically consider the problem of designing short-range pairwise interaction potentials between particles on a planar surface to induce self-assembly of a desired lattice.   The main results to be presented are new methods---a heuristic \emph{geometric method} as well as a robust \emph{trend optimization method} for the design of isotropic interaction potentials that lead to high quality honeycomb lattices as the system of particles is cooled.  The geometric method is also extended to the case of anisotropic potentials which allows for the formation of more exotic kagome lattices.    Another contribution of this work is the development of tools for objective assessment of quality of lattices, that mimic intuitive human perception of lattice quality.

Rechtsman \textit{et al} in \cite{rechtsman2006a} have already demonstrated computational methods for finding solutions to the inverse self-assembly problem.  They used a simulated annealing optimization procedure to find potentials that lead to the self-assembly of particles into square and honeycomb lattices.  To be sure, the intent of \cite{rechtsman2006a} was to demonstrate that the inverse problem of potential design for the purpose of inducing the formation of a target lattice can be solved in practice, and to carefully verify that the potentials they proposed do indeed lead to the target lattices through Monte Carlo simulation.  Hence, the computational effort required to obtain the potentials was only a marginal consideration in their work.  Presently, we consider the straightforward simulated annealing method of \cite{rechtsman2006a} as a baseline method with which we may compare the new methods presented here.  When compared with the baseline simulated annealing method, the optimization procedure described in this paper leads to a hundredfold speed-up in the generation of potentials, as well as the formation of higher quality lattices. Furthermore, the procedure for finding potentials is more robust, and the resultant potentials form the target lattices more robustly with respect to variations in the initial conditions of the particles.   As will be demonstrated, the chief reason for the marked speed-up over the simulated annealing method is the facility of the trend optimization method to optimize objective functions that are both noisy and expensive to evaluate.


The organization of the paper is as follows.  In Section \ref{sec:TheSelfAssemblyProblem}, we precisely define the self-assembly problem, and summarize the method for generating potentials devised previously by \cite{rechtsman2006a} that will serve as a baseline method for purposes of comparison.   In Section \ref{sec:LatticeQualityMetrics}, we establish objective metrics for measuring lattice quality so that reasonable comparisons between the methods can be made.  

We approach the self-assembly problem by framing it as an optimization problem in which the desired potential optimizes a suitably chosen objective function.  This approach requires that we choose both an optimization scheme and an objective function to be optimized.  Section \ref{sec:ObjectiveFunctions} describes three objective functions that will be used, while Section \ref{sec:OptimizationMethods} describes in detail the trend optimization method.  A discussion on the  hierarchical nature of the trend method is also provided.  We proceed in Section \ref{sec:MethodsforGenerationofPotentials} to list the five solution methods to be compared and describe their implementation.   The final comparison of the methods is presented graphically in the plots of Section \ref{sec:Results}.  In Section \ref{sec:AnisotropicPotentials}, we present an extension of the geometric method to anisotropic potentials that lead to self-assembly of the kagome lattice.

All molecular dynamics simulations performed during the design and testing of the potentials were executed on the CITerra high performance computing cluster housed in the Division of Geophysical and Planetary Sciences at Caltech using the LAMMPS software package~\cite{plimpton1995a} from Sandia Laboratories.
\medskip

\noindent
{\bf Acknowledgments.} We would like to thank Andrzej Banaszuk, Ronald Coifman, Yannis Kevrekidis, Alison Marsden, Matthew West, Jos\'e Miguel Pasini, and Igor Mezi\'c for their interest and helpful comments. This work was supported in part by DARPA DSO under AFOSR contract FA9550-07-C-0024.

\section{The Self Assembly Problem\label{sec:TheSelfAssemblyProblem}}

The availability of laboratory methods to tune interaction potentials between components, and hence influence the structure of the resulting self-assembled configurations, motivates the use of self-assembly to produce materials with desired structural properties.  The specific self-assembly problem addressed in this paper is defined as follows:\\

\medskip

\parbox{0.9\textwidth}{
\textbf{Definition: The Self-Assembly Problem}\newline
\textit{Design a radially symmetric pairwise interaction potential, $V_\mathsf{HC}(r)$, so that when a system of particles interacting with each other in the plane through this potential is cooled, the particles form a honeycomb lattice.}\\}\\

\medskip

The purpose of the present paper is to compare methods for generating potentials that solve the self-assembly problem.   The methods for generating potentials are compared using three criteria:

\begin{description}
\item{\textbf{C1.}} The computational effort required by the method to produce the potential;
\item{\textbf{C2.}} The quality of the lattices formed by the potential;
\item{\textbf{C3.}} The robustness of the quality of the lattices formed to variations in the initial conditions of the particles.
\end{description}

Certainly, for a given method we expect to see trade-offs between these criteria.  For example, a faster method may lead to a potential that produces lattices of poorer quality.

\medskip

Laboratory techniques for tuning interaction potentials motivated Rechtsman \textit{et al}~\cite{rechtsman2006a} to consider physically realizable potentials as basis functions for the desired potential, where the basis functions contain parameters that allow for tuning the shape of the total potential obtained from their sum.  For the case when the desired final configuration of particles is the honeycomb lattice, \cite{rechtsman2006a} proposed an interaction potential consisting of the sum of a Lennard-Jones potential, an exponentially decaying potential, and a Gaussian-shaped potential, parameterized in the following way:
\begin{equation}
 V_{\mathsf{HC}}(r;a_0,a_1,a_2,a_3) = \frac{5}{r^{12}} - \frac{a_0}{r^{10}} + a_1 \, e^{-a_2 r} - 0.4 \, e^{-40(r-a_3)^2},
 \label{eq:VHC}
 \end{equation}
where $a_0$, $a_1$, $a_2$, and $a_3$ are four free parameters that can be tuned to adjust the shape of the potential.  

\begin{figure}[ht]
\begin{center}
   \subfigure[\footnotesize \label{fig:perfectlattice} A perfect honeycomb lattice.]{
        \includegraphics[width=4.5cm]{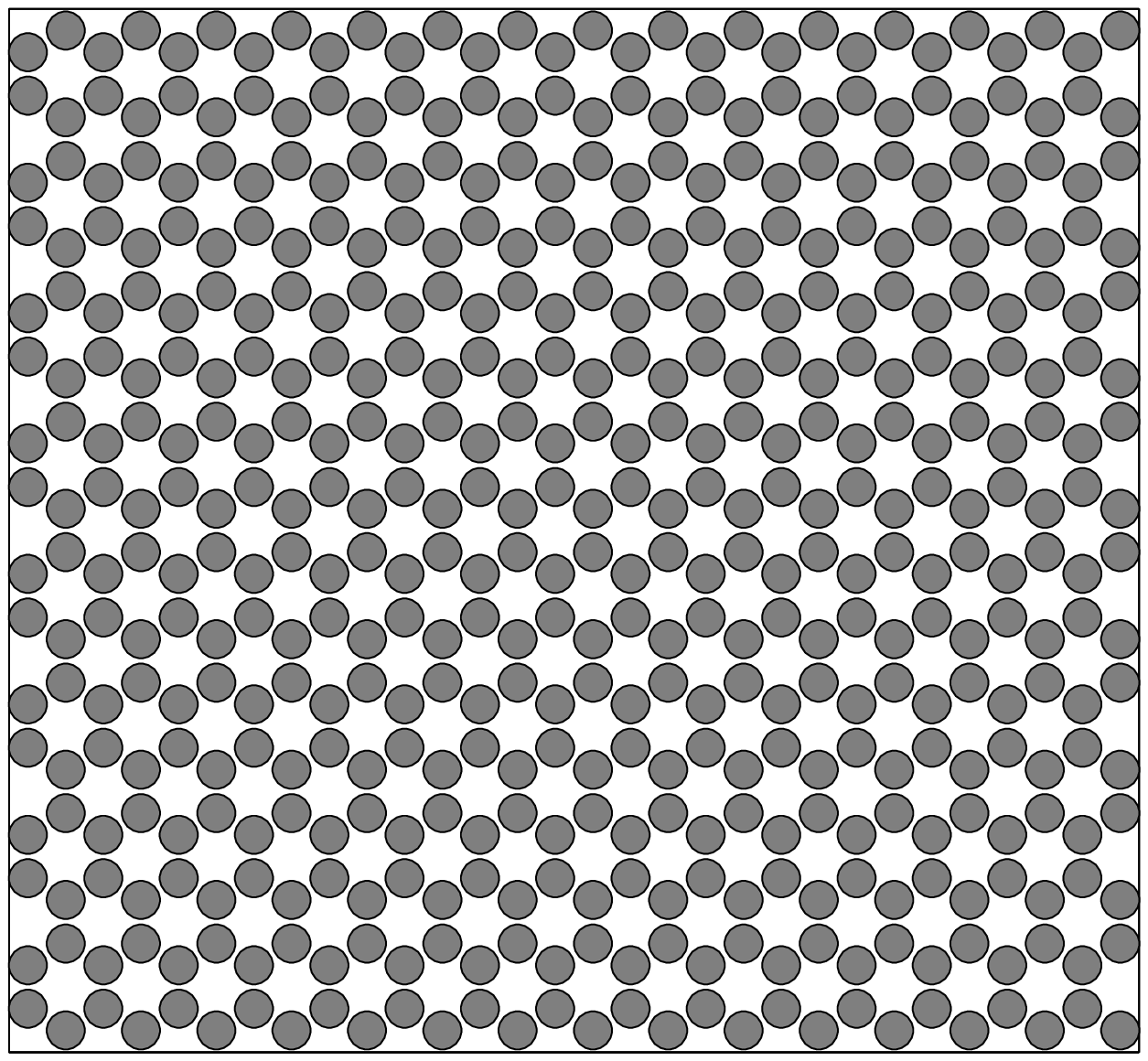}}        
           \subfigure[\footnotesize \label{fig:rechtpotential} The potential proposed in \cite{rechtsman2006a}.]{
        \includegraphics[width=5.2cm]{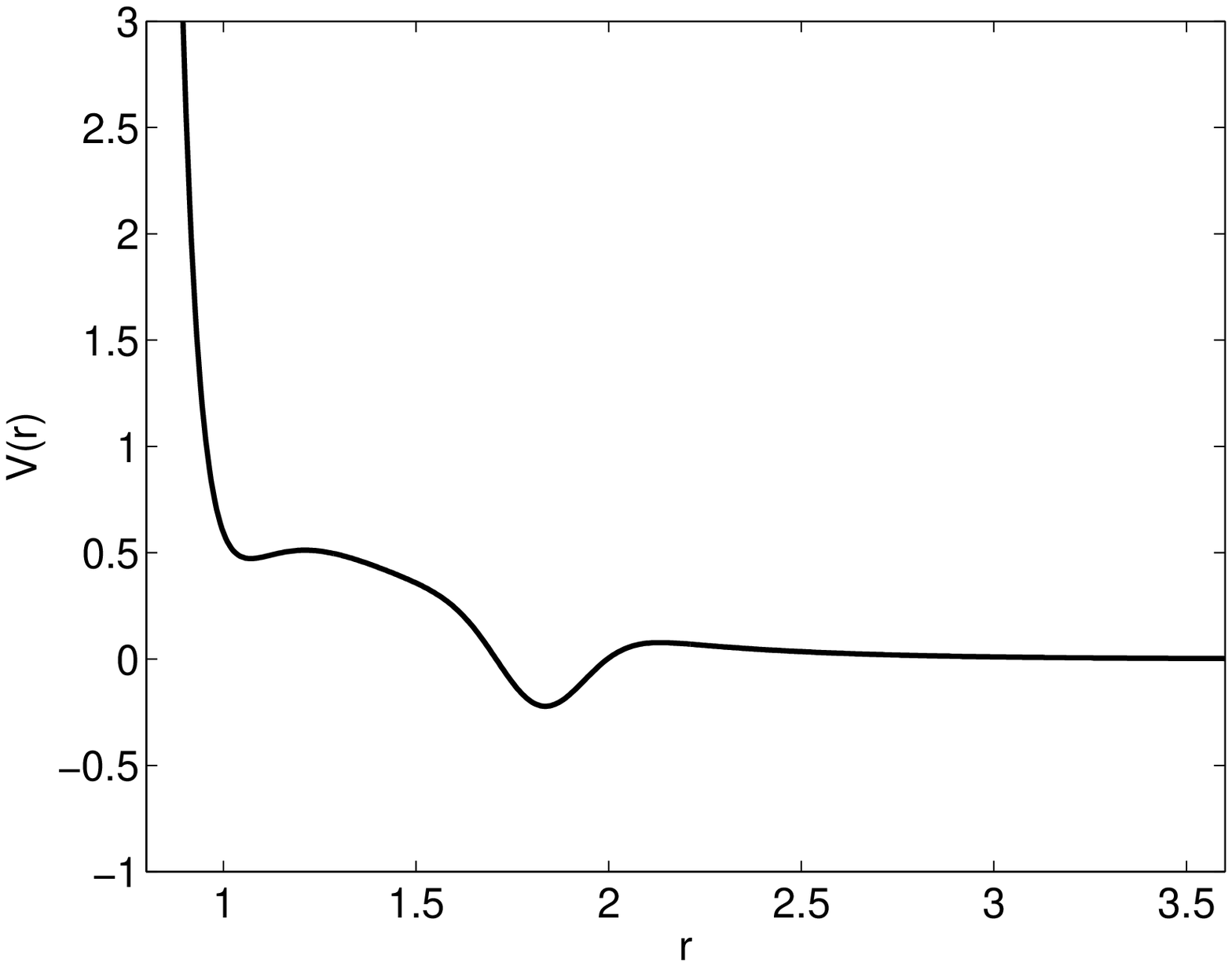}}   
          \subfigure[\footnotesize \label{fig:rechtlattice} A sample lattice obtained using the potential shown in (b).]{
        \includegraphics[width=4.5cm]{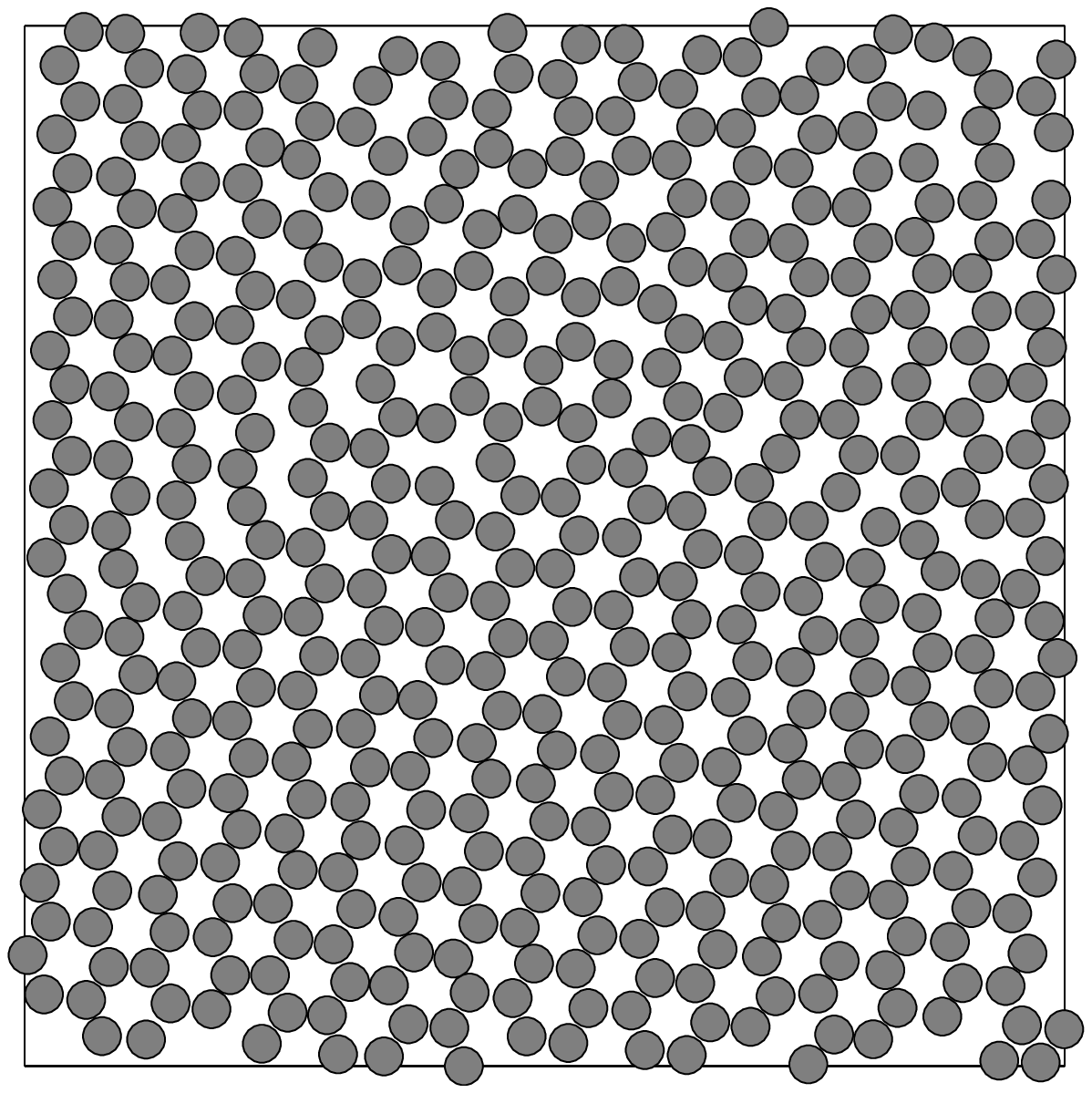}}             
  \caption{\label{fig:rechtsolution}\footnotesize  The self-assembly problem entails finding an interaction potential that leads to the formation of a honeycomb lattice (a).  \cite{rechtsman2006a} used a simulated annealing optimization procedure to generate a potential (b).  A sample lattice formed using this potential is shown in (c) and exhibits defects that result from the finite duration of the cooling simulation.} 
\end{center}
\end{figure}

A solution (there are many) to the self-assembly problem has been provided by \cite{rechtsman2006a}: namely, choosing $a_0=5.89$, $a_1=17.9$, $a_2=2.49$, and $a_3=1.823$ in the expression for $V_\mathsf{HC}$.  A sample lattice obtained when cooling a system of particles using these parameters for the interaction potential is shown in Figure \ref{fig:rechtlattice}.  The honeycomb lattice is the dominant structure in the lattice, although there are still visible defects that arise due to the finite duration of the cooling schedule.  The cooling simulation used to obtain this lattice, as well as all other simulations referred to in this paper, were performed using periodic boundary conditions. 

\begin{figure}[ht]
\begin{center}
   \subfigure[\footnotesize \label{fig:rechtsmangood484}$a_{0} = 5.89$, $a_{1}=17.9$, $a_{2}=2.49$, $a_{3}=1.823$.]{
        \includegraphics[width=3.8cm]{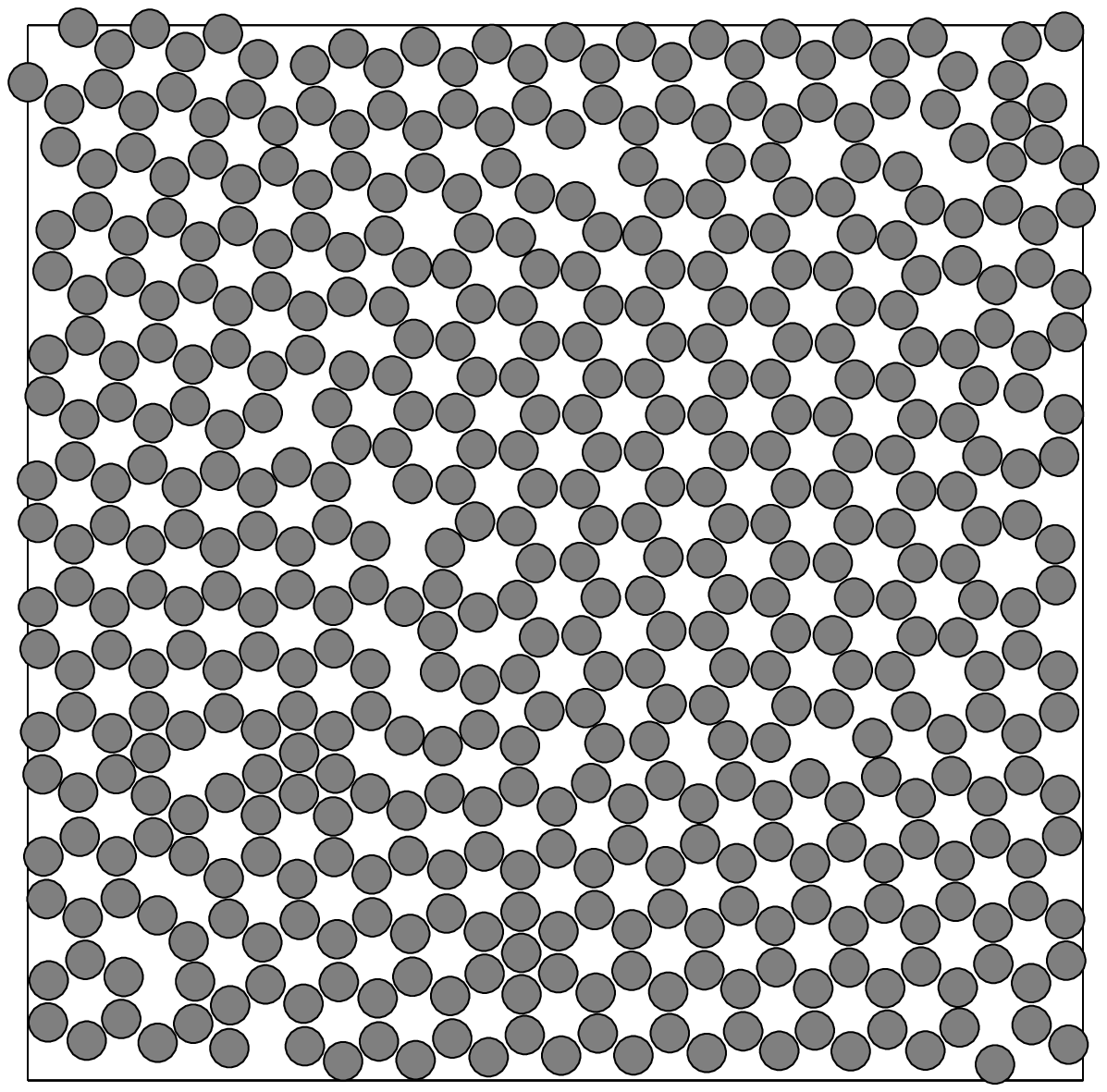}}
      \subfigure[\footnotesize \label{fig:rechtsmanbad} $a_{0} = 5.89$, $a_{1}=17.9$, $a_{2}=2.49$, $a_{3}=1.89$.]{
        \includegraphics[width=3.8cm]{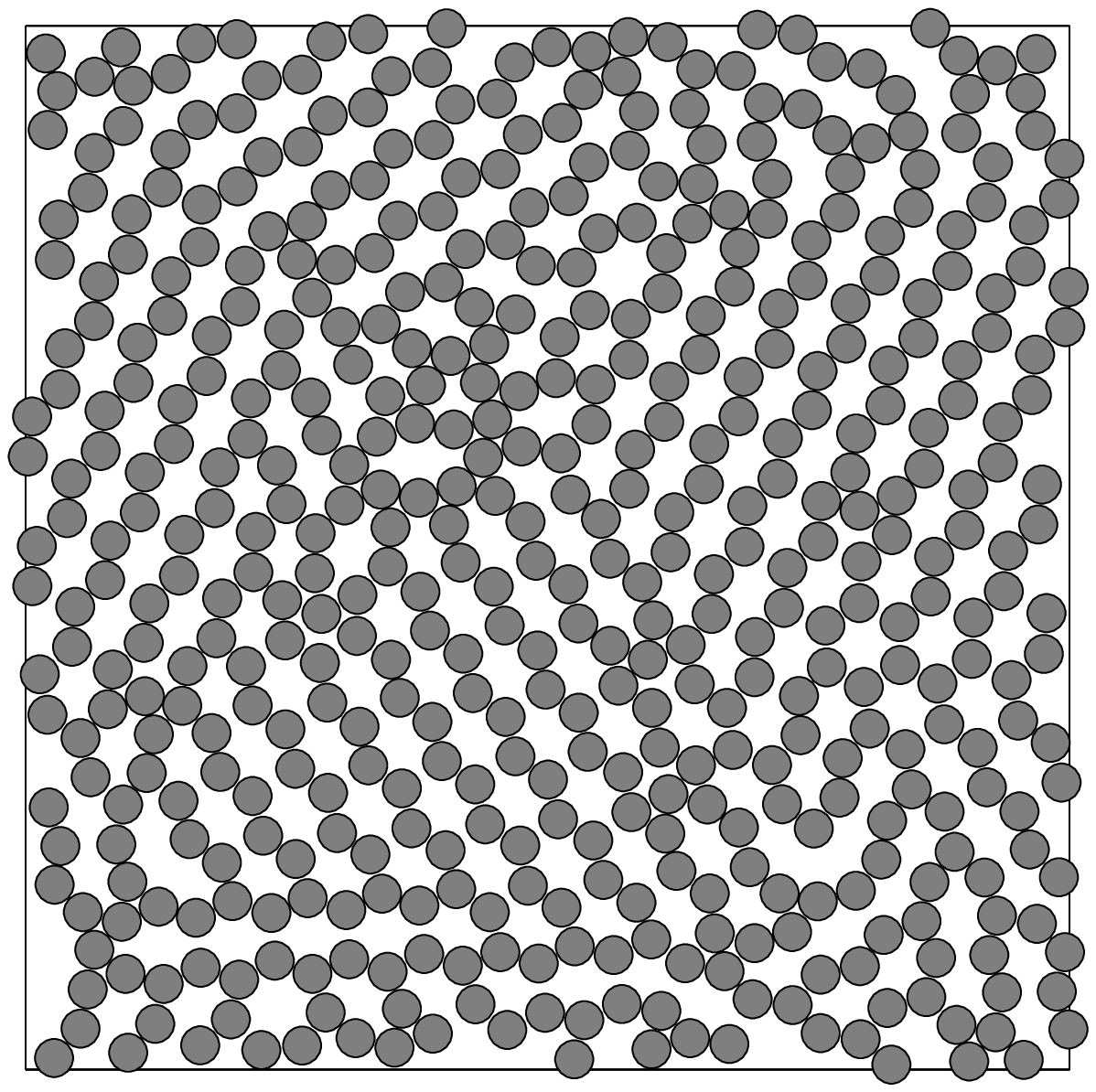}}         
 \subfigure[\footnotesize \label{fig:amorphous1} $a_{0} = 5.0$, $a_{1}=17.0$, $a_{2}=2.0$, $a_{3}=1.5$.]{
        \includegraphics[width=3.8cm]{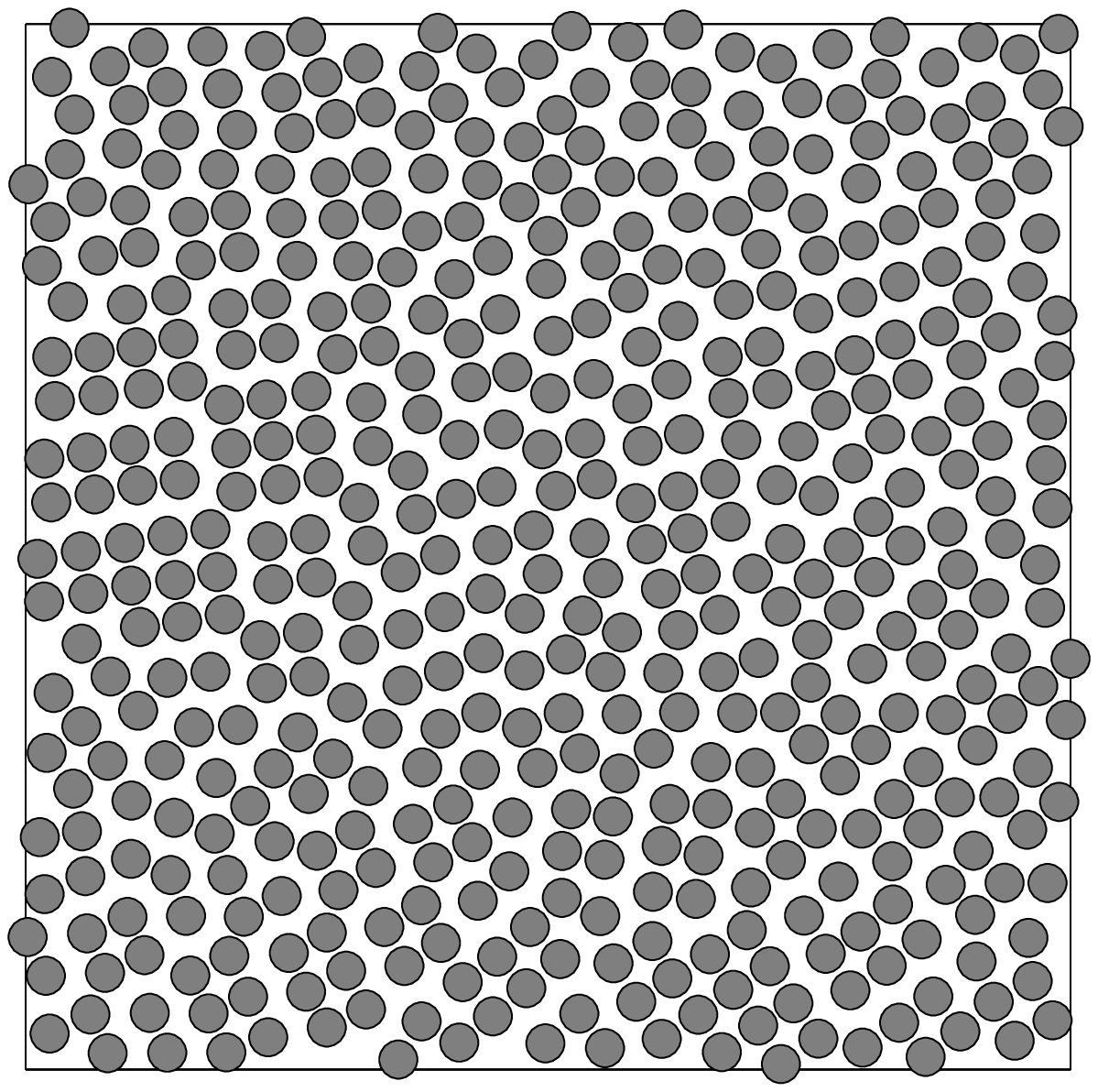}}     
         \subfigure[\footnotesize \label{fig:amorphous2} $a_{0} = 6.0$, $a_{1}=18.0$, $a_{2}=3.0$, $a_{3}=2.0$.]{
        \includegraphics[width=3.8cm]{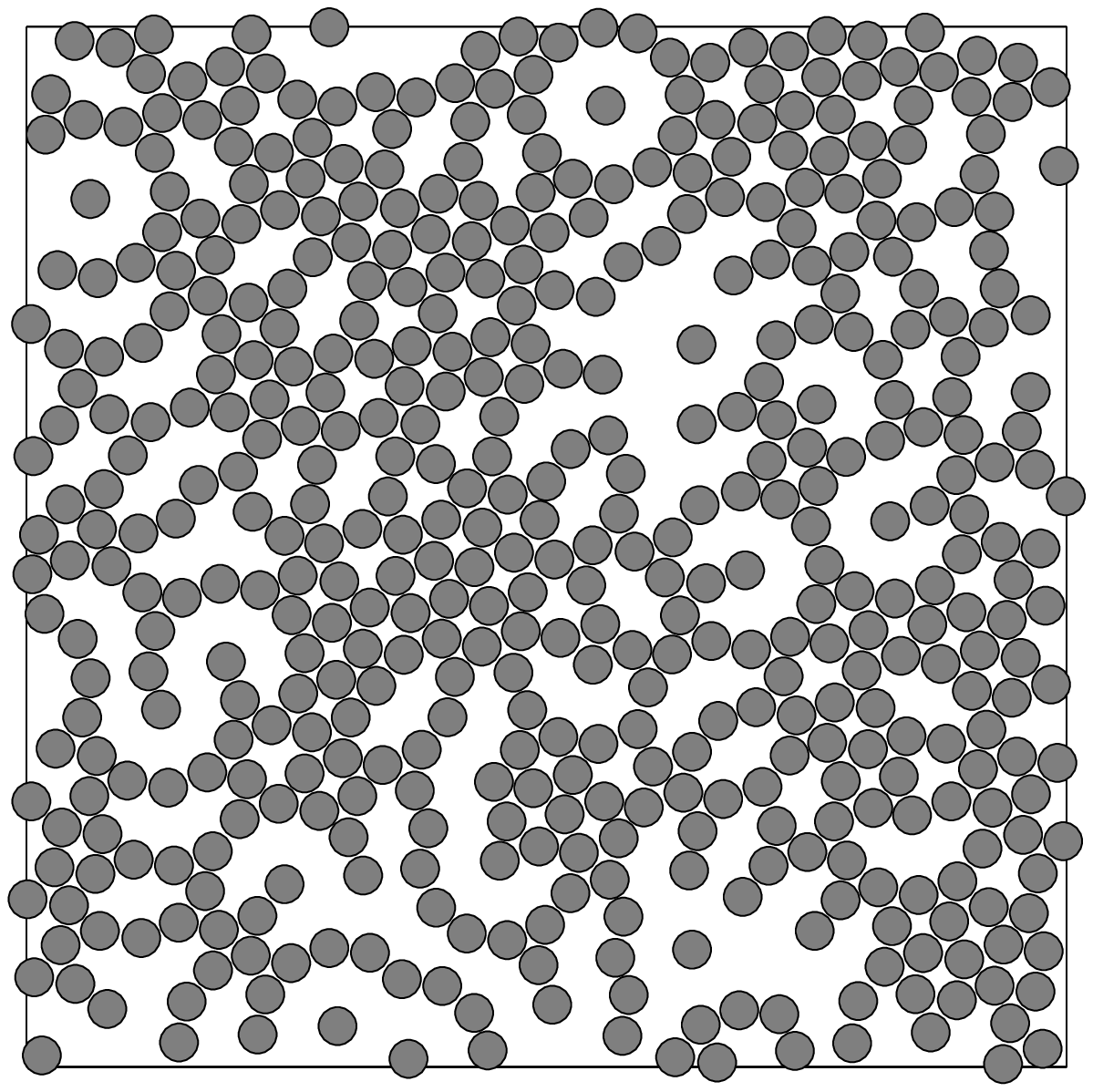}}               
  \caption{\label{fig:lattices}\footnotesize The lattices shown here are obtained by slowly cooling 484 particles using the interaction potential $V_{\textsf{HC}}(r;a_{0},a_{1},a_{2},a_{3})$ provided in Equation \ref{eq:VHC} with the parameter values indicated.  Figure \ref{fig:rechtsmangood484} uses the parameters proposed by  \cite{rechtsman2006a}, while Figure \ref{fig:rechtsmanbad} indicates how fragile the honeycomb lattice is with respect to slight changes in the parameters.  Figures \ref{fig:amorphous1} and \ref{fig:amorphous2} indicate the range of structures obtained for generic values of the parameters.} 
\end{center}
\end{figure}

A reasonable question to ask at this point is: ``How difficult is the self-assembly problem?"  Experience shows that although easy to state, the self-assembly problem is difficult to solve in that solutions that lead to the honeycomb lattice are difficult to find and possibly very fragile.  For instance, adding small perturbations to the parameters in \cite{rechtsman2006a} for the honeycomb potential leads to the configuration in Figure \ref{fig:rechtsmanbad} in which the honeycomb structure is less pronounced.   The amorphous configurations shown in Figures \ref{fig:amorphous1} and \ref{fig:amorphous2} indicate the range of structures that are possible for generic values of the parameters.

We have identified the honeycomb lattice as the target lattice in the self-assembly problem because it the simplest lattice structure that represents a non-trivial case for self-assembly.  The triangular lattice is easily and robustly assembled by a straightforward Lennard-Jones interaction potential.   This fact can be readily demonstrated through direct simulation; however, it is interesting that a rigorous mathematical proof that the triangular lattice is the global energy minimizer and ground state of a particle system with a pairwise Lennard-Jones interaction potential was provided only recently by Theil in \cite{Theil}.  Generating a potential that robustly induces the formation of a square lattice in particle simulations is also a simple matter using the geometric method to be presented later.  Producing the honeycomb lattice using only a radially symmetric potential is more troublesome chiefly because regions of the domain tend to form the triangular lattice -- a lattice that competes strongly with the honeycomb lattice.  This is due to the fact that the triangular and honeycomb lattices have identical distances to nearest neighbors, and the triangular lattice is formed by simply adding a particle to the center of each hexagonal cell in the honeycomb lattice.       

For concreteness, we constrain the parameter space over which we search for parameter values in $V_\textsf{HC}(r;a_{0},a_{1},a_{2},a_{3})$ such that 
\begin{align} 
4.0<a_{0}<8.0, \hspace{1cm} 0<a_{1}<30.0, \hspace{1cm} 0<a_{2}<3.8,\hspace{0.6cm}\text{and}\hspace{0.6cm} 1.25<a_{3}<2.25\,.
\label{eq:searchspace}
\end{align} 

These ranges yield a large class of function shapes over which the search is performed.
\medskip

Rechtsman and co-workers developed two computational algorithms to find potentials that lead to the self-assembly of particles into a given target lattice~\cite{Rechts2005,rechtsman2006a}.  

The first optimization scheme chooses the shape of the potential so that the energy difference between the target lattice configuration and the configuration of the competitor lattices is maximized. The optimization is performed while ensuring mechanical stability by allowing only real phonon frequencies.  This method is purely static in that it seeks to ensure that the target lattice configuration is energetically the most preferred final state;  the method does not incorporate information about the dynamics of the particles as they tend \emph{towards} this final configuration.  

The second optimization scheme considered by Rechtsman \textit{et al}  concentrates on choosing potentials so as to maximize the stability of the target lattice near its melting point, while also requiring stability of the the target lattice with respect to changes in density, and mechanical stability by ensuring that the phonon frequencies are real.  After placing particles into the target lattice configuration, a short molecular dynamics simulation is performed at a fixed temperature just below the melting temperature of the lattice.  The deviation of the the final configuration from the initial target lattice is computed using the Lindemann parameter, 
\begin{equation}
\textsf{LP}:=\sqrt{\frac{1}{N} \sum_i \left( {\bf r}_i - {\bf r}_i^{(0)} \right)^2 -  \left( \frac{1}{N} \sum_i \left( {\bf r}_i - {\bf r}_i^{(0)}  \right) \right)^2          }\,,
\label{LP}
\end{equation}
where  $N$ is the number of particles, and ${\bf r}_i^{(0)}$ and ${\bf r}_i$ are the initial and final positions of particle $i$, respectively. A simulated annealing procedure is used to choose parameters in the expression for $V_\textsf{HC}(r;a_{0},a_{1},a_{2},a_{3})$ that minimize the Lindemann parameter.   

After generating a potential using these methods, Rechtsman \textit{et al} carefully checked using Monte Carlo simulations that particles starting in a random initial configuration do indeed self-assemble into the target lattice. The determination as to whether or not the target lattice was formed, was made by visual inspection of the final configuration and deciding if the configuration has few enough defects to be considered a lattice, as well as checking the long range order by visually inspecting simulated Bragg diffraction patterns.  

\section{Lattice Quality Metrics\label{sec:LatticeQualityMetrics}}

As described in Section \ref{sec:TheSelfAssemblyProblem}, one of the criteria used to compare methods for generating interaction potentials is the quality of the resulting lattices.  In the work of \cite{rechtsman2006a}, a simple visual check of the resulting lattice was used to determine if the desired lattice (with a few defects perhaps) was obtained, and since their purpose was to show that the desired lattices \emph{can} be obtained, this visual check was sufficient.  The purpose of the present paper is to compare several methods according to the lattice quality criterium, thus we must first introduce objective methods for assessing lattice quality.   
 
A prevalent metric for lattice quality is the \emph{structure factor}, a quantity that assesses long range spectral order in the lattice using diffraction patterns.  We have found that this quality metric is inadequate for our purposes, firstly because the structure factor does not provide a scalar value for quality, and secondly because we have observed that long range order is not necessarily strongly correlated with visual perception of lattice quality -- a lattice with glaring defects may still exhibit a high degree of long range order, for example, while a lattice that has weak long range order due to a grain boundary may have excellent local lattice structure.  Hence, we have developed two lattice quality metrics that mimic nearly as possible the visual assessment of lattices. 
When the human eye looks at a lattice and makes a judgement with respect to lattice quality, the emphasis is on order within sub-regions of the entire lattice. A lattice that consists of two perfectly formed sub-lattices that have a domain wall where they meet will be judged by the eye to be quite well-formed. Thus, although long range order is important, a determination of local ordering is crucial for assessing lattice quality in a manner similar to the eye. The metrics that we use to quantify lattice quality have this local feature. 

The two lattice quality metrics we present here are called the \textit{Template Measure} and the \textit{Defect Measure}.  In both cases, a lower value of the metric corresponds to a higher quality lattice.

\subsection{Template Measure\label{sec:TemplateMeasure}}

The \emph{Template Measure} uses a small segment of the target lattice as a template with which to locally compare nearby lattice particle positions for each particle in the given lattice. In the honeycomb lattice, a suitable template may be one hexagonal cell composed of six particles, or points. For each particle in a given lattice configuration, the first point in the template is pinned to the particle, and the template is then rotated to find the best fit to other nearby particles.   As the template is rotated, each point in the template is paired with the nearest particle in the given lattice.  The angle of rotation of the template that produces the least deviation between the template points and lattice particles is considered the best fit.  Once this best fit position of the template has been found for each particle in the lattice, the Template Measure (\textsf{TM}) is obtained by summing the deviation in the positions of the template points and lattice particles from the best fit for each particle in the lattice:
\begin{equation}
\mathsf{TM}
:=\frac{100}{N}\sum_{p=1}^{N}  \min_{\theta}\,\left[\,\,\sum_{i=2}^c \left( {\bf r}_{i,p}^\theta - {\bf r}_{i,p}^{\theta,\text{template}} \right)^2\,+\,n_{p,\text{extra}}\,\right],
\label{TM}
\end{equation}
where  the index $p$ ranges over all $N$ particles in the given lattice, $\theta$ is the angle of rotation of the template, ${\bf r}_{i,p}^{\theta,\text{template}}$ is the position of the $i^{th}$ point in the template when the template is attached to particle $p$ and rotated by angle $\theta$, ${\bf r}_{i,p}^\theta$ is the position of the particle in the given lattice that is closest to ${\bf r}_{i,p}^{\theta,\text{template}}$, and $c$ is the number of points in the template ($c=6$ for the honeycomb cell template).   Notice that since the first point in the template is pinned to the given lattice particle their positions are equal,  that is $\mathbf{r}_{1,p}^\theta = \mathbf{r}_{1,p}^{\theta,\text{template}}$, and hence the sums over the template points need not consider this first template point.   The extra term, $n_{p,\text{extra}}$, is a count of any extra particles in the given lattice that fall inside the hexagonal template, but are not paired with any of the template points.  In this way, the best fit of the template seeks not only to match the particle locations, but also the void within the hexagon.  This `opacity' of the template ensures that defects that arise due to the formation of the triangular lattice that has a particle located at the center of the hexagon will be penalized.    The prepended scaling factor of $100/N$ is not strictly necessary in the Template Measure, but is included for convenience so that the resulting lattice quality values can be interpreted as a measure of the defectiveness \textit{per} particle, and have a magnitude in a range from zero to roughly 100.

\begin{figure}[ht]
\begin{center}
   \subfigure[\footnotesize \label{fig:TMa}]{
        \includegraphics[width=4cm]{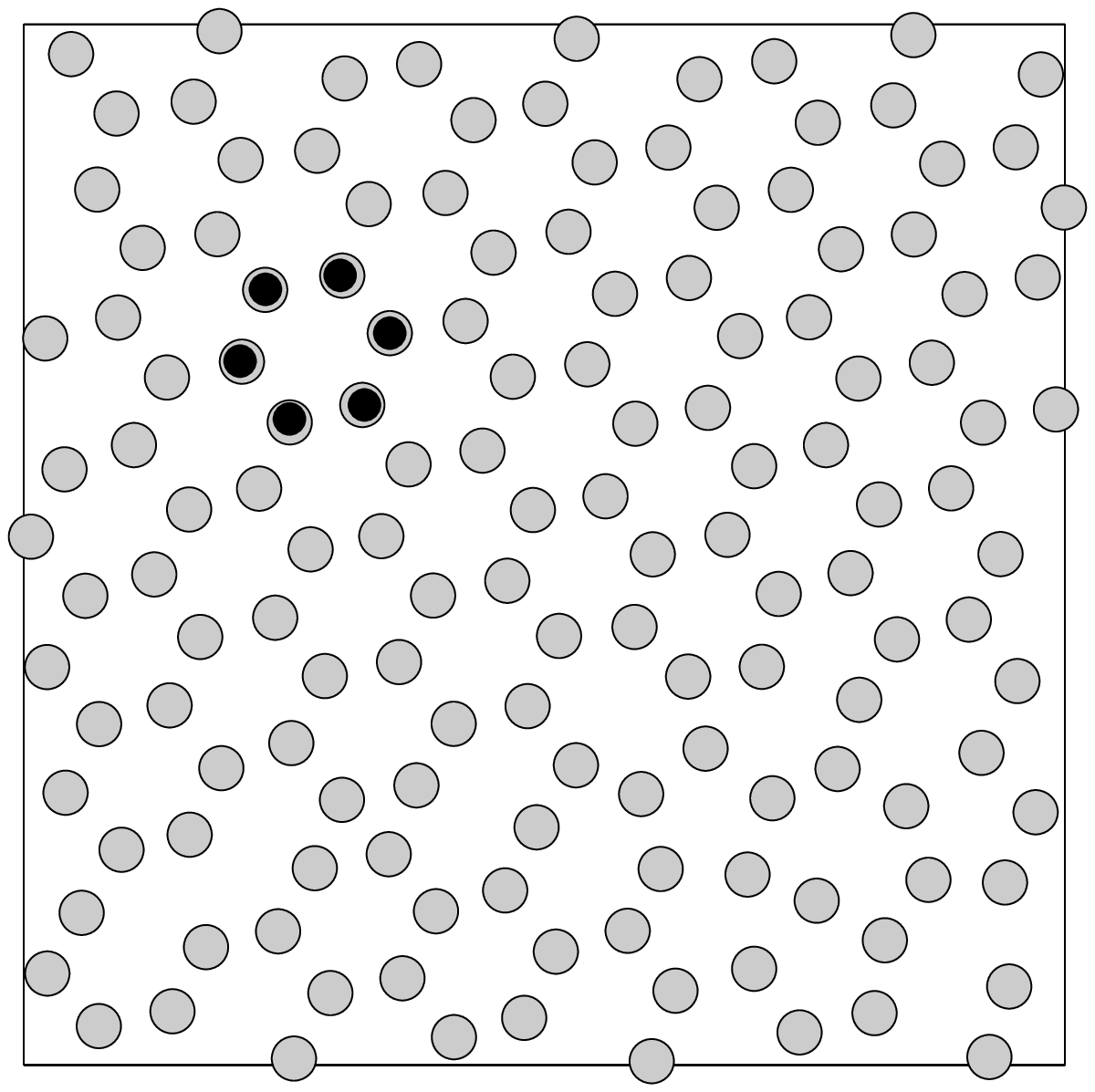}}         
   \subfigure[\footnotesize \label{fig:TMb}]{
        \includegraphics[width=4cm]{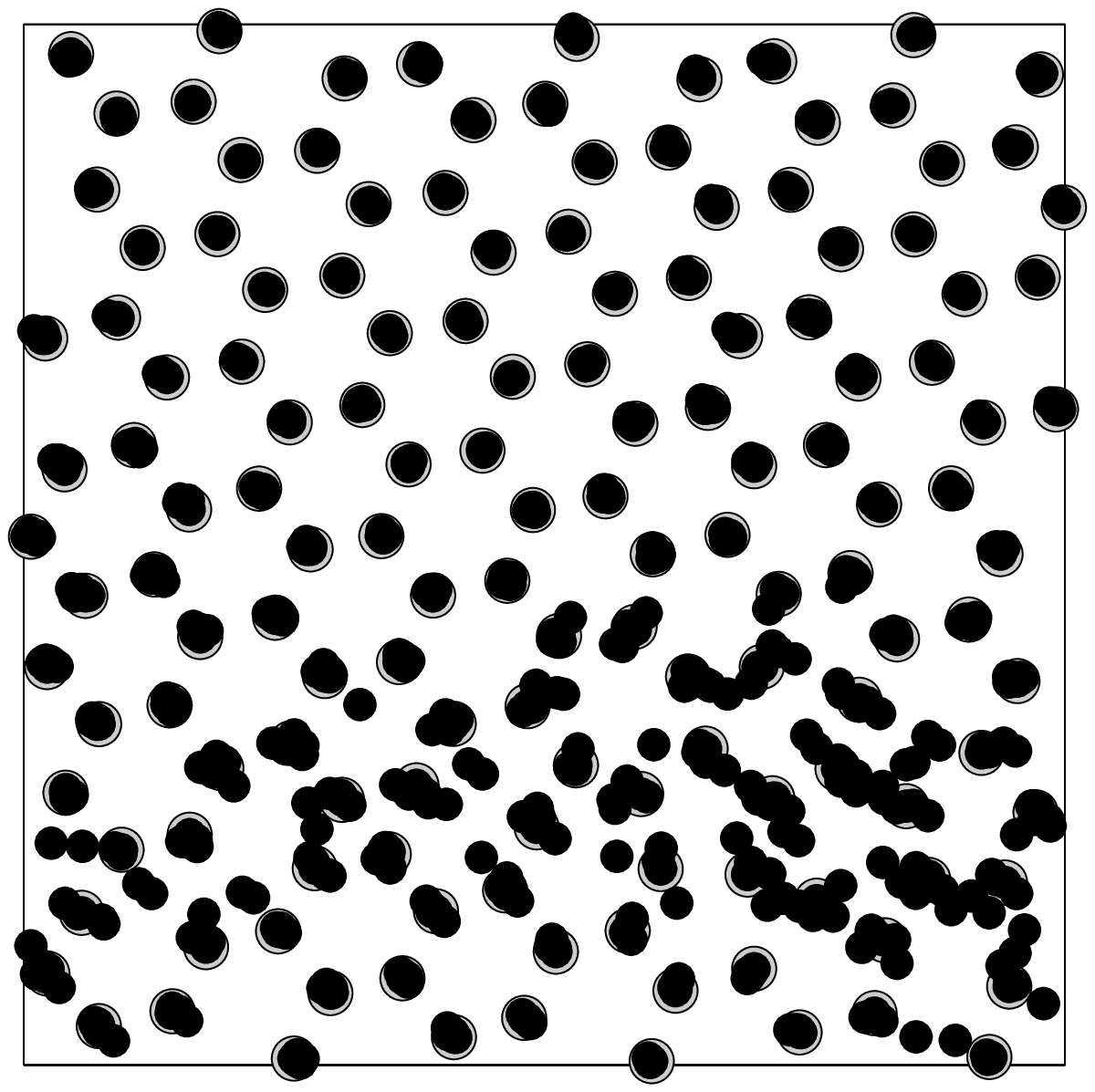}}
  \caption{\label{fig:TM}\footnotesize Illustration of the Template Measure. (a) A template consisting of a single hexagonal cell is pinned to a particle in the given lattice, and rotated to the position which minimizes the distance between points in the template and the nearest particles in the lattice. (b)  Fitting the template cell to each particle in the lattice and rotating to find the best fit, quickly reveals the locations of defects.}
\end{center}
\end{figure}

An illustration of how the Template Measure is implemented in practice is provided in Figure~\ref{fig:TM}.  In Figure \ref{fig:TMa}, a single hexagonal honeycomb cell template is attached to a particle in a honeycomb lattice and rotated until a best fit with the surrounding particles is achieved.  Repeatedly attaching the template to each particle in the lattice and realigning as in Figure \ref{fig:TMb} quickly reveals the locations of the defects.

\subsection{Defect Measure\label{sec:DefectMeasure}}

\begin{figure}[ht]
\begin{center}
  \subfigure{
        \includegraphics[width=3cm]{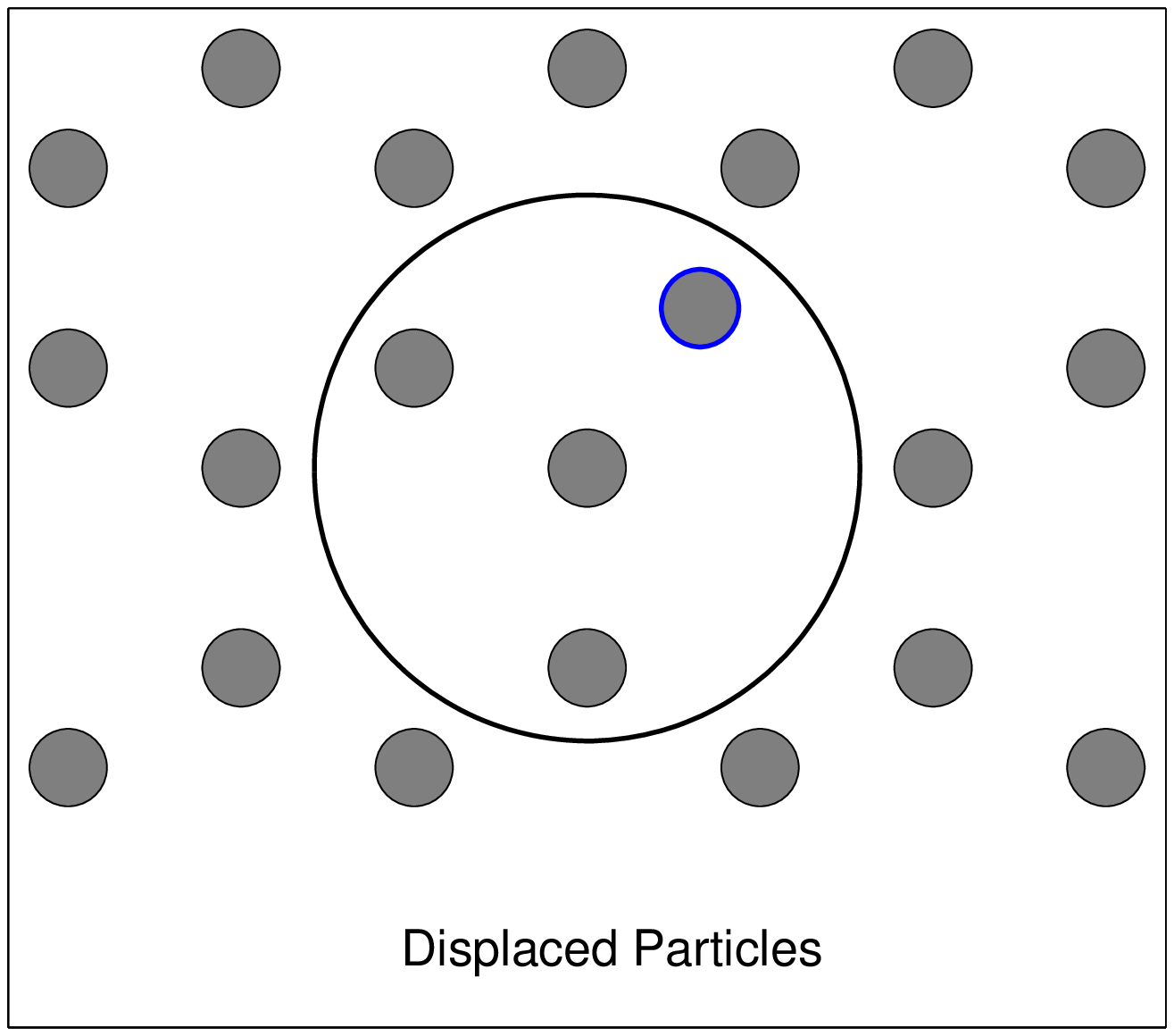}}         
   \subfigure{
        \includegraphics[width=3cm]{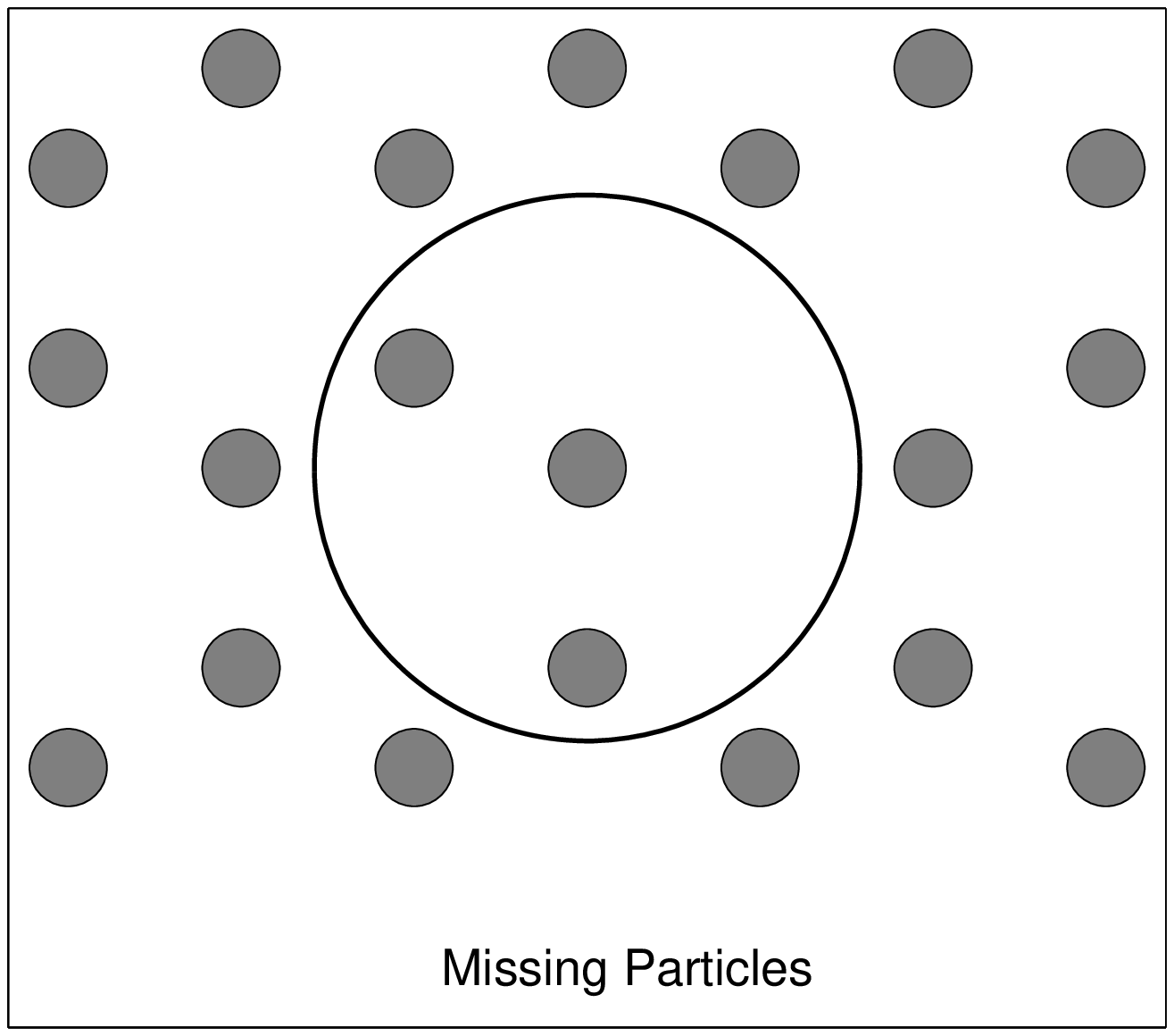}}
           \subfigure{
        \includegraphics[width=3cm]{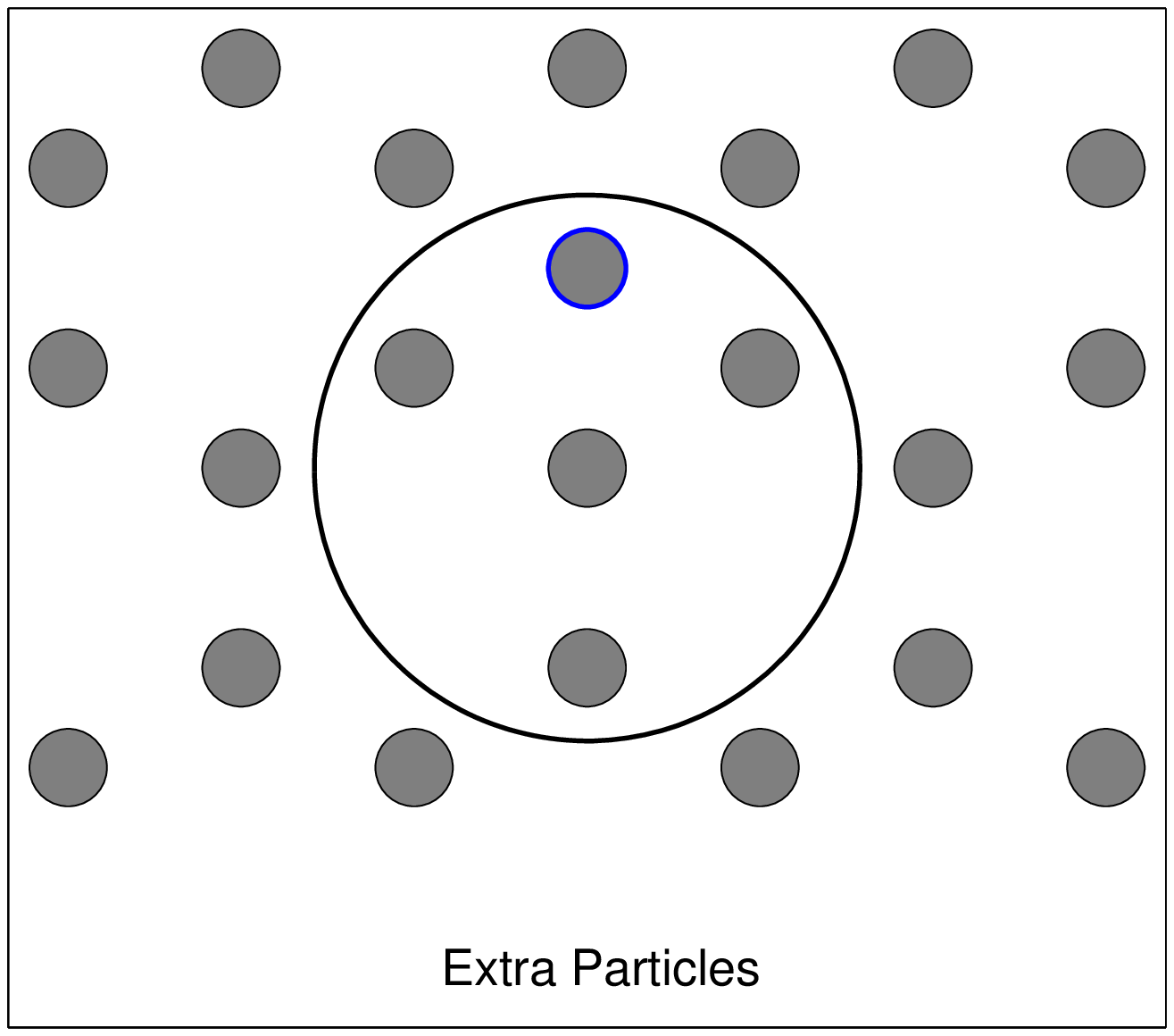}}\\
           \subfigure{
        \includegraphics[width=3cm]{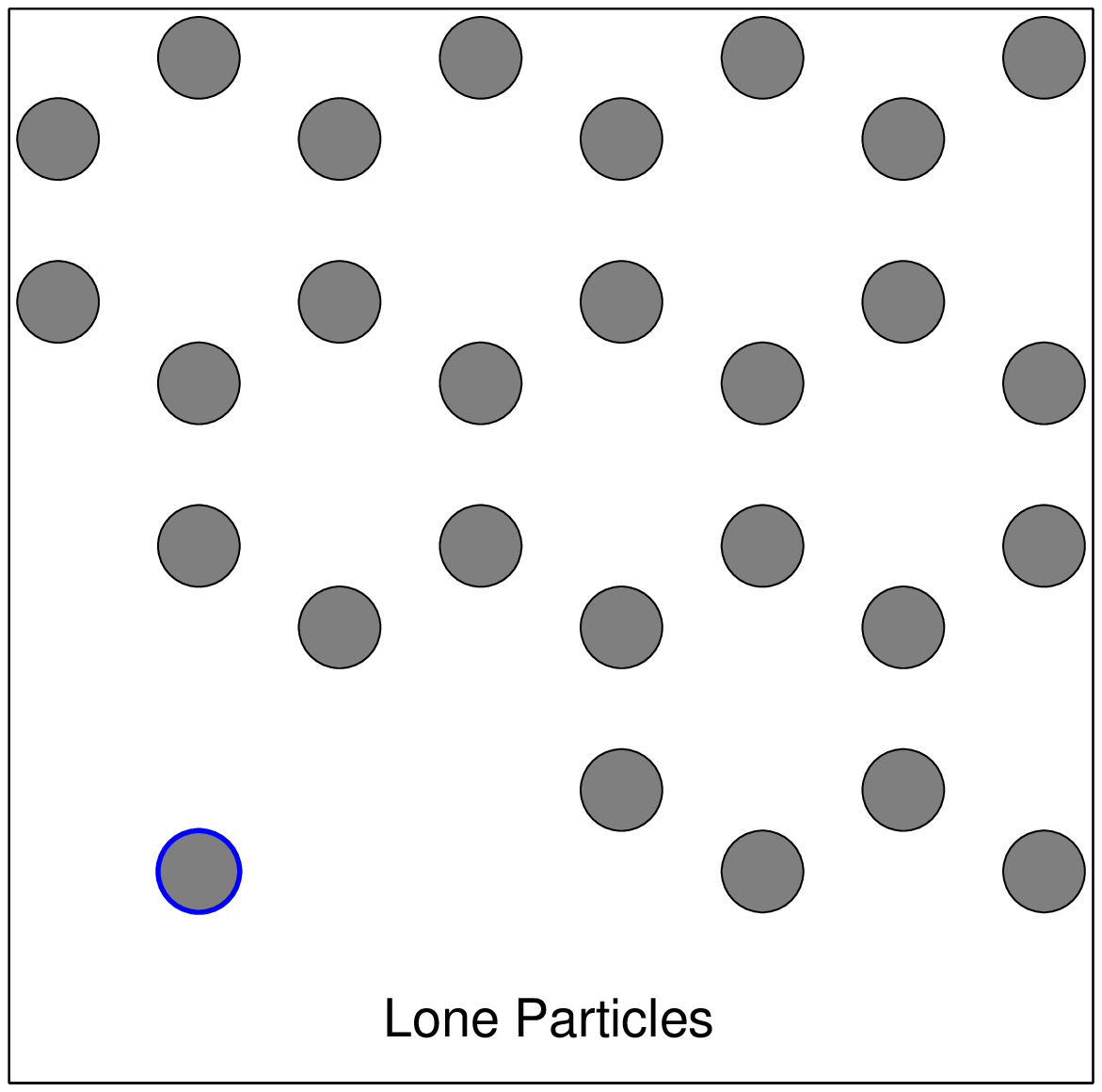}}
           \subfigure{
        \includegraphics[width=3cm]{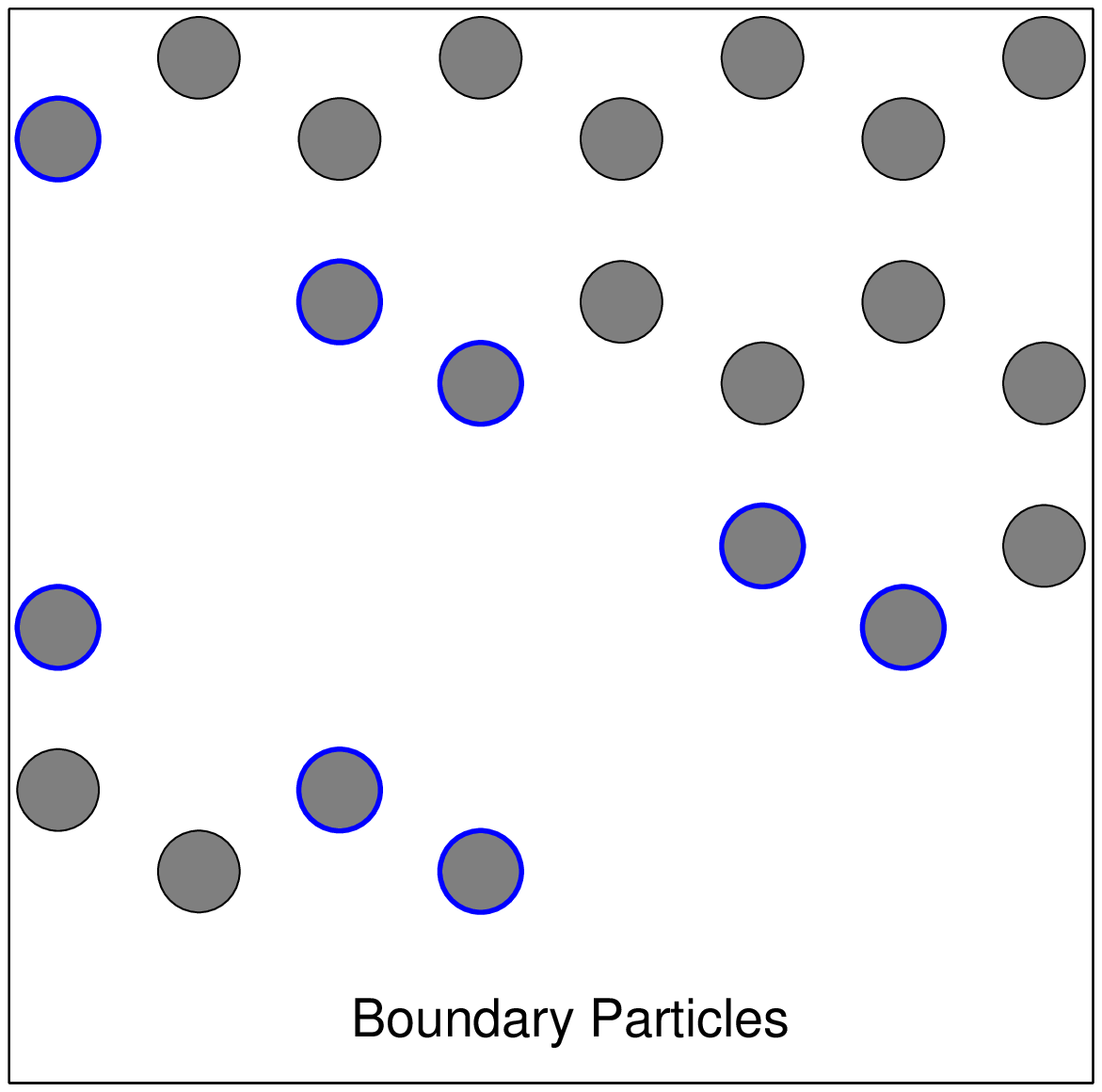}}

  \caption{\label{fig:DM}\footnotesize To compute the Defect Measure, the number of defects in a circular neighborhood of each particle are weighted and summed.  The types of defects used in computing the Defect Measure are displaced, missing, and extra particles, as well as lone and boundary particles.}
\end{center}
\end{figure}

The second lattice quality metric we present is the \emph{Defect Measure}.  The Defect Measure provides a weighted count of all the defects in the local neighborhood of each particle in the given lattice. The types of defects considered are shown in Figure~\ref{fig:DM}, and include displaced, missing, and extra particles. Note that all of the possible defects, including global defects, are taken into account by the types of defects shown. For example, extended grain boundaries are taken into account by contributions to the Defect Measure from locally displaced, missing, and extra particles.

Recall that in the perfect honeycomb lattice, the distance to nearest neighbors is unity, while the distance to second nearest neighbors is $\sqrt{3}$.  In order to calculate the Defect Measure, we consider only a small circular region around each particle of radius $(1+\sqrt{3})/2 \approx 1.366$ so that only three nearest neighbors, each at unit distance, should be included.  Then, using the locations of all particles actually located within this circular region for the given lattice, the contributions from each of the defects is calculated, and then summed with the specified weighting, for each particle in the given lattice to provide a measure of the quality of the lattice as a whole. The weights attributed to each type of defect may be chosen according to the desired properties for the self-assembled lattice. These weights may emphasize correct local densities, correct alignment of particles, correct inter-particle distances, or any other customized weighting. 

Accordingly, the Defect Measure (\textsf{DM}) is given by 
\begin{eqnarray*}
\mathsf{DM}
& := &\frac{100}{N}\sum_{p=1}^{N}\Bigg[ \omega_\text{displaced}  \left( \sum_{j} \chi(d_{pj}) \cdot (d_{pj}-1.0)^2 \right) \\
& & \\ 
& & + \; \omega_\text{missing}\cdot  n_{p,\text{missing}} + \omega_\text{extra}\cdot  n_{p,\text{extra}} +   \omega_\text{lone}\cdot\eta_{p,\text{lone}}  +  \omega_\text{boundary}\cdot\eta_{p,\text{boundary}}   \Bigg]
\label{eq:DM}
\end{eqnarray*}
where the $\omega$'s are the weights attributed to each type of defect; the $n_{p}$'s are the integer number of missing and extra particles within the small circular region surrounding particle $p$; the $\eta_{p}$'s are indicator functions that equal unity if particle $p$ is a lone or boundary particle and zero otherwise; the index $p$ ranges over all the $N$ particles in the given lattice; the index $j$ ranges over the particles contained within the small circular region surrounding particle $p$; and $d_{pj}$ is the positive distance between particle $p$ and particle $j$.  As with the Template Measure, the scaling factor of $100/N$ is included so that the quality values are normalized by the number of particles and fall within a convenient range. In the first term, $\chi(\cdot)$ is a smooth cutoff function that decreases from 1 to 0 at the outer edges of the circular region, or more precisely, as its argument increases from  1.275 to 1.366.  This smooth cutoff function ensures that the Defect Measure remains continuous with respect to motion of the particles and as the number of particles entering the circular region of each particle fluctuates.  

More explanation of the Defect Measure and its applications can be found in \cite{Grubits}, where there is also a discussion of other lattice quality metrics.  

\medskip

For the purposes of this paper, we use the following weight values for the Defect Measure:
\begin{equation*} 
\omega_\text{displaced} = 1.0\,, \hspace{0.5cm}
 \omega_\text{missing} =  1.5\,,  \hspace{0.5cm}
 \omega_\text{extra} = 0.8\,,  \hspace{0.5cm}
  \omega_\text{lone} = 2.0\,,  \hspace{0.5cm}
   \omega_\text{boundary} = 0.1\,.
\end{equation*} 

\begin{figure}[ht]

\begin{center}
   \subfigure[\footnotesize TM = 131.6\hspace{2cm} DM = 162.8\label{fig:random}]{
        \includegraphics[width=3.8cm]{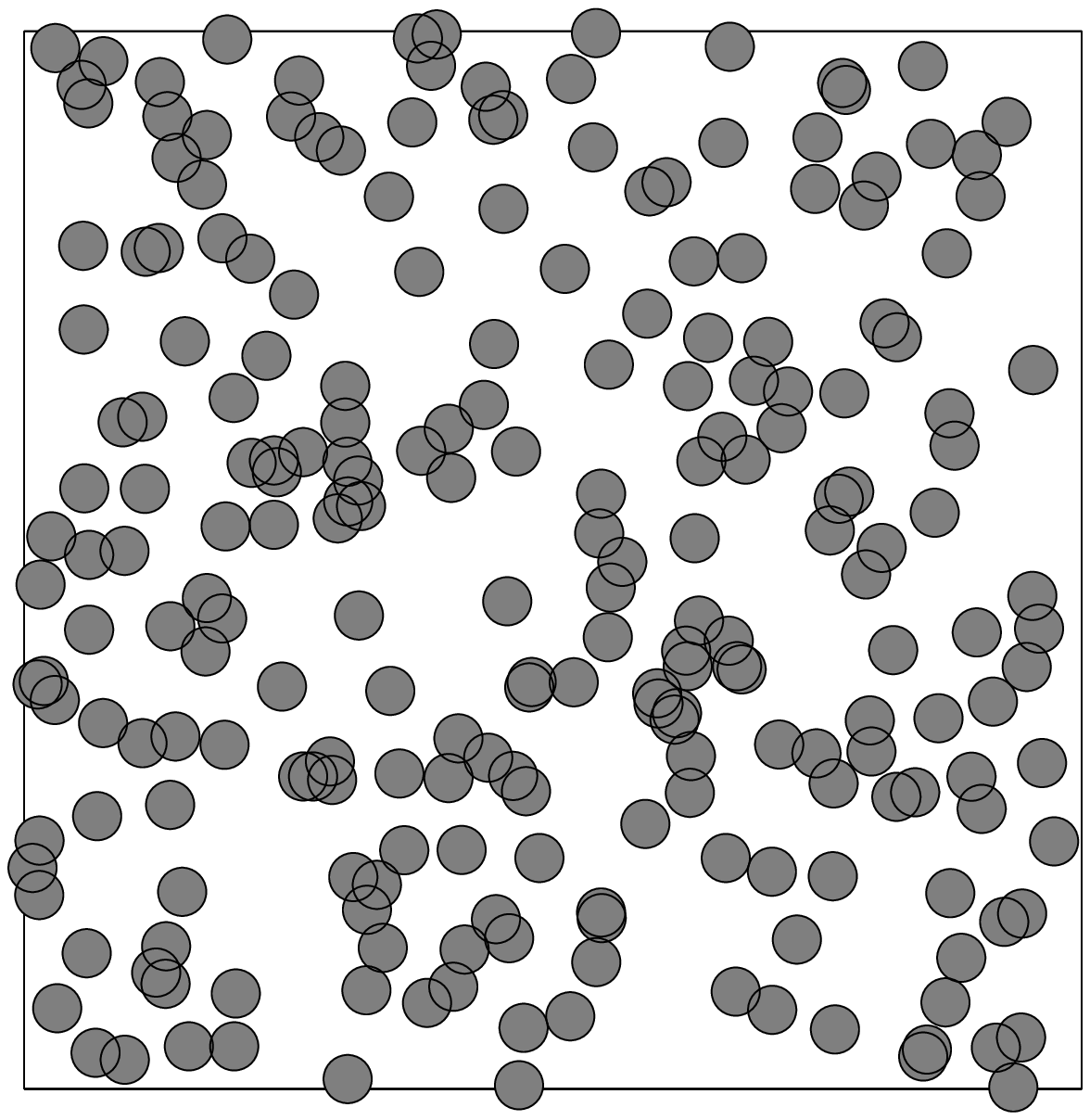}}        
        \hspace{0.01cm} 
   \subfigure[\footnotesize TM = 63.6\hspace{2cm} DM = 44.2\label{fig:latB2}]{
        \includegraphics[width=3.8cm]{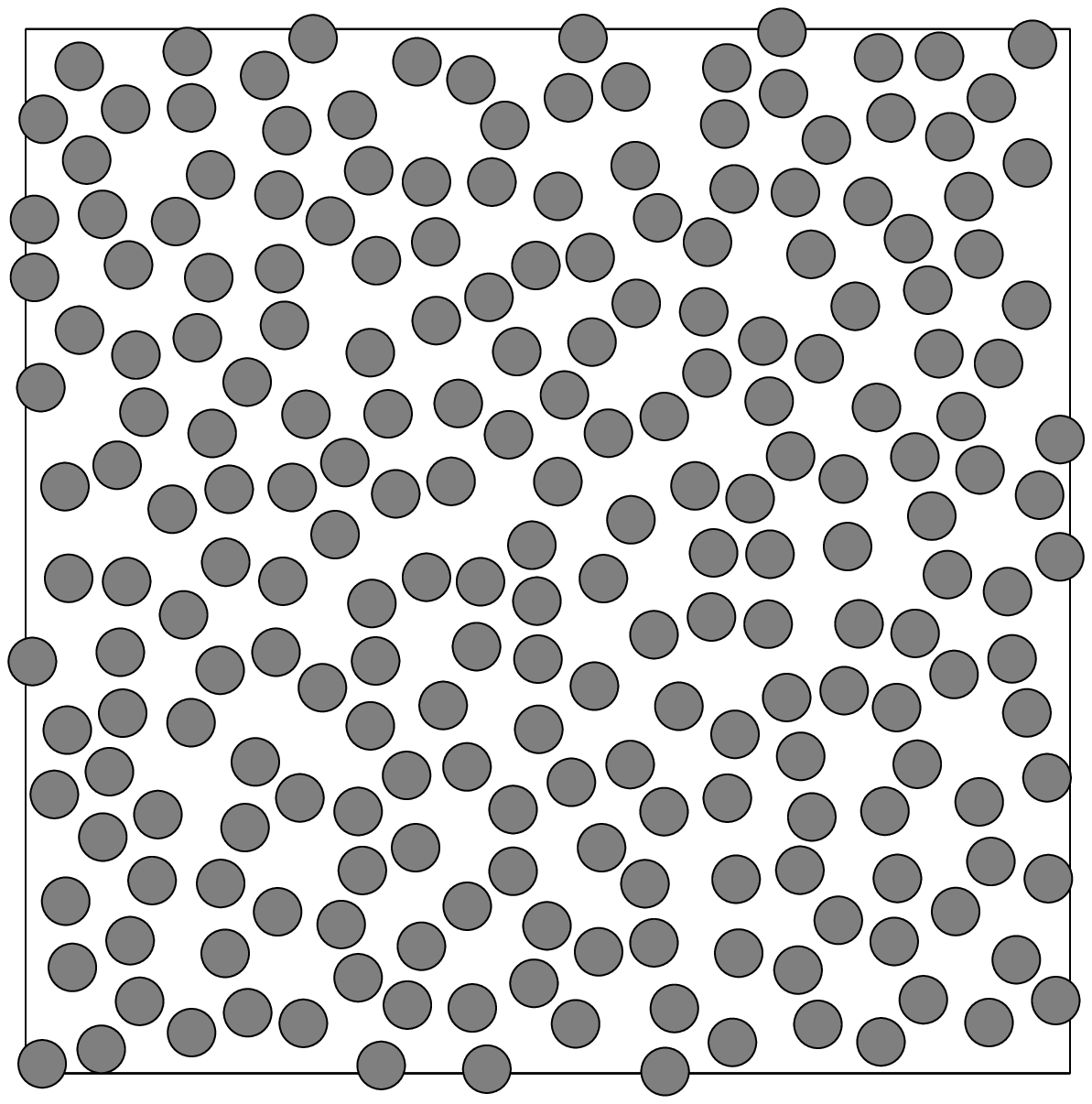}}        
        \hspace{0.01cm} 
  \subfigure[\footnotesize TM = 33.9 \hspace{2cm} DM = 17.5\label{fig:latB45}]{
        \includegraphics[width=3.8cm]{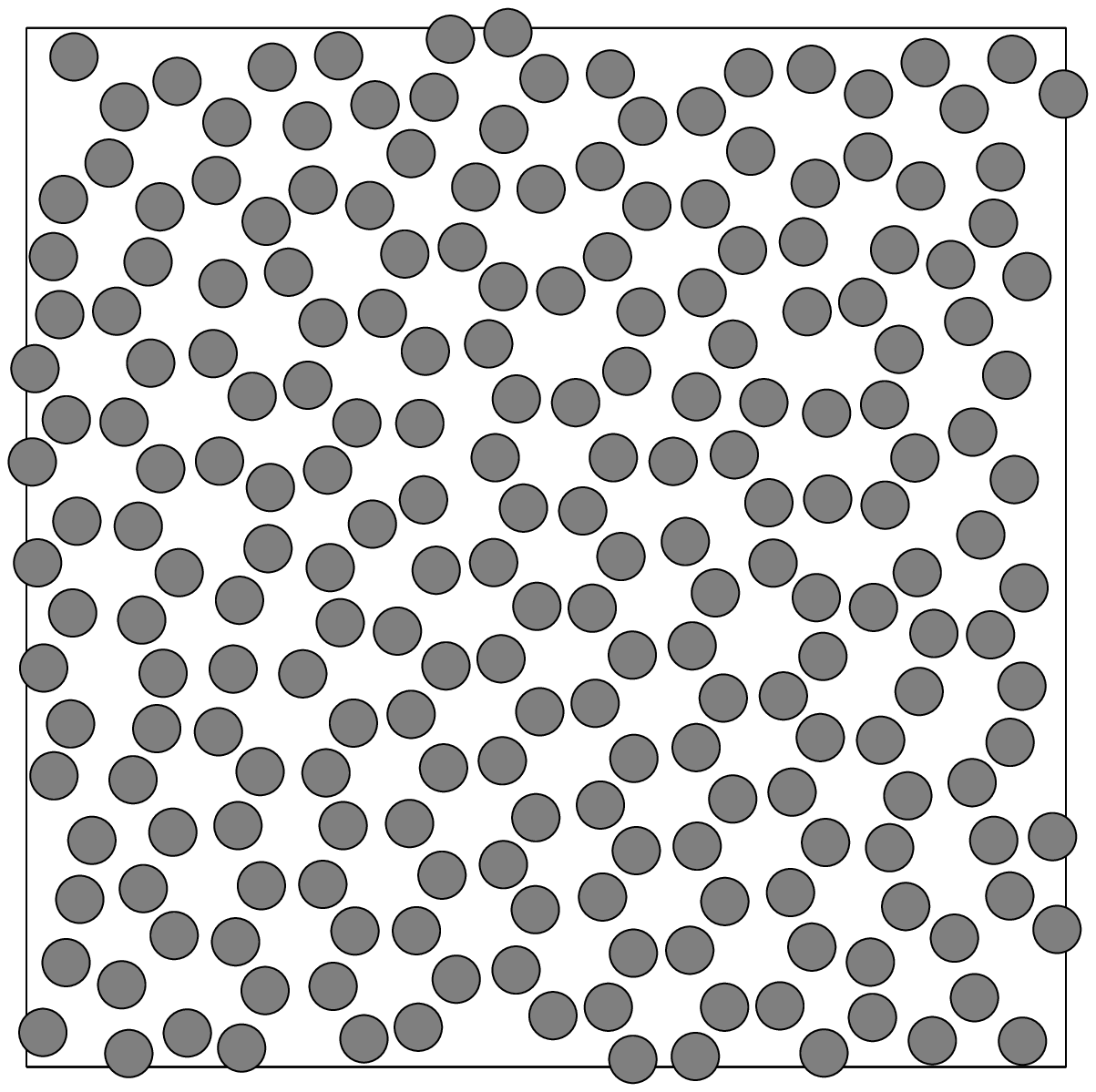}} \\
    \subfigure[\footnotesize TM = 19.9 \hspace{2cm} DM = 12.4\label{fig:latB51}]{
        \includegraphics[width=3.8cm]{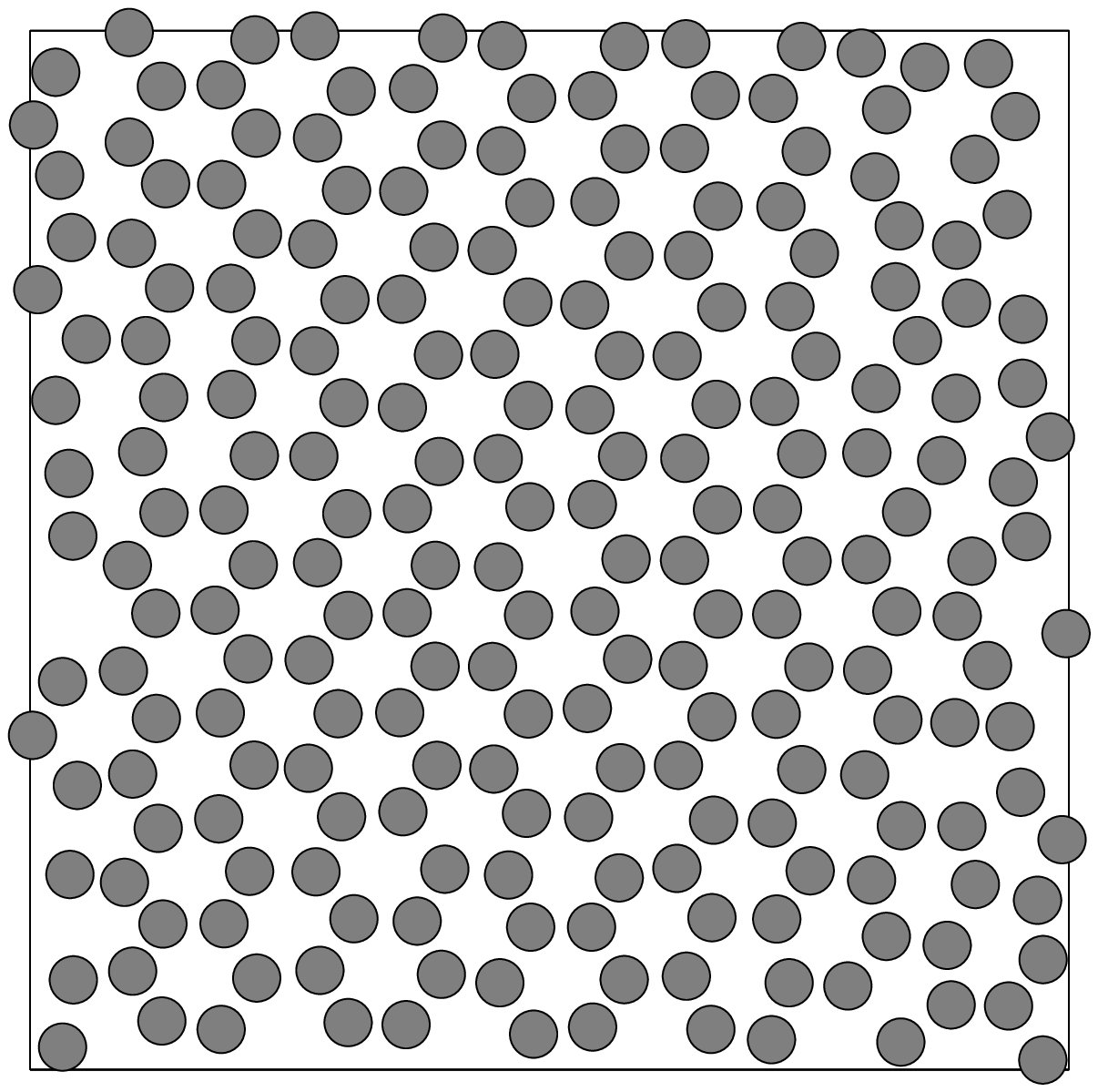}}    
            \hspace{0.01cm} 
    \subfigure[\footnotesize TM = 10.2 \hspace{2cm} DM = 4.3\label{fig:latB101}]{
        \includegraphics[width=3.8cm]{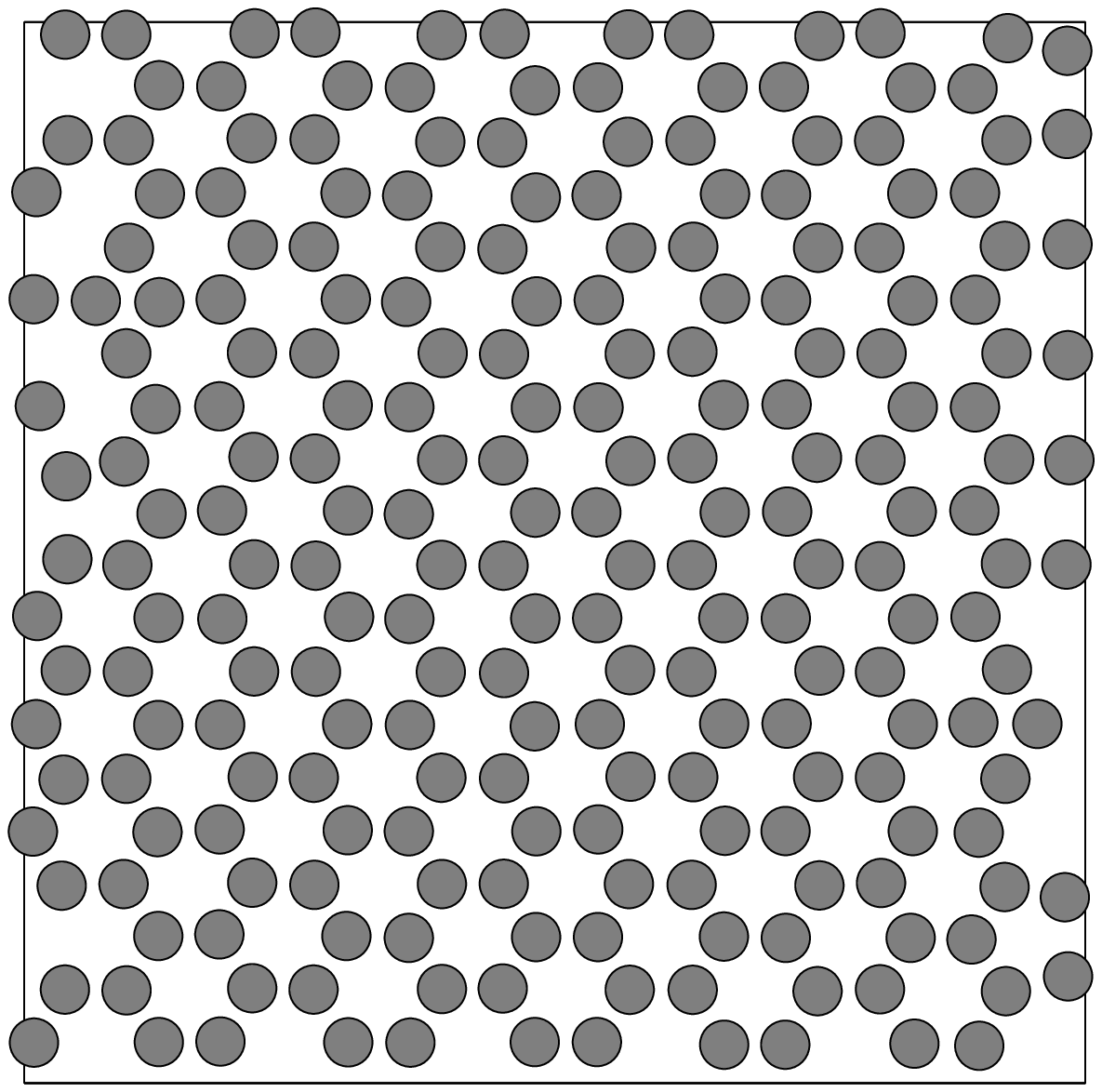}}      
          \caption{\footnotesize Sample values of the lattice quality metrics are shown for a range of lattices.  In each figure, the value of the Template Measure (TM) and the Defect Measure (DM) are provided.  Both measures assign a lower value to a lattice of higher quality.\label{fig:lat}}
\end{center}
\end{figure}

Sample values of both the Template Measure and the Defect Measure for four lattices of varying quality are shown in Figure~\ref{fig:lat}.  

\section{Objective Functions\label{sec:ObjectiveFunctions}}

A natural way in which to find solutions to the self-assembly problem is to search for a potential that optimizes an appropriate objective function.   The objective function must be well-chosen so that 1) optimizing the objective correlates well with the formation of the desired lattice, and 2) the objective function is not prohibitively expensive to evaluate.  In this section, we present the objective functions that will be used in the potential generation methods under comparison.

\subsection{The Lindemann Parameter Objective Function\label{sec:TheLindemannParameter}}

As noted in Section~\ref{sec:TheSelfAssemblyProblem}, the Lindemann parameter is a measure of how much a lattice configuration has deviated from the perfect lattice configuration after a short simulation.  Consequently, the Lindemann parameter cannot be used as a metric for lattice quality \textit{per se} since it measures how far a lattice has deviated when initially placed in the perfect target lattice configuration, whereas a lattice quality metric must assess the quality of a lattice obtained after a slow cooling process from an \emph{arbitrary} initial condition.  The Lindemann parameter can nevertheless be used as an objective function that must be minimized in order to find potentials that lead to formation of the target lattice -- an approach that was first proposed and developed by \cite{rechtsman2006a}.  Intuitively, minimizing the Lindemann parameter produces potentials that stabilize the honeycomb lattice to thermal agitation.  Used as an objective function, the Lindemann parameter is advantageous since it is much faster to evaluate than other objective functions that require much longer molecular dynamics simulations.

As implemeted in this paper, the Lindemann parameter objective function is computed for a given interaction potential by placing 72 particles in the honeyomb lattice formation, and then performing a brief simulation at a temperature very near the melting temperature of the lattice. The use of 72 particles allows for the construction of a lattice consisting of 30 honeycomb cells that form an infinite honeycomb lattice when the 72-particle configuration is used to tile the plane.  The value of the Lindemann parameter is calculated using the initial and final lattice configurations.  The wall clock time to compute the Lindemann parameter on a single CPU is approximately 2 seconds.

\medskip

Optimizing the Lindemann parameter is, however, an \emph{indirect} method in that the quanitity being optimized is not the quantity that will be used to determine the final quality of the potential. A more direct approach is to explicitly optimize the lattice quality metrics. Although the quality metrics require a slow cooling simulation and are consequently more expensive to evaluate, optimization of the quality metrics guarantees optimization of lattice quality. Indeed, an important observation made in this paper is that optimizing the Lindemann parameter is only moderately correlated with lattice quality -- potentials can be found that produce a low value of the Lindemann parameter, yet when tested in a slow cooling simulation produce lattices of poor quality.   

\subsection{Quality Metric Objective Functions\label{sec:QualityMetrics}}

We also consider direct evaluation of the Template Measure and Defect Measure quality metrics as objective functions.  To evaluate these objective functions, we start with a regular Cartesian grid of 64 particles with a spacing that provides a particle density equal to the density of the honeycomb lattice.  These particles are initialized with a temperature well above the lattice melting point and then this temperature is slowly reduced until the particles freeze into a lattice configuration. The quality metrics are computed using only the final lattice configuration. Compared with the computation of the Lindemann parameter, these cooling simulations are expensive to carry out; a single call to a quality metric objective function on a single CPU takes approximately 70 seconds.  

\medskip

It should be noted that the computational expense associated with evaluation of the objective functions arises almost entirely from the molecular dynamics simulations.  Computation of the actual value of the Lindemann parameter, the Template measure, or the Defect measure after the final configuration has been obtained, requires less than a tenth of a second.

\subsection{Properties of the Objective Functions\label{sec:PropertiesoftheObjectiveFunctions}}

The objective functions presented have important features that influence the effectiveness of the optimization schemes employed.  One of the chief contributions of this paper is the use of a trend optimization scheme that is better suited to these objective functions over a standard simulated annealing optimization procedure.

The fact that the objective functions require evaluation times on the order of seconds and minutes implies that any optimization method that requires many thousands of evaluations to search the four-dimensional parameter space will necessitate a computation time measured in hours and days.  The time taken to run an optimization algorithm will be spent almost entirely evaluating the objective functions, and overhead computation required by the optimization scheme in choosing the next location at which to evaluate the objective, for example, is practically negligible in comparison.  The trend optimization method we propose is particularly well-suited for this situation in which the objective functions are expensive to evaluate.

\begin{figure}[ht]
\begin{center}
   \subfigure{
        \includegraphics[width=3.8cm]{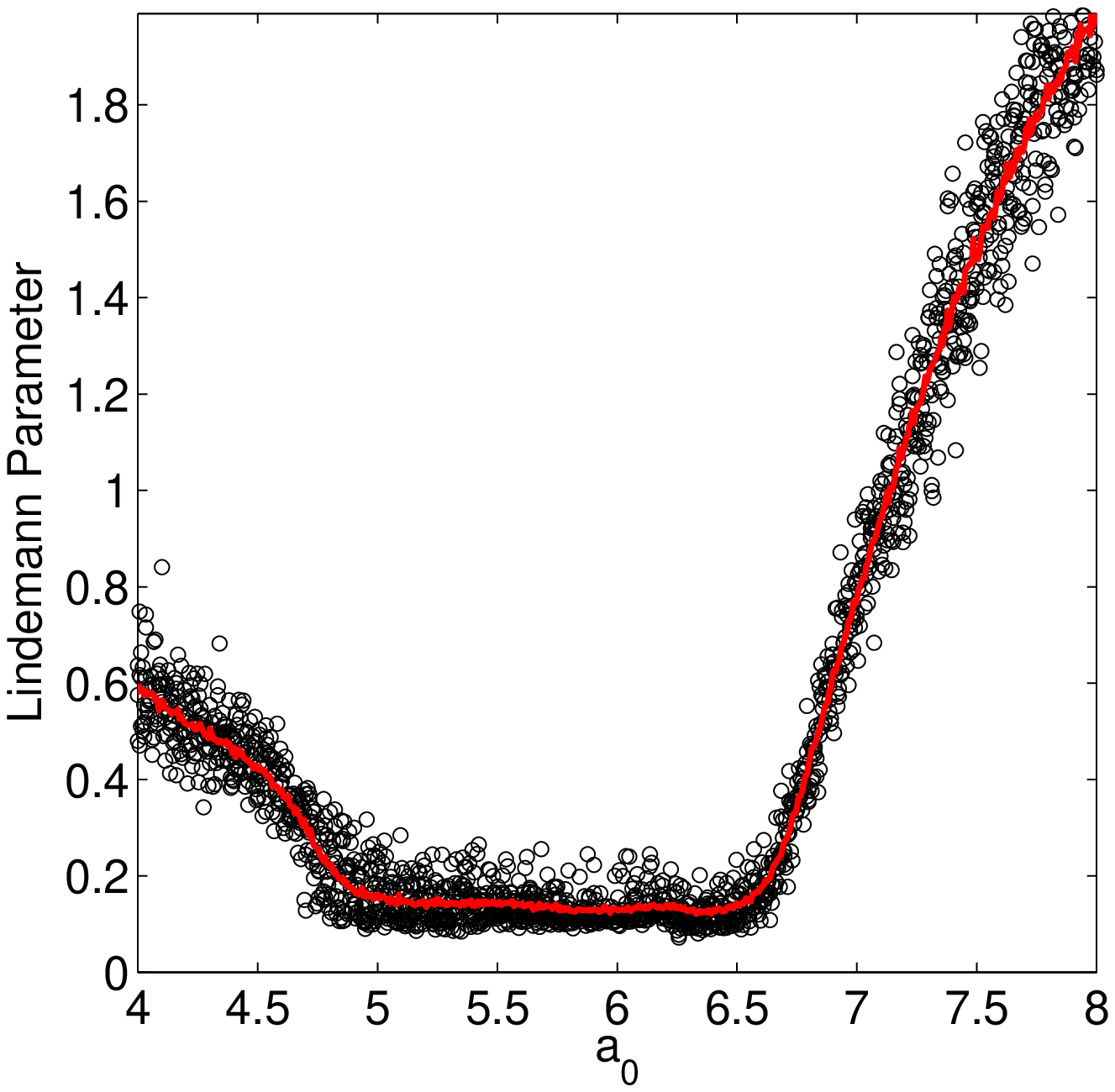}}         
        \hspace{0.1cm}
   \subfigure{
        \includegraphics[width=3.8cm]{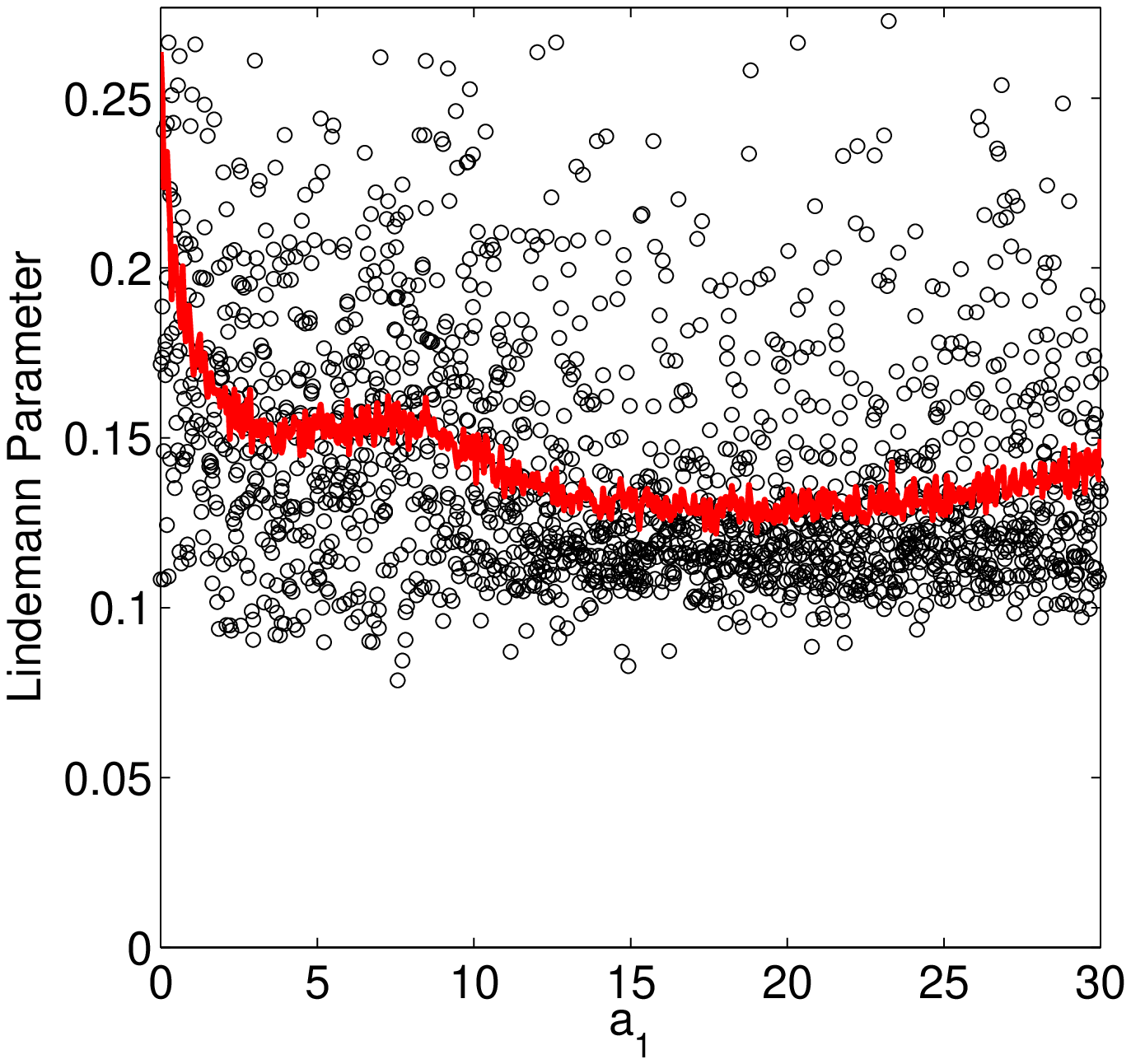}}                    
   \subfigure{
        \includegraphics[width=3.8cm]{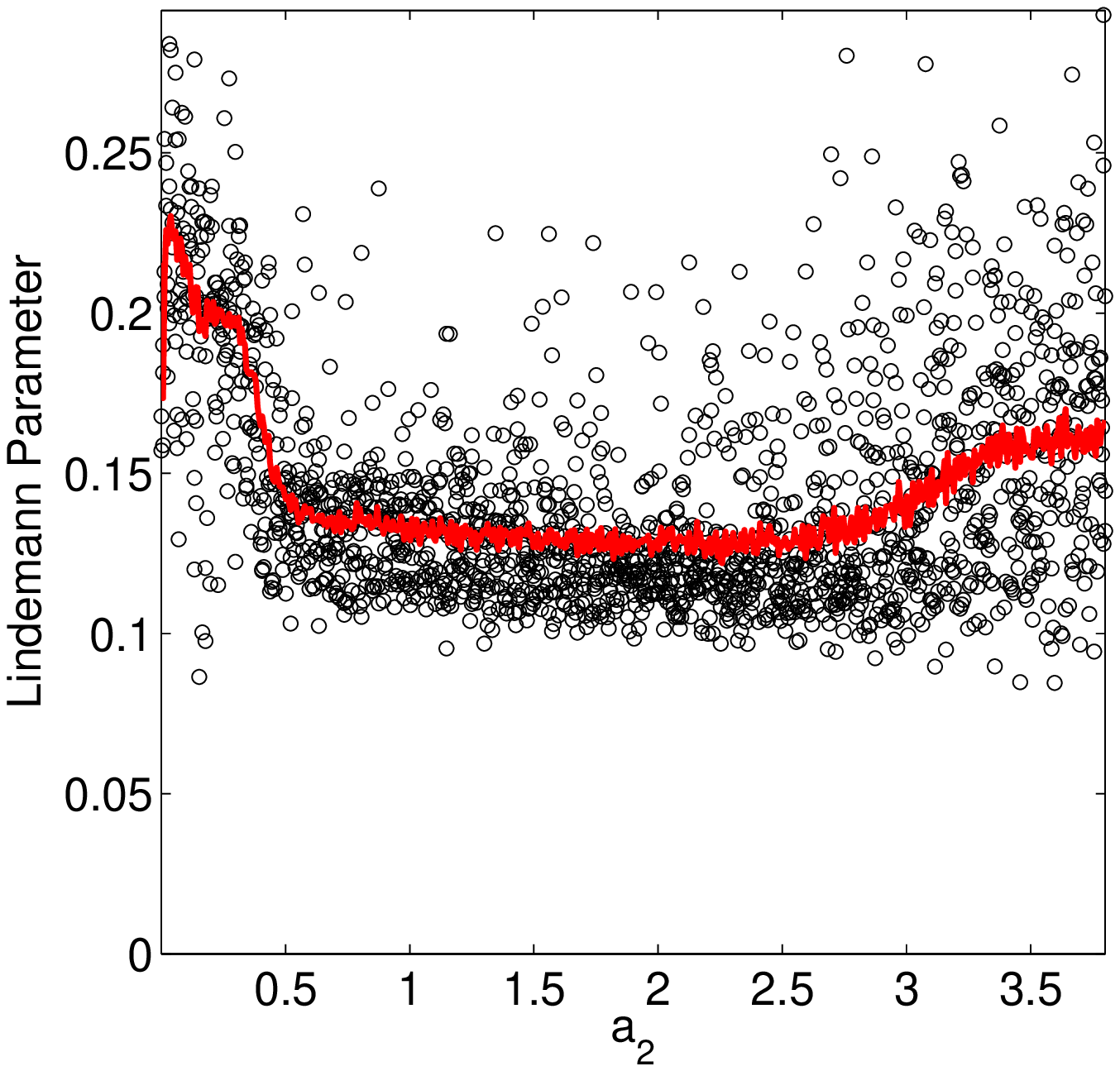}}     
         \hspace{0.1cm}
    \subfigure{
        \includegraphics[width=3.8cm]{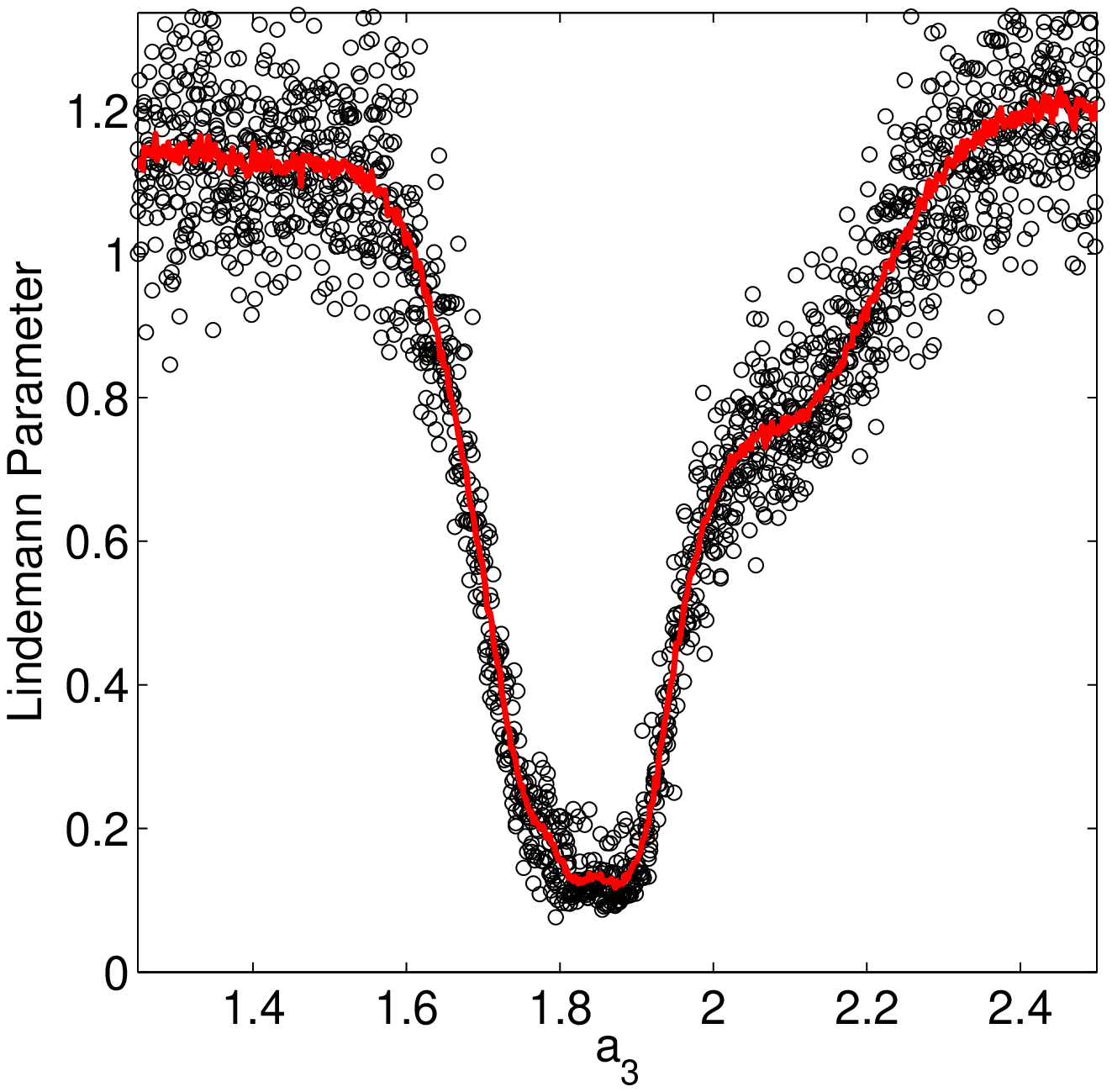}}       
   \end{center}                     
\caption{\label{fig:noisescan}\footnotesize Each circle in the figures above corresponds to a sample evaluation of the the Lindemann parameter objective function.  The parameter shown on the $x$-axis is varied over the indicated domain, while the remaining parameters are held fixed at the values provided by Rechstman \textit{et al} ($a_0$=5.89, $a_1$=17.9, $a_2$=2.49, and $a_3$=1.823).  The vertical axis in each plot represents the value of the Lindemann parameter. Note that the vertical scale in each case is different.  The red line is computed by averaging over 100 trials at each location, and reveals a smooth trend in the data.
}
\end{figure}

Figure \ref{fig:noisescan} displays repeated evaluations of the Lindemann parameter objective function as each of the parameters in $V_\textsf{HC}(r;a_{0}, a_{1}, a_{2}, a_{3})$ is varied in turn, while the remaining parameters are held fixed at the values provided by Rechstman \textit{et al} (namely, $a_0$=5.89, $a_1$=17.9, $a_2$=2.49, and $a_3$=1.823).    Each circle in the plots corresponds to a single evaluation of the the Lindemann parameter objective function.  The red line in each plot represents a trend line that is computed by averaging over 100 samples taken at each of 600 regularly spaced intervals along the axis of the varied parameter.  

Our first observation is that due to randomness induced by the initial conditions and simulation at constant temperature, the objective function does not produce repeatable values for a fixed set of parameter values.  Repeated evaluation of the objective functions for the same parameter values leads to a wide range of output values.   For this reason, the objective functions are not strictly \emph{functions}, although we continue to use this term.  More correctly, due to the randomness introduced by the initial conditions and the simulation with a thermostat, we can think of the objective functions as assigning a one dimensional probability distribution to each set of values that define the potential.  Thus, a call to the objective function returns a value drawn from the probability distribution specified by the parameters.   We shall see that the trend optimization method exploits the fact that the expectation values of the probability distributions vary smoothly with respect to changes in the parameters of the potential.

Examining the evaluations of the objective functions reveals that they are noisy with respect to variations in the parameters.  The functions are not smooth and certainly no derivatives of the objectives are available.  Nevertheless, averaging over many evaluations for a fixed set of parameter values reveals that the objective functions do possess a smooth and slowly varying trend.  As indicated in Figure \ref{fig:noisescan}, the smoothness of the average values (shown in red) indicate that, when averaged, the objective functions do admit a sensible notion of a minimum.  \\

In summary, the salient features of the objective functions in the self-assembly problem are the following:
\begin{enumerate}
\item They are expensive to evaluate;
\item They are highly variable---repeated evaluations of the objective functions for the same input potential yields a broad range of values;
\item They are highly non-smooth with respect to changes in the parameters;
\item They have a smooth trend when averaged over many evaluations.
\end{enumerate}

Objectives with these features are often encountered in optimal design problems in which evaluation of the objective function for a specific set of parameter values requires completion of an actual laboratory experiment, or the execution of a computationally expensive simulation using random initial conditions~\cite{booker1996a,myers1995a}.  Optimization of these objective functions is clearly not tractable using standard gradient-based methods.  In this paper, we implement an optimization scheme ideally suited for objective functions of this type that has allowed us to construct an efficient procedure for solution of the self-assembly problem, an approach that we call \emph{trend optimization}. 

\section{Optimization Methods\label{sec:OptimizationMethods}}

In the previous section, various objective functions have been introduced that can be used in the search for lattice-forming potentials.  In this section, we present two methods for finding optimal values of these objectives.  First, we briefly discuss simulated annealing which was used by \cite{rechtsman2006a} in the baseline approach.  Second, we present in detail the trend optimization approach.

\subsection{Simulated Annealing\label{sec:SimulatedAnnealing}}

Simulated annealing mimics the ability of a thermal process to search a configuration space to find the ground state.  We begin the search by evaluating the objective at an initial point, $x_{0}$, in the parameter space and denote the value of the objective function at this location by $E_{0}$.  The search then proceeds by jumping to new locations in a random walk.  The probability, $p$, of making the jump from $x_{i}$ to a new location $x_{i+1}$ is determined by
\begin{align} 
p = \begin{cases} 
1&\text{if } E_{i+1}  \leq  E_i\,,\\ 
e^{-(E_{i+1} - E_i)/(T)}&\text{if }E_{i+1}  >  E_{i}\,.\end{cases}
\end{align} 

The small probability of moving to a location where the objective is in fact higher than the current value, allows for the search to escape from a local minimum.   In analogy with true simulated annealing in metals, the temperature, $T$, is slowly reduced so that at first the search eagerly traverses the search space by easily escaping from local minima, and then settles to a specific local minimum as the temperature decreases.  This simulated annealing optimization scheme was implemented using the GNU Scientific Library \texttt{GSL\_SIMAN\_SOLVE(\,)} routine~\cite{galassi2009a}.  

When applied to the self-assembly problem, the simple simulated annealing method described here fails to converge to a sensible minimum because of noise in the objective functions.  Understandably, the method gets stuck in ephemeral local minima that appear and disappear due to noise in the objectives.  Convergence can be obtained if the objectives are sufficiently smoothed.   Smoothing necessitates averaging over at least 20 independent simulations and thus requires added computational expense, and even then a large proportion of optimizations fail to converge within a reasonable number of objective evaluations. 

Admittedly, the simulated annealing method described here represents a very simple approach, and more complex methods involving adaptive search, for example, could be employed.  The straightforward simulated annealing approach serves, therefore, as a modest but easily understood baseline method.  We have investigated the simulated annealing method using many different cooling regimens including adaptive methods, and have found little improvement in each case.

\subsection{Trend Optimization\label{TrendOptimization}}

Trend optimization is well-suited for problems in which the objective functions are expensive to evaluate, derivatives of the objectives are not available, and the objectives are noisy yet exhibit a simple underlying trend when averaged over many evaluations. 

In the trend approach, we use a few well-distributed evaluations of the objective function to generate a smooth global approximation to the averaged objective function.  This smooth approximating surface, referred to as the \textit{surrogate}, attempts to capture the underlying smooth trend in the noisy objective evaluations.  The optimization then proceeds by finding the optimal values of this surrogate function that is computationally cheap to evaluate.  Trends in the surrogate quickly reveal regions of the parameter space in which the optimal parameters are most likely to be found and hence the search is greatly accelerated.   The insight provided by the smooth surrogate function then informs the choice of parameters at which subsequent evaluations of the objective should be made.

Booker \textit{et al}~\cite{booker1999a} combined the speed and facility with which the trend approach zooms in on regions of optimal parameter values with the rigorous convergence guarantees of patterned search methods developed previously by Torczon~\cite{torczon1997a}, to produce what is known in the literature as the \textit{Surrogate Management Framework}.  In this two-pronged approach, the trend method is used in the \textit{global search} step of the patterned search method to accelerate the search, while the use of a \textit{polling} step on a patterned conceptual grid provides the guarantees of convergence.  In this seminal paper, Booker \textit{et al} also applied the Surrogate Management Framework approach to the optimal design of a helicopter rotor blade with thirty-one design variables.  More recently, Audet \textit{et al}~\cite{audet2003a,audet2006a} have provided a generalization of Torczon's pattern search method, which they refer to as Mesh Adaptive Direct Search, and have applied it to optimization of the chemical treatment of discarded potliners to minimize release of toxic waste in the production of aluminum~\cite{audet2008a}.   Mesh Adaptive Direct Search has since been incorporated into the Surrogate Management Framework by Marsden \textit{et al} in \cite{marsden2008a}.

The range of design problems to which the trend optimization approach has been applied is starting to grow.  Marsden \textit{et al} have used the Surrogate Management Framework in the optimal design of airfoils to reduce noise generated in the trailing turbulent flow~\cite{marsden2007a}, and a computational framework has been provided in \cite{marsden2008a} for optimizing design of surgeries for improved blood flow and cardiovascular geometry.  Siah \textit{et al}, have used Kriging surrogate models to design optimal configurations and shapes of automobile antennae to minimize electromagnetic coupling~\cite{siah2004a}, and Raza \textit{et al} have compared methods for generating surrogate functions in a design problem seeking the optimal arrangement of fuel rods in a liquid metal reactor~\cite{raza2007a}.

A unifying theme in all these applications is the parsimonious way in which trend optimization is able to optimize expensive, noisy objective functions.   Moreover, trend optimization is robust to noise in that the trends approximate the general shape of the objective function with smooth surfaces that quieten the noise and anomalous evaluations  of the objective function. Hence, trend-based approaches survey the parameter landscape for large depressions, and are not distracted by superficial deep spikes that may arise due to noise.  Because the surrogate prioritizes regions that consistently perform well, rather than a single ``flash-in-the-pan'' evaluation, the parameter values returned by the trend are robust to uncertainties and are more likely to reliably reproduce near-optimal values of the objective upon repeated evaluation.  

Trend optimization can be performed in a coarse-to-fine hierarchical manner by recursively building a hierarchy of trend-fitting surfaces. Each successive iteration of the procedure yields a new trend that focuses on the most optimal region of the search space.  Successive trend surfaces utilize all objective function evaluations obtained in previous iterations to more accurately model the objective function.  After an initial global trend is developed, the search area is refined to the area surrounding the global minimum of the surrogate  -- recall that since the surrogate is smooth and cheap to evaluate, the global minimum of the surrogate can easily be found.  Refinement of the search area helps to ensure that subsequent function evaluations are chosen in locations that are most relevant and promising.  As the search area becomes more refined, successive iterations may use a larger basis of fitting functions, or use more sophisticated trend construction methods to more accurately pinpoint the location of the minimum.  These features of iterative trend optimization yield a hierarchy of coarse to fine trends that enable the method to initially make large strides toward the optimal value, and then to focus ever more tightly on the exact location of the optimal value.

In our discussion thus far, it remains to be described how the surrogate functions are generated from a small number of function evaluations.  This topic lies within the province of \textit{data approximation} and fills a large body of literature.  Needless to say, there are a great number of interpolation and fitting approaches available.  Popular methods include polynomial interpolation, splines, Kriging, distance-based interpolation, linear and nonlinear regressions, radial basis functions, neural networks, and kernel-based approaches~\cite{cressie,giunta1998a,simpson1998a}. For a thorough survey, please see the monograph by Hastie~\cite{hastie2001a}.  

Reviewing the Lindemann parameter objective evaluations depicted in Figure~\ref{fig:noisescan}, we plainly see that an interpolating scheme is not appropriate for the noisy objective functions of the self-assembly problem.  For this reason, we have chosen the \textit{ridge regression} method for generating surrogate functions that is particularly well-suited for noisy data in high dimensions.  Ridge regression was originally developed by Tikhonov (hence the method is sometimes referred to as Tikhonov regularization) for ill-conditioned linear regression problems~\cite{tikhonov1977a}.   His approach was to introduce diagonal stabilization to regularize the linear interpolation system.  The added regularization improves the conditioning, and introduces smoothing.     

\begin{figure}[ht]
\begin{center}
   \subfigure{
        \includegraphics[width=3.8cm]{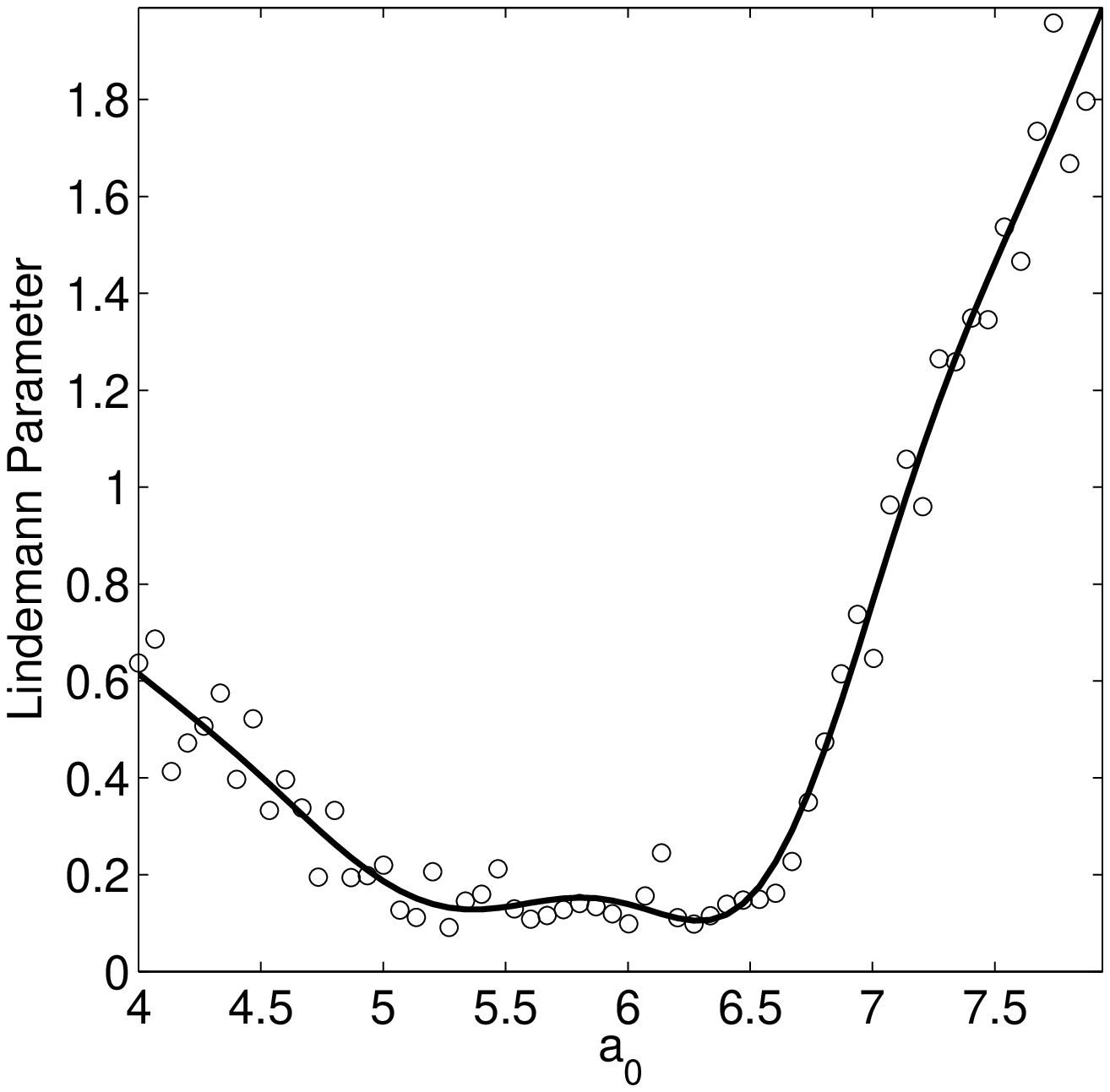}}         
        \hspace{0.1cm}
   \subfigure{
        \includegraphics[width=3.8cm]{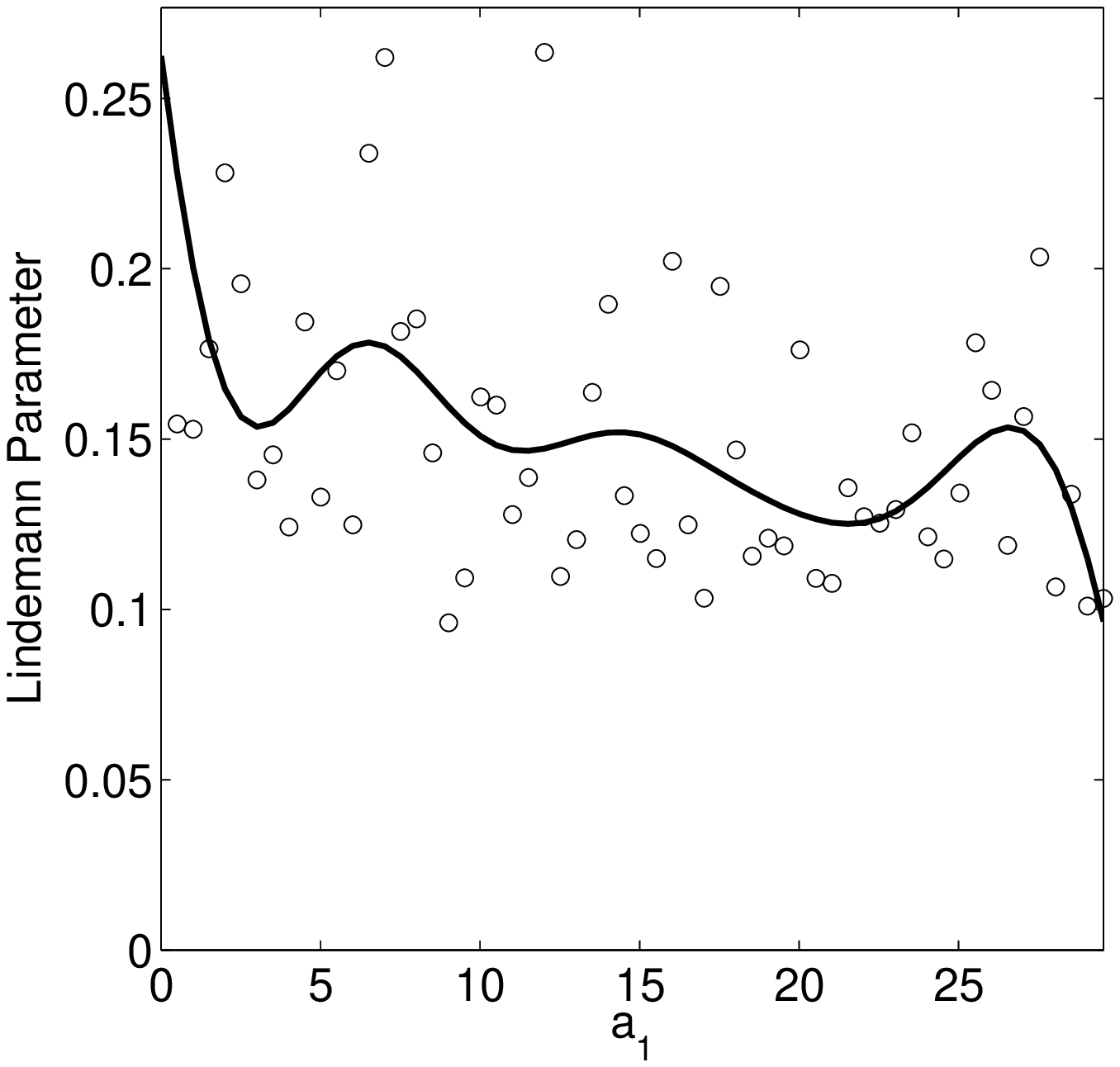}}                    
   \subfigure{
        \includegraphics[width=3.8cm]{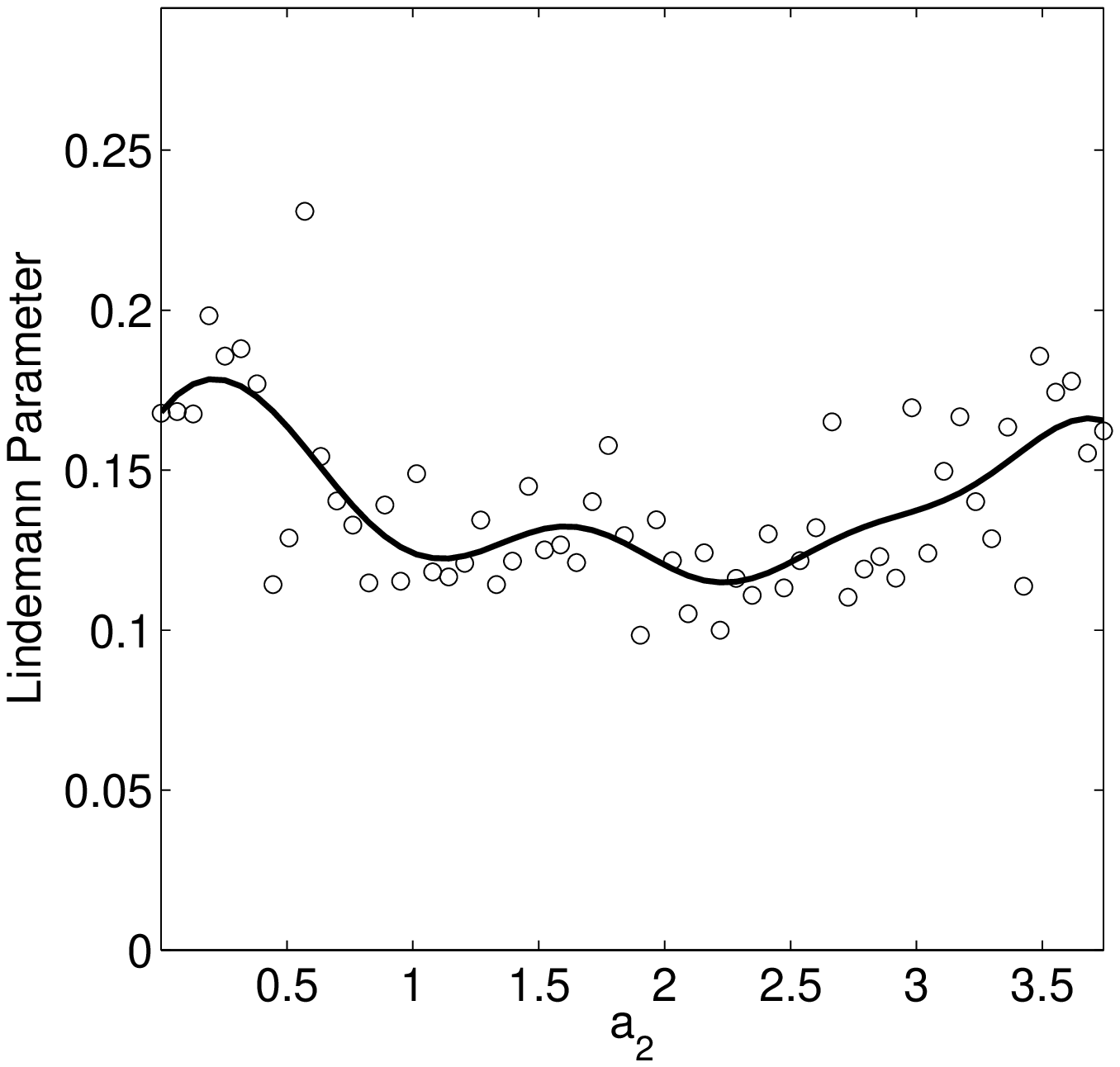}}     
         \hspace{0.1cm}
    \subfigure{
        \includegraphics[width=3.8cm]{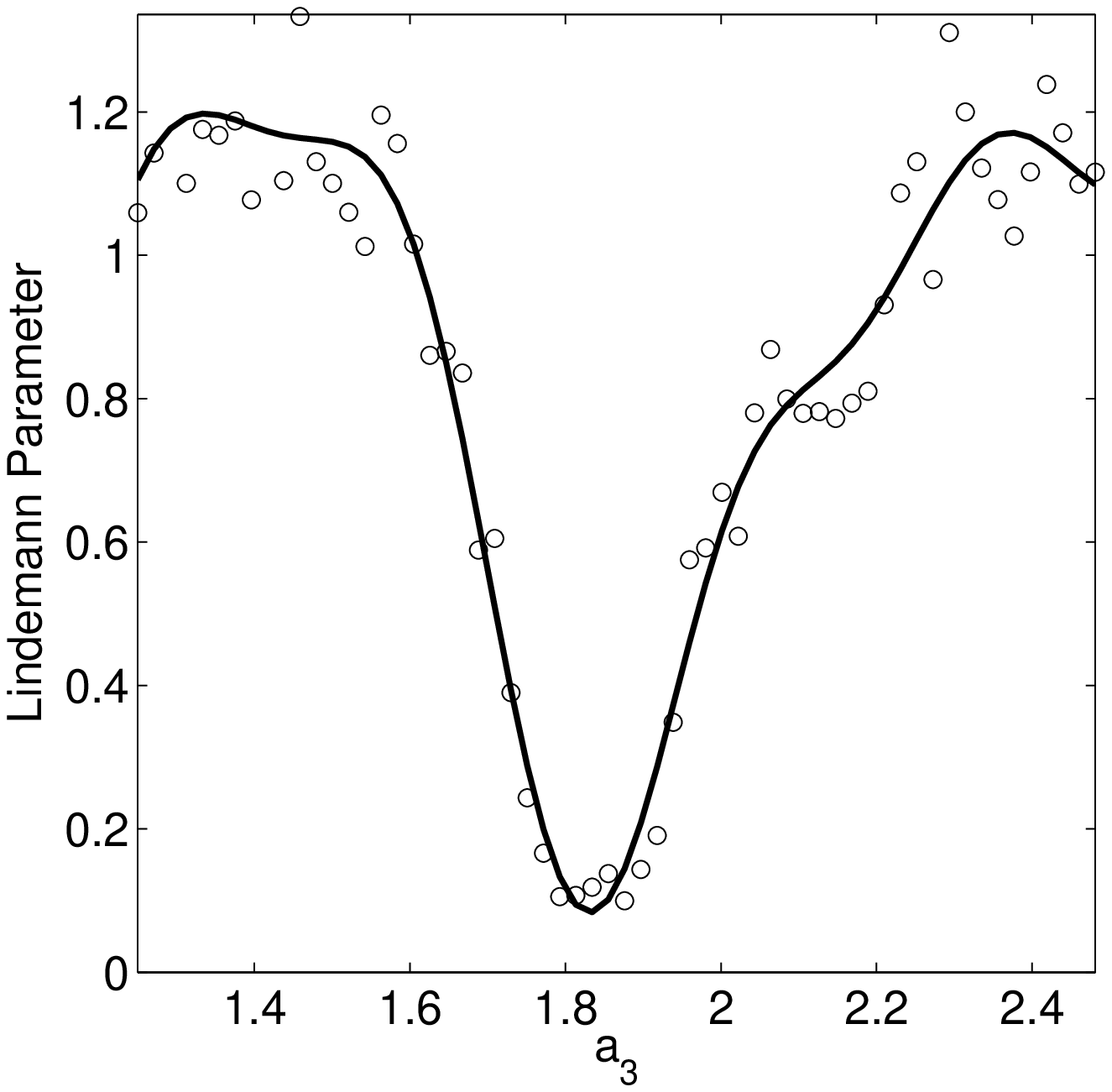}}       
   \end{center}                     
\caption{\label{fig:ridge}\footnotesize Generating surrogate functions for noisy data.  Each circle represents an evaluation  of the Lindemann parameter as in Figure \ref{fig:noisescan}.  The solid trend line is generated using Gaussian radial basis functions and ridge regression. 
}
\end{figure}

In the current context, we are provided with a vector of $M$ noisy measurements, $\mathbf{y} = [y_{1}, \cdots,y_{M}]$, of the objective function at the vector of locations $\mathbf{x}=[x_{1}, \cdots,x_{M}]$. We want to find a smooth function that best represents the smooth trend in this data.  This trend, denoted $T(x)$, is constructed as the weighted sum of basis functions:
\begin{align*} 
T(x) = \sum_{k=1}^{M}c_{k}\Phi(x,x_{k}) \,,
\end{align*} 
where we must now solve for the vector of coefficients $\mathbf{c} = [c_{1},\cdots,c_{M}]$.  Gaussian radial basis functions are chosen for the basis since they are well-suited for regression problems in high dimensions~\cite{dyn1986a,fasshauer2007a,gutmann2001a}.  To be specific, we choose radial basis functions of the form
\begin{align*} 
\Phi(x,x_{j}) =  \phi(\|x-x_{j}\|_{2})
\end{align*} 
where
\begin{align*}
\phi(r)=e^{-(\epsilon r)^{2}}\,,\hspace{0.5cm}r\in \mathbb{R}\,.
\end{align*}
Proceeding in the standard manner for linear regression, the vector of coefficients, $\mathbf{c}$, in the expression for the trend are obtained by solving the linear system
\begin{align} 
\mathbf{A}\mathbf{c}  = \mathbf{y}  
\label{eq:noreg}
\end{align} 
where the elements of the square symmetric matrix $\mathbf{A}$ are given by 
\begin{align*} 
\mathbf{A}_{ij} := \Phi(x_{i},x_{j})\,. 
\end{align*} 
This procedure assumes that $\mathbf{A}$ is full rank.
In ridge regression, the conditioning of $\mathbf{A}$ is improved by adding diagonal regularization.  The linear system in \eqref{eq:noreg} above is replaced by
\begin{align} 
\label{eq:reg}
\left( \mathbf{A} +  \frac{1}{2\omega}\mathbf{I}\right) \mathbf{c} = \mathbf{y}   
\end{align} 
in which $\omega \in \mathbb{R}$, and $ \mathbf{I}$ is the identity matrix.  Regularization is obtained at the expense of introducing the new free parameter $\omega$.  In the traditional usage of ridge regression, the analyst must choose an optimal value of $\omega$ that balances the need for improved conditioning with the desire to keep the departure from the least-squares solution small.  In our context, since we are not immediately concerned with conditioning, we use $\omega$ to control the amount of smoothing introduced.  Smaller values of $\omega$ yield larger smoothing, while larger values of $\omega$ ensure increased pointwise accuracy to the noisy data.  In practice, this approach is straightforward and robustly produces smooth trends to noisy objective functions.  In Figure~\ref{fig:ridge}, an illustration of smooth trends generated using ridge regression are given for noisy evaluations of the Lindemann parameter.  It must be remembered though that these surrogates are constructed as one-dimensional curves for illustration, whereas in the full self-assembly problem the surrogate functions are smooth four-dimensional hypersurfaces.

The ridge regression method can be extended to include adaptive control of the amount of smoothness introduced in response to local conditions.   This \textit{local ridge regression} approach is implemented by replacing Equation~\eqref{eq:reg} with
\begin{align*} 
\left( \mathbf{A} + \textbf{diag}\left[\frac{1}{2\omega_{1}},\cdots,\frac{1}{2\omega_{M}}\right]\right) \mathbf{c} = \mathbf{y}   
\end{align*} 
where the vector of $M$ free parameters $[\omega_{1},\cdots,\omega_{M}]$ can be chosen independently to adjust the amount of smoothing local to each measurement.  In this way, a low value of $\omega$ can be chosen to locally increase the smoothing in a region in which the noise in the data is high, and similarly, a large value of $\omega$ can be chosen for measurements in regions in which there is high confidence in the data and noise is low.  This facility is not currently implemented in our trend optimization algorithm, but is mentioned here to indicate the flexibility that the ridge regression method affords.

As previously mentioned, there are many candidate approaches for trend fitting besides ridge regression that could be uses in the self-assembly problem.  For instance, simple quadratic fitting is very easily implemented and guarantees convexity of the surrogate.  In practice, this method also works remarkably well for the objectives of the self-assembly problem.

\subsubsection*{Trend Optimization Algorithm used in the Self-Assembly Problem}

In using the trend optimization approach for the design of the potentials in the self-assembly problem, we have implemented it without coupling to a Direct Search method.  Consequently, we lose rigorous guarantees of convergence, but in practice trend optimization alone performs remarkably well and reliably converges to an optimal value, as will be demonstrated over repeated trials.  

Throughout the computations, we scale the four-dimensional parameter search space to form a unit hypercube.  We use the hierarchical approach described above with three levels of recursion.  Hence, during each optimization three trend surfaces will be constructed at finer and finer resolution.  The basic steps of the procedure are as follows.

Let $U_{1}$ denote the four-dimensional unit hypercube of the parameter space to be searched, and let $H$ be the number of levels in the trend hierarchy (we use $H=3$). 

\begin{description} 
\item{\texttt{FOR}} each iteration in the hierarchical trend optimization method, indexed by $k={1,\cdots, H}$:
\begin{description} 
\item{\texttt{Step 1.} \textbf{Generate Sample Locations:}}\\
Select $M$ points, $\left[x_{1}^{k},\cdots,x_{M}^{k}\right]$, from $U_{k}$ using Latin Hypercube Sampling~\cite{mckay1979a}.
\item{\texttt{Step 2.} \textbf{Evaluate the Objective:}}\\
Evaluate the objective at each of the $M$ locations $x_{i}^{k}$, and store each corresponding result in $y_{i}^{k}$.
\item{\texttt{Step 3.} \textbf{Build the Trend:}}\\
Use all data $\left\{(x_{i}^{j},y_{i}^{j})\,:\,j=1,\cdots,k\,;\,i=1,\cdots,M\right\}$ obtained during the optimization so far to construct the trend surface $T_{k}$ via ridge regression.
\item{\texttt{Step 4.}  \textbf{Optimize the Trend:}}\\
Quickly find $x_{*}^{k}$, the location in the parameter space that globally minimizes the trend surface $T_{k}$.
\item{\texttt{Step 5.}  \textbf{Refine the Search Domain:}}\\
Generate a new search domain, $U_{k+1}$, by reducing the size of the current search domain, $U_{k}$, by a factor of 2 along each dimension centered about the point $x_{*}^{k}$.  
\end{description}
\item{\texttt{END}}
\end{description}  

After $H$ iterations, declare $x_{*} := x_{*}^{H}$ as the parameter location that minimizes the objective function.  If desired, the objective function can be evaluated repeatedly at $x_{*}$, and then averaged, to generate $y_{*}$, the expected value of the objective at $x_{*}$.  

\medskip

The most computationally expensive step is \texttt{Step 2}, the evaluation of the objective function.  In comparison, optimization of the surrogate performed in \texttt{Step 4} is extremely fast.  To effect \texttt{Step 4}, we simply evaluated the surrogate at 5000 points randomly distributed throughout the search space and selected the point that produces the lowest value of the surrogate.  The surrogate provides repeatable values so there is no need to average over many evaluations of the surrogate.

At the completion of the algorithm, the total number of required objective function evaluations is $M\cdot H$, that is, $M$ objective evaluations during each of the $H$ trend iterations.  An important point is that, at each iteration, the $M$ objective evaluations required by \texttt{Step 2} are \emph{independent}, meaning, therefore, that \emph{these $M$ evaluations can be computed in parallel}.  In our optimization source code, we have parallelized \texttt{Step 2} so that the total wall clock time required for the entire optimization algorithm to complete is $H\cdot T_\text{obj}$, where $T_\text{obj}$ is the CPU time required to run the optimization with a single objective function evaluation.  An important point, therefore, is that if the number of available computing processors is greater than $M$ (and we consider any overhead communication costs between the $M$ processors as negligible), then the execution time of the algorithm is independent of $M$.  Thus, trend optimization provides an effective method to harness parallel computation resources for fast global optimization.  

Since in our specific implementation of the trend algorithm we recursively generate three surrogate functions, and since the evaluation of even the most expensive objective function is approximately a minute; the trend algorithm terminates after just three minutes of computation on 40 processors having used information gathered from 120 objective evaluations.  The CITerra cluster at Caltech has 4096 processes that could conceivably be used to find an optimal solution; however, we have observed that trend optimization provides reliable convergence when $M$ (the number of objective evaluations made at each step) is $\sim 40$ or above, so that a far more modest number of processors is actually required.    In summary, the trend optimization scheme as we have applied it to the self-assembly problem is able to provide parameters that optimize the most expensive objective functions in just over three minutes when run on a cluster of 40 parallelized processors.

It should be noted that the hundredfold speed-up obtained by trend optimization over simulated annealing in the time taken to generate potentials (as stated in the introduction), is assessed using the \emph{total} CPU time summed over all processors, and not the wall clock time.  The speed-up is attributed to the robust manner in which trend optimization accelerates search of a noisy objective with a smooth trend.  Hence, the speed-up attributed to the facility with which the trend optimization admits parallelization of the computation represents an additional time savings over and above the hundredfold speed-up.

\section{Methods for Generating Potentials\label{sec:MethodsforGenerationofPotentials}}

The five methods for generating potentials that we compare in this paper are described below. The first is a heuristic geometric method that requires no computation. The other four methods utilize an optimization procedure. 

\subsection{Geometric Method\label{sec:GeometricMethod}}

The geometric method (abbreviated as \textsf{GM}) that we present here is an optimization-free procedure that exploits differences between the geometry of the desired target lattice and competitor lattices that we hope to discourage.  The design of the potential is based on four main principles: 
\begin{description} 
\item{\textbf{GM1.}} The potential must have local minima located at each of the radial distances to nearest neighbors in the desired lattice and nowhere else;
\item{\textbf{GM2.}} The potential may include local maxima at radial distances at which competitor lattices have nearest neighbors but the target lattice does not;
\item{\textbf{GM3.}} When the target and competitor lattices have nearest neighbors at the same radial distances, the energy levels of the local minima identified in \textbf{GM1} are chosen to energetically prefer the target lattice;
\item{\textbf{GM4.}} A Lennard-Jones potential is spliced into the potential at the origin to provide a hard core potential.  
\end{description} 

\textbf{GM1} and \textbf{GM2} use information about the geometric structure of the target and competitor lattices to stabilize the target lattice, and to discourage a competitor lattice that has different distances to nearest neighbors.  \textbf{GM3} uses energetics to discriminate between lattices that have identical distances to nearest neighbors and differ only in the numbers of particles at those distances.  

\begin{figure}[ht]
\begin{center}
 \subfigure[\footnotesize \label{fig:polypotentials}]{
        \includegraphics[height=4.5cm]{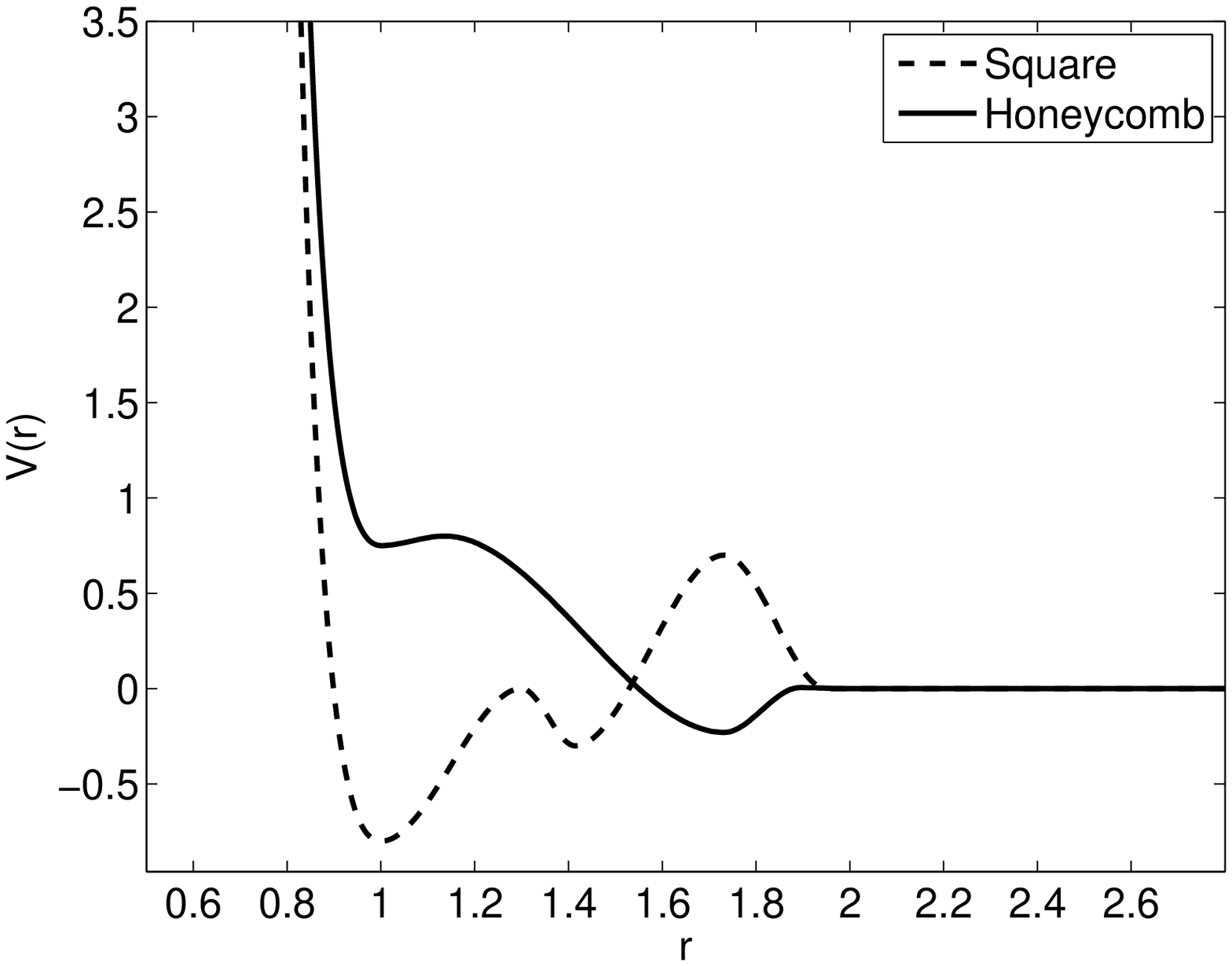}}  
        \hspace{0.4cm}
   \subfigure[\footnotesize \label{fig:polysq}]{
        \includegraphics[height=4.5cm]{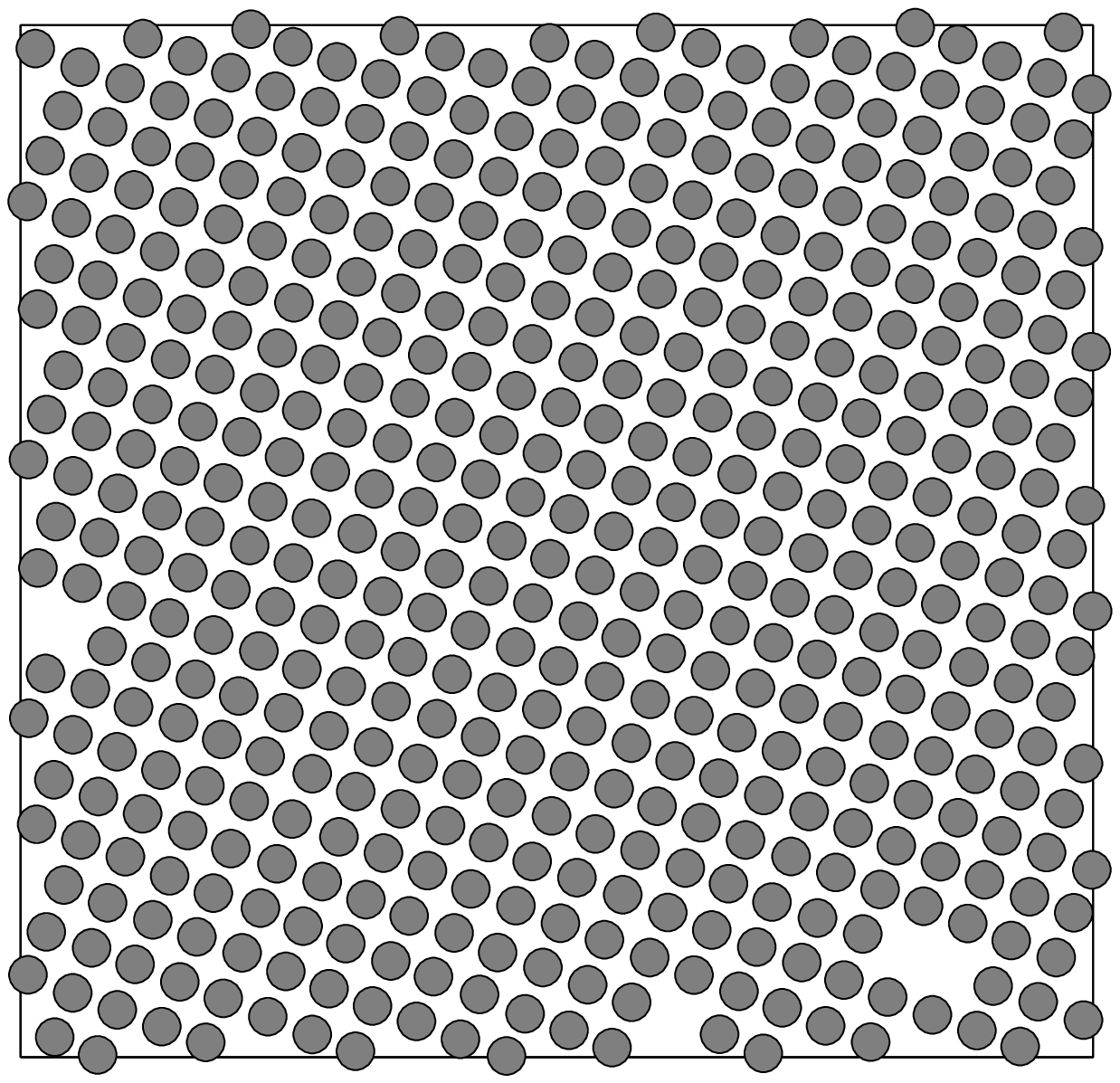}}       
          \hspace{0.1cm}  
   \subfigure[\footnotesize \label{fig:polyhc}]{
        \includegraphics[height=4.5cm]{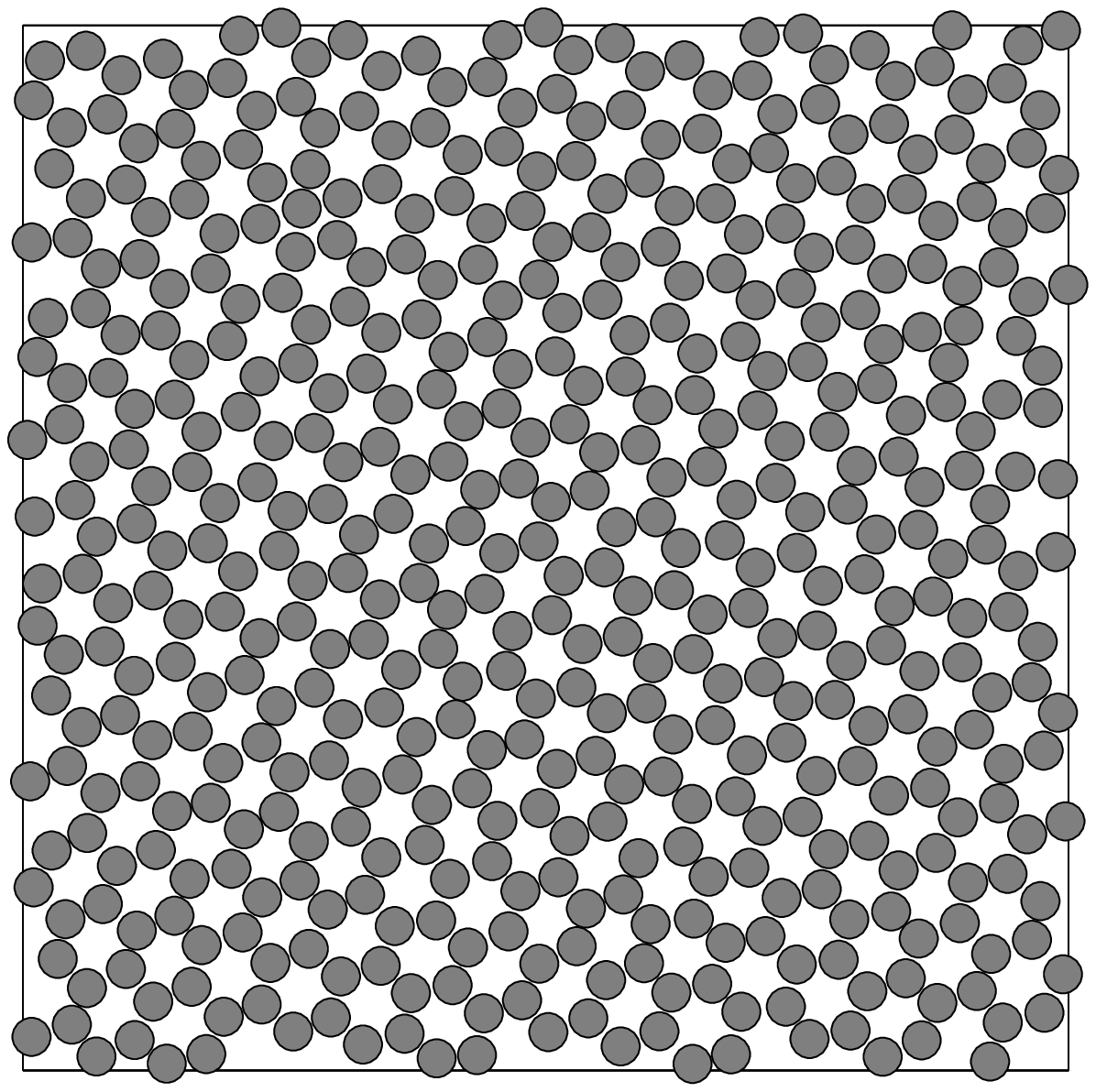}}
  \caption{\label{fig:poly}\footnotesize Potentials developed using the heuristic geometric method for both the square and honeycomb lattices are shown in Figure (a).  Figures (b) and (c) show typical lattices obtained using these potentials.}
\end{center}
\end{figure}

When the target lattice is the square lattice, and the identified competitor is a triangular lattice, the fact that these lattices have different distances to nearest neighbors makes this method of potential generation a very robust approach.  By explicit construction, the locations of the local minima and maxima discourage the triangular lattice and stabilize the square lattice.  The shape of the potential generated for this case is shown in Figure~\ref{fig:polypotentials}.  Notice in particular that the potential has a very simple form with local minima at distances of 1.0 and $\sqrt{2}$ to encourage the square lattice, and a local maximum located at $\sqrt{3}$ to discourage the triangular lattice.  In simulation this potential performs remarkably well.  The potential robustly forms the square lattice, and defects (except for voids) are seldom observed.  A sample square lattice obtained with this potential is shown in Figure~\ref{fig:polysq}.

The honeycomb lattice represents a more difficult self-assembly problem since both this target lattice and the competitor triangular lattice have identical distances to nearest neighbors.  The only difference between these lattices that must be exploited is the \emph{different number of neighbors at each distance}.  Hence, in \textbf{GM3}, the relative heights of the local minima are chosen to ensure that the target lattice represents a lower energy state than the competitor lattice and is thus more likely to form upon cooling.  

The application of the \textsf{GM} method for the specific case of constructing a potential that favors the honeycomb lattice and discourages the triangular lattice proceeds as follows:
\begin{enumerate} 
\item Recognize that both the triangular and honeycomb lattices have nearest neighbors at a distance of 1.0, and second nearest neighbors at a distance of $\sqrt{3}$.  The number of neighbors at those distances (the lattice coordination numbers) for the honeycomb lattice are 3 and 6, while the triangular lattice has 6 and 6.  
\item Assign local minima at distances of 1.0 and $\sqrt{3}$.
\item Since the triangular lattice has more particles at distance 1.0 than the target lattice, raise the first minimum at 1.0 with respect to the height of the minimum at $\sqrt{3}$.
\item Splice in a 6--12 Lennard-Jones type repulsive potential at the origin to simulate a rigid core.
\item Use cubic splines to piece together the repulsive core at the origin and the local minima at their various locations and heights. 
\end{enumerate}  

One other consideration that we include when designing this potential is that we are careful not to make the local maximum between the first and second minima too high.  This ensures that particles that are stuck in the triangular formation have increased likelihood of escaping to the lower potential well of the honeycomb lattice.  

The  cubic splines are constructed so as to have zero derivative at the local minima, thus ensuring that the potential is continuously differentiable on the whole positive real line.  A sample potential constructed using this method for the generation of honeycomb lattice potentials is shown in Figure~\ref{fig:polypotentials}.   Notice that the potential has local minima located at distances of 1.0 and $\sqrt{3}$, with the first minimum located above the second minimum.   Since the triangular lattice has more particles located at distance 1.0 than the honeycomb lattice, this ensures that the triangular lattice is a higher energy state than the honeycomb lattice.  A sample lattice produced using this potential is shown in Figure~\ref{fig:polyhc}.  The fact that this heuristic geometric method is able to produce honeycomb lattices of this quality without having to use any computation and optimization is quite remarkable and until now has been overlooked in the self-assembly literature.

We refer to the geometric method as a heuristic method since we do not provide formulas or algorithms for choosing the relative heights of the local minima and maxima.  The only prescribed constraints are the locations of the local minima that, by construction, ensure stability of the target lattice.  In practice, we find that any sensible choice of the heights that avoids sharp gradients in the potential works reasonably well.  In the implementation of our computer code, the user need only input the location and heights of the desired minima and maxima, and the spline fitting and splicing of the Lennard-Jones potential are computed automatically, which makes the approach very simple to execute.   Given the simplicity of the geometric method, the ease with which it is implemented, and the relatively high quality of the resulting lattices, it represents a ``rough-and-ready'' approach that anyone needing to design a potential would do well to consider before pursuing more computationally expensive schemes.

\medskip

Whereas the geometric method utilizes only the geometries of the static target lattice and competitor lattice, the optimization methods discussed next incorporate information about particle dynamics by optimizing parameters in the potential with respect to dynamic particle simulations. 

\subsection{Baseline Method\label{sec:BaselineMethod}}

The baseline simulated annealing method of \cite{rechtsman2006a} uses the simulated annealing optimization procedure with the Lindemann parameter as the objective function.  As mentioned previously, the simulated annealing procedure fails to converge to a meaningful minimum unless the objective function is averaged over many trials. We refer to this method with the abbreviation \textsf{SA-LP20} to indicate that the simulated annealing approach is applied to the Lindemann parameter averaged over 20 independent evaluations.  The large number of samples required to sufficiently reduce the noise in the objective function, means that it is impractical to apply the simulated annealing optimization procedure to the more expensive quality metric objective functions.  The simulated annealing procedure is initiated with initial guess solutions chosen randomly and uniformly from the parameter space described by the inequalities in line \eqref{eq:searchspace} of Section \ref{sec:TheSelfAssemblyProblem}.

\subsection{Trend Optimization Methods}

The remaining three methods that we consider utilize the trend optimization method applied to the Lindemann parameter, the Template Measure, and the Defect Measure as objective functions.  These methods are abbreviated \textsf{T-LP}, \textsf{T-TM}, and \textsf{T-DM} respectively.  Since the trend optimization method is well-suited to noisy objective functions, the objectives do not need to be averaged over many runs when evaluated.   

\bigskip
A summary of all the potential generation methods under study is provided in Table \ref{table:MethodSummary}.

\begin{table}
\begin{tabular}{|l|l|l|r|}
\hline
\multicolumn{4}{|c|}{Summary of Potential Generation Methods} \\
\hline
\textbf{Method} & \textbf{Optimization} & \textbf{Objective} & \textbf{Cost} (s)\\
\hline
\textsf{GM} & \multicolumn{2}{|c|}{Geometric method} & 0 \\
\hline
\textsf{SA-LP20} & Sim. anneal. & Lindemann (20 sample avg.) & 40\\
\hline
\textsf{T-LP} & Trend & Lindemann & 2\\
\hline
\textsf{T-TM} & Trend & Template Measure  & 70\\
\hline
\textsf{T-DM} & Trend & Defect Measure & 70\\
\hline
\end{tabular}
\caption{\label{table:MethodSummary} \footnotesize Each of the five methods for generating potentials is listed here with the associated computational cost of evaluating the objective function.  Notice that the Geometric Method is not an optimization-based method and incurs no computational cost.}
\end{table}

\section{Results\label{sec:Results}}

In this section, we compare the effectiveness of the proposed methods for generating interaction potentials that lead to the self-assembly of the honeycomb lattice.  In particular, in accordance with the comparison criteria enumerated in Section~\ref{sec:TheSelfAssemblyProblem}, we seek answers to the following questions:  
\begin{enumerate}  
\item How much computational effort is required to generate the potentials?
\item What is the quality of the lattices generated by the potentials?  
\item How reliably do the potentials form high quality lattices given uncertainty in the initial conditions of the particles?   
\end{enumerate} 

\begin{figure}[ht]
\begin{center}
   \subfigure[\footnotesize  Simulated annealing optimization of the Lindemann parameter (\textsf{SA-LP20}). \label{fig:optSA-LP20}]{
        \includegraphics[width=5.5cm]{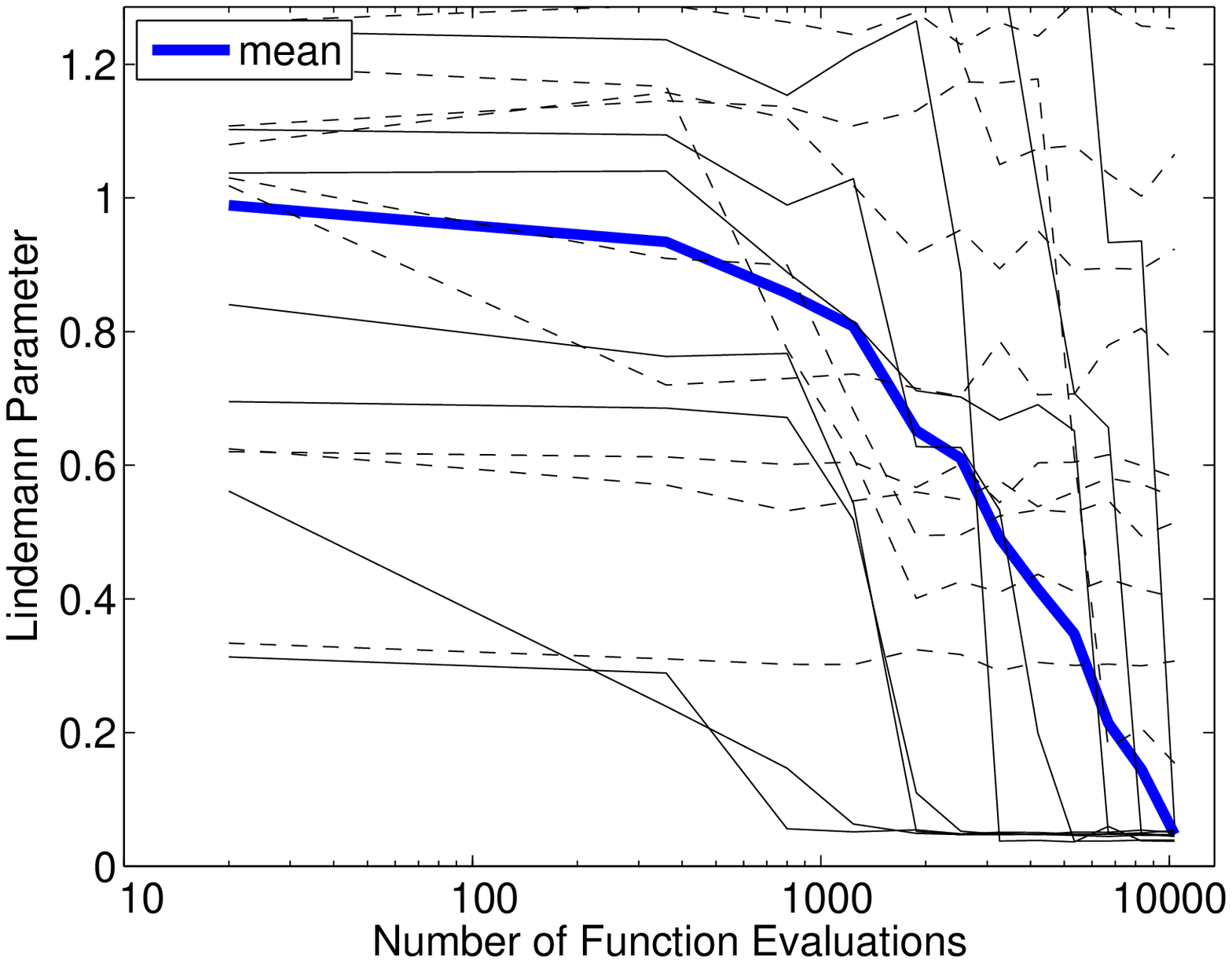}}     
        \hspace{0.2cm}  
           \subfigure[\footnotesize  Trend optimization of the Lindemann parameter (\textsf{T-LP}).\label{fig:optT-LP}]{
        \includegraphics[width=5.5cm]{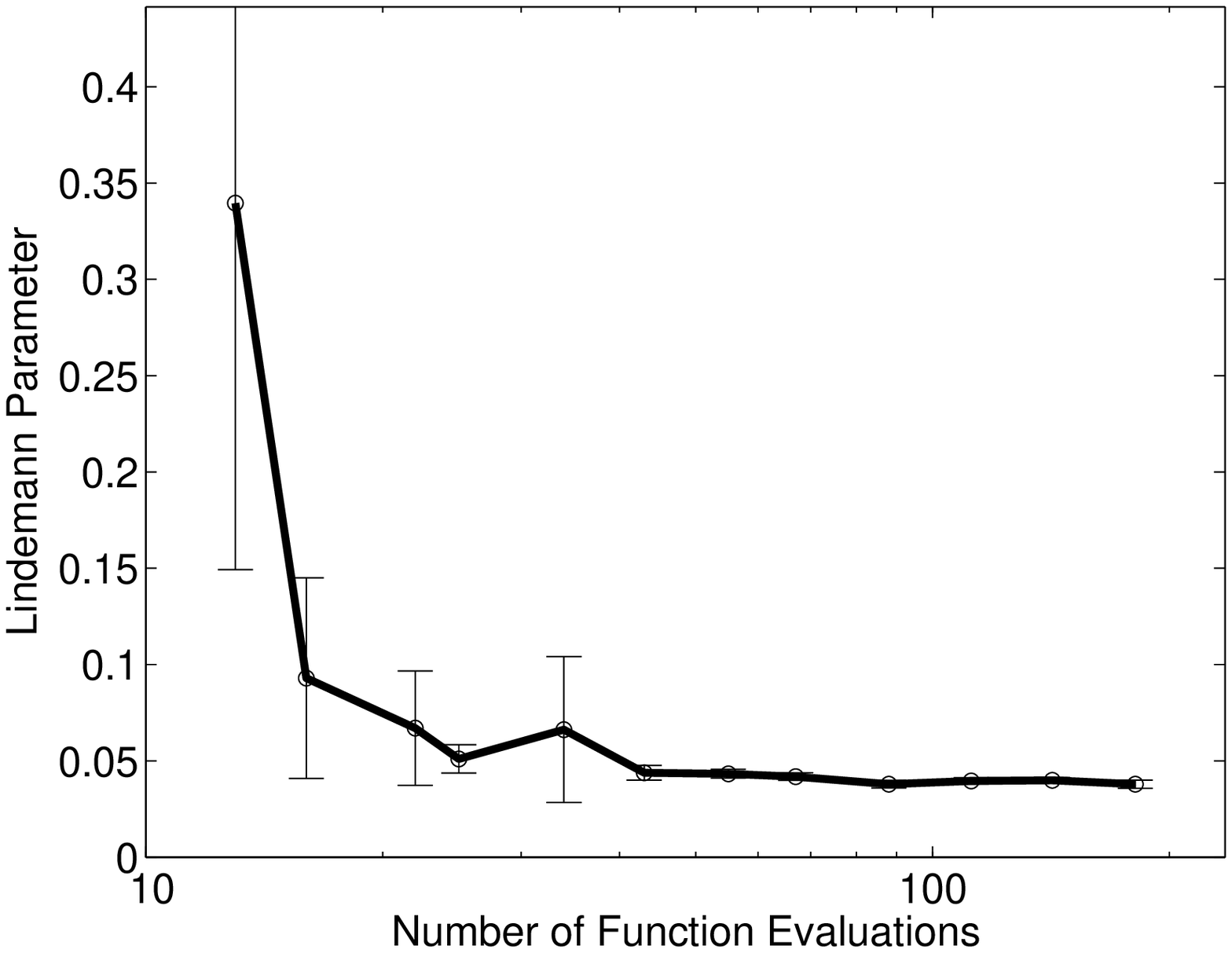}}     \\  
           \subfigure[\footnotesize  Trend optimization of the Template Measure (\textsf{T-TM}).\label{fig:optT-TM}]{
        \includegraphics[width=5.5cm]{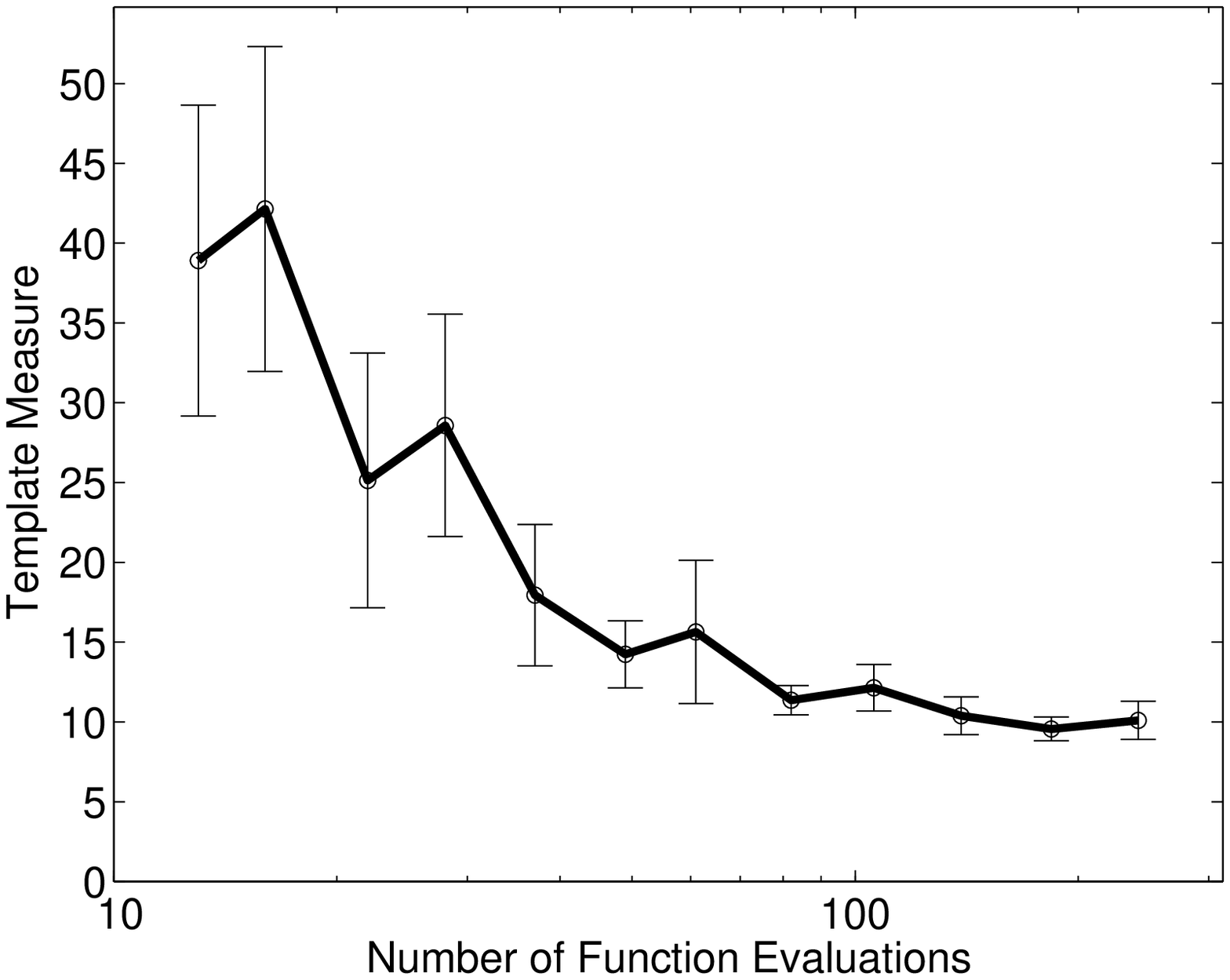}}  
          \hspace{0.2cm}           
   \subfigure[\footnotesize  Trend optimization of the Defect Measure (\textsf{T-DM}).\label{fig:optT-DM}]{
        \includegraphics[width=5.5cm]{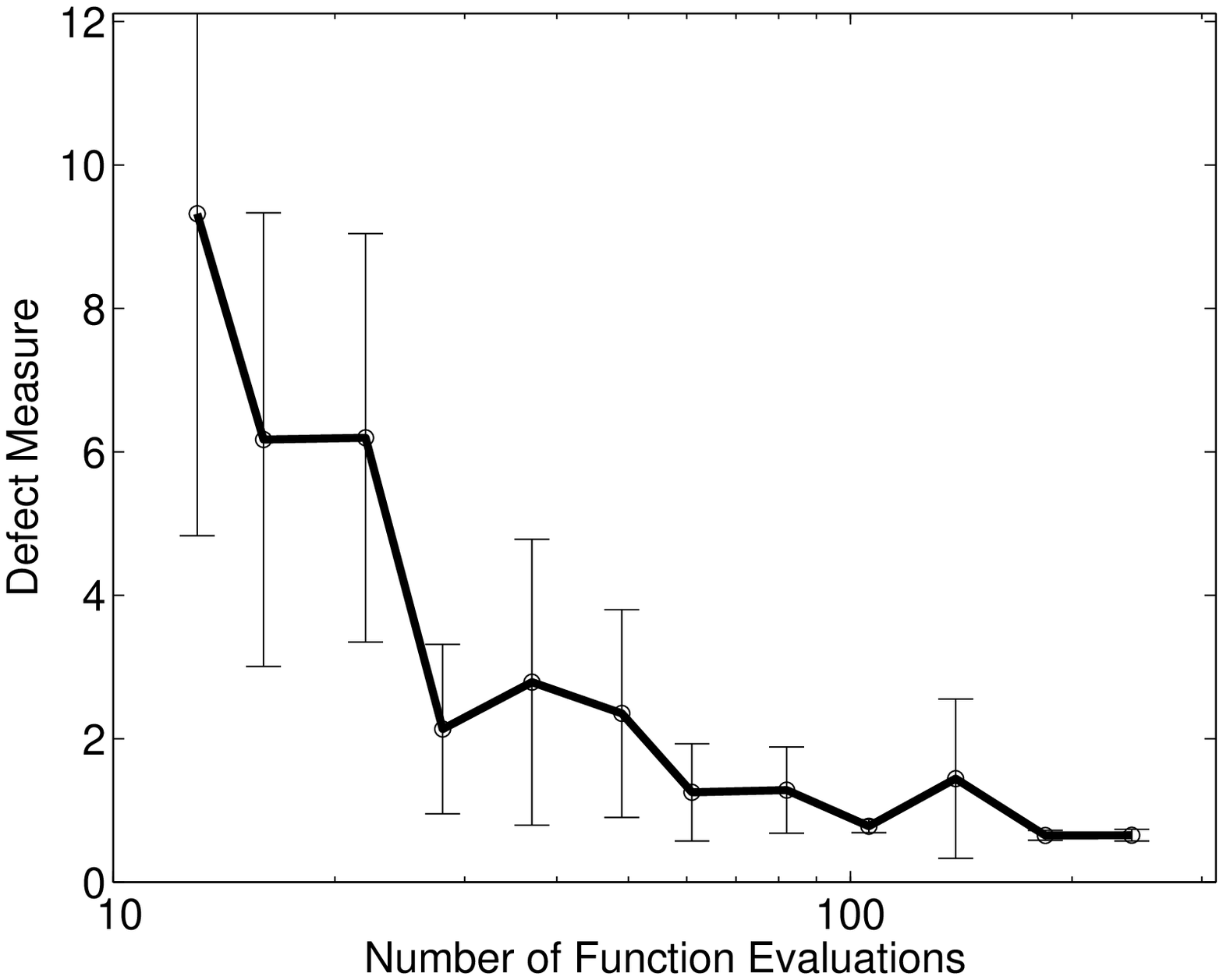}}     
      \caption{\label{fig:optimizations}\footnotesize Averaged objective function values versus number of objective function evaluations for four optimization methods are shown here. In the results for simulated annealing shown in (a), dashed lines indicate optimizations that failed to find an optimal value.  The single bold line represents the mean of the optimizations that did converge to an optimal value (indicated by solid lines). Most notably, trend optimization reliably converges to a minimum value of the Lindemann parameter with a one-hundredfold reduction in the number of required objective function evaluations when compared with the simulated annealing approach.}
\end{center}
\end{figure}

\begin{figure}[ht]
\begin{center}
	      \subfigure[ \footnotesize  Quality of lattices generated using simulated annealing measured with the Template Measure. \label{fig:satemplate}]{
        \includegraphics[width=6.0cm]{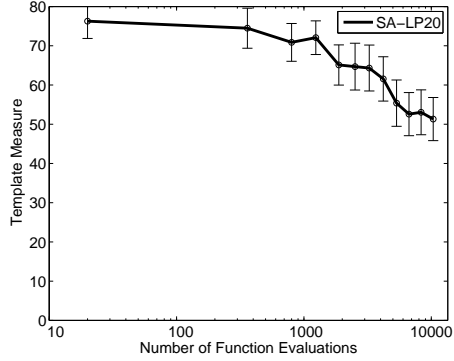}}     
        \hspace{0.4cm}  
           \subfigure[\footnotesize  Quality of lattices generated using simulated annealing measured with the Defect Measure.\label{fig:sadefect}]{
        \includegraphics[width=6.0cm]{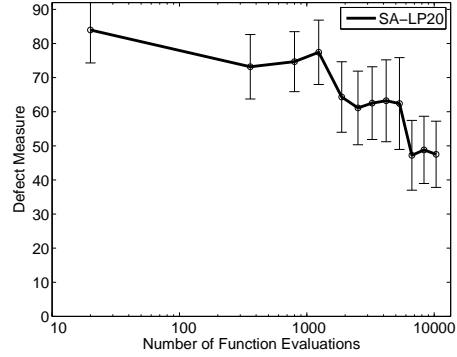}}   
   \subfigure[\footnotesize  Comparison of lattice quality using the Template Measure. \label{fig:template}]{
        \includegraphics[width=6.0cm]{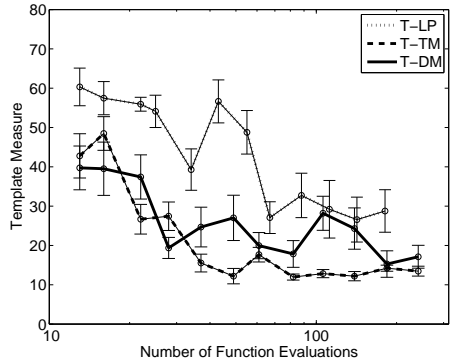}}     
        \hspace{0.4cm}  
           \subfigure[\footnotesize  Comparison of lattice quality using the Defect Measure.\label{fig:defect}]{
        \includegraphics[width=6.0cm]{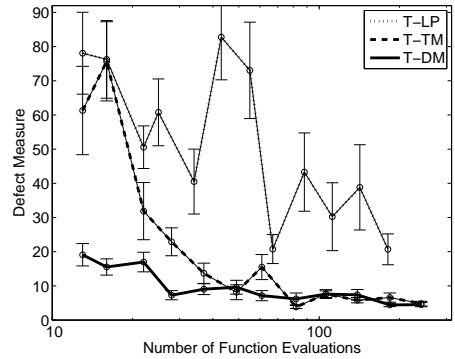}}       
\caption{\label{fig:metrics}\footnotesize Here we compare the quality of the lattices produced using the four optimization methods.  In (a) and (b), the quality of the lattices produced using the simulated optimization method is shown.  It should be noted that only the nine potentials generated by the convergent simulated annealing trials in Figure \ref{fig:optSA-LP20} were used to produce these plots.  Figures (c) and (d) show the quality of lattices produced using the three trend optimization methods.  Using trend optimization directly on the quality metrics reliably produces high quality lattices.  Optimization of the Lindemann parameter is only moderately correlated with improved lattice quality.  In these Figures, recall that values of the Template Measure and the Defect Measure for purely random configurations are 132 and 161, respectively, and that the computation-free Geometric Method produces lattice quality scores of 18 and 12, respectively.}
\end{center}
\end{figure}

Answers to these questions, as well as a comparison of the five potential generation methods discussed in Section \ref{sec:MethodsforGenerationofPotentials}, are summarized in the suite of plots shown in Figures \ref{fig:optimizations} and \ref{fig:metrics}.  In brief, Figure \ref{fig:optimizations} shows the superior effectiveness of trend optimization over simulated annealing in minimizing the respective objectives as a function of the number of objective evaluations required, while Figure \ref{fig:metrics} shows the quality of the lattices obtained by the potentials that were generated by the optimizations.   

\medskip

The geometric method (\textsf{GM}) is the simplest method to implement.  It does not require any objective optimization to search for parameters and consequently does not incur any computational cost.    In order to evaluate the quality of the lattices produced by the geometric method, one-hundred cooling simulations were performed using the honeycomb potential shown in Figure~\ref{fig:polypotentials}.  After each simulation, the Template Measure and the Defect Measure were computed.  Averaging over all these runs yielded the following scores:
\begin{enumerate}	
\item Template Measure =   17.9  with standard deviation 3.4,
\item Defect measure = 11.8 with standard deviation 4.5.
\end{enumerate} 
Comparison with the lattice qualities obtained using the optimization-based methods as shown in Figure~\ref{fig:metrics} indicates that these quality scores are exceptionally good, and that assiduous computational optimization is required to produce potentials that produce lattices of higher quality. 

\medskip

The four optimization-based methods (\textsf{SA-LP20},\,\textsf{T-LP},\,\textsf{T-TM}, and \textsf{T-DM}) were each run for an increasing number of allowed objective function evaluations.  For each number of objective function evaluations, each trend optimization method was executed in twenty independent trials, with each trial generating parameters for the interaction potential that seek to minimize the respective objective function.  In Figure \ref{fig:optimizations}, the average of the values of the objective function obtained in the twenty trials is plotted for each number of function evaluations (the error bars indicate the standard deviation over the twenty trials for each number of function evaluations).   

After all the optimizations are completed and the potentials have been generated, the quality of each potential must be tested.  For each number of function evaluations, and for each of the twenty independent trials, the potential produced by a method was quality tested by running twenty cooling simulations on a system of 225 particles, and then measuring the quality of the final lattices using both the Template Measure and the Defect Measure.  For a single trend optimization method, this requires

{\scriptsize
\begin{align*} 
&\left(\frac{\text{\scriptsize\hspace{0.1cm}  12 different numbers of function evaluations}}{\text{\scriptsize\hspace{0.1cm} each trend method}}\right)
\times\left(\frac{\text{\scriptsize\hspace{0.1cm} 20 independent trials}}{\text{\scriptsize\hspace{0.1cm} each number of function evaluations}}\right)
\times\left(\frac{\text{\scriptsize\hspace{0.1cm} 20 cooling simulations}}{\text{\scriptsize\hspace{0.1cm} each independent trial}}\right)\\
&=\left(\frac{\text{\scriptsize\hspace{0.1cm} 4800 cooling simulations}}{\text{\scriptsize\hspace{0.1cm} each trend method}}\right)\,. 
\end{align*}}
In other words, 240 independent optimization trials are required to produce a single curve in Figure \ref{fig:optimizations}, and 4800 cooling simulations are required to produce a single curve in Figure \ref{fig:metrics}.  The cooling simulations were initialized with a temperature approximately 1.5 times the melting temperature of the lattice and then slowly cooled using a Nos\'{e}-Hoover thermostat to less than ten percent of the melting temperature. At the completion of each cooling simulation, both the Template Measure and the Defect Measure were computed to measure the quality of the resulting lattice.  Averaging over the 20 cooling simulations yields two quality scores for each potential -- one for each quality metric.  Results for the Template Measure are shown in Figure~\ref{fig:template} while results for the Defect Measure are shown in  Figure~\ref{fig:defect}.

\medskip

After reviewing the results in Figure~\ref{fig:optimizations}, we see that the trend approach clearly provides a faster and more robust method for optimizing the objective functions.  Figure \ref{fig:optSA-LP20} shows the results from twenty simulated annealing optimizations for different randomly chosen starting points in the search domain, and indicates that the success of the method is highly variable. Simulated annealing often fails to find optimal parameter values and remains stuck in local minima even after 10,000 evaluations of the objective.  Of the twenty simulated annealing optimization trials performed, only nine were able to find an optimal value of the Lindemann parameter less than 0.15.  The bold line in Figure \ref{fig:optSA-LP20} represents the mean over only these nine best-performing trials.  Moreover, many of these simulated annealing optimizations that do converge require more than 6,000 objective evaluations.  In contrast, we see from Figure~\ref{fig:optT-LP}  that trend optimization reliably finds optimal values after sixty evaluations of the Lindemann parameter.  We conclude that when trend optimization is used to optimize the Lindemann parameter, we obtain a one-hundred-fold reduction in computation time over the simulated annealing method of \cite{rechtsman2006a}, and that the optimization is more robust.  The speed-up can be attributed to the fact that the objective function is noisy yet has a simple trend -- two properties for which trend optimization is ideally suited. 

Furthermore,  the accelerated search of the trend method makes it possible to use trend optimization with the more expensive quality metrics as objective functions.  Doing so provides the extra guarantee that the generated potentials produce high-quality lattices.  As indicated in Figures~\ref{fig:optT-TM} and \ref{fig:optT-DM}, trend optimization reliably finds optimal values of the lattice quality objectives in less than 120 function evaluations, although it must be remembered that these objectives are approximately 35 times more expensive to evaluate than the Lindemann parameter.  Nevertheless, the total CPU time required to perform these optimizations is still less than the time taken by simulated annealing to optimize the Lindemann parameter.

Figure~\ref{fig:metrics} shows the quality of the lattices generated by the potentials produced by the various methods, as measured using both quality metrics.  Several important observations are to be made after reviewing these plots.  First, optimization of the Lindemann parameter leads to lattices of modest and unreliable quality, indicating that the Lindemann parameter and lattice quality are only moderately correlated.  Second, even when simulated annealing is successful in optimizing the Lindemann parameter, the corresponding potential may yield lattices of poor quality.  This occurs because simulated annealing may find minima corresponding to very narrow wells that do not provide robustness against uncertainty in initial conditions. In contrast, since trend optimization seeks out the general trend over the entire search space, this method finds minima that more robustly lead to high quality lattices. Finally, the two lattice quality metrics are reasonably consistent in that a potential generated by optimizing one of the quality measures is also considered high-quality in the other measure.

\begin{figure}[ht]
\begin{center}
        \includegraphics[height=6.0cm]{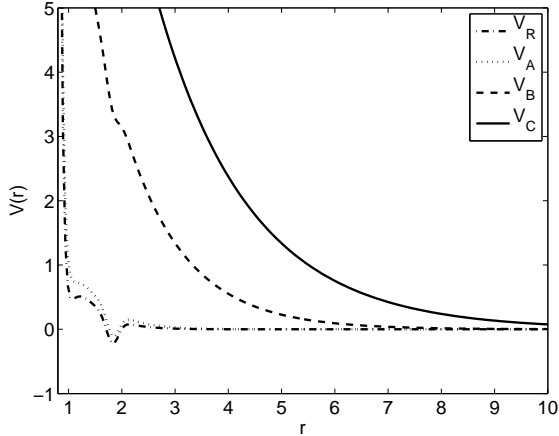}
\caption{\label{fig:potentials}\footnotesize Potentials for the generation of honeycomb lattices. $V_{R}$ is the potential previously provided by \cite{rechtsman2006a}.  Trend optimization generates potentials similar to \cite{rechtsman2006a} (labelled $V_{A}$), but also uncovers a new family of solutions that have a more repulsive shape and no local minima (sample potentials in this family are labelled $V_{B}$ and $V_{C}$).  Remarkably, these repulsive potentials robustly form large regions of honeycomb lattice without defects.}
\end{center}
\end{figure}

Figure \ref{fig:potentials} shows the range of potentials generated by the trend optimization methods. The potential provided by \cite{rechtsman2006a} is also shown for reference. Most striking is that trend optimization uncovers an entirely new family of potentials not previously considered.  In this family of potentials, the exponential term is dominant and consequently, their shape does not admit a local minimum.  The repulsive shape of these potentials leads to higher quality lattices since particles do not get stuck in local minima associated with local potential wells.  In simulation, we observe that the defects continue to move through the configuration until they leave the domain entirely, or annihilate one another through collisions.

\begin{figure}[ht]
\begin{center}
           \subfigure[\footnotesize \label{fig:hugehc}]{
        \includegraphics[height=7.5cm]{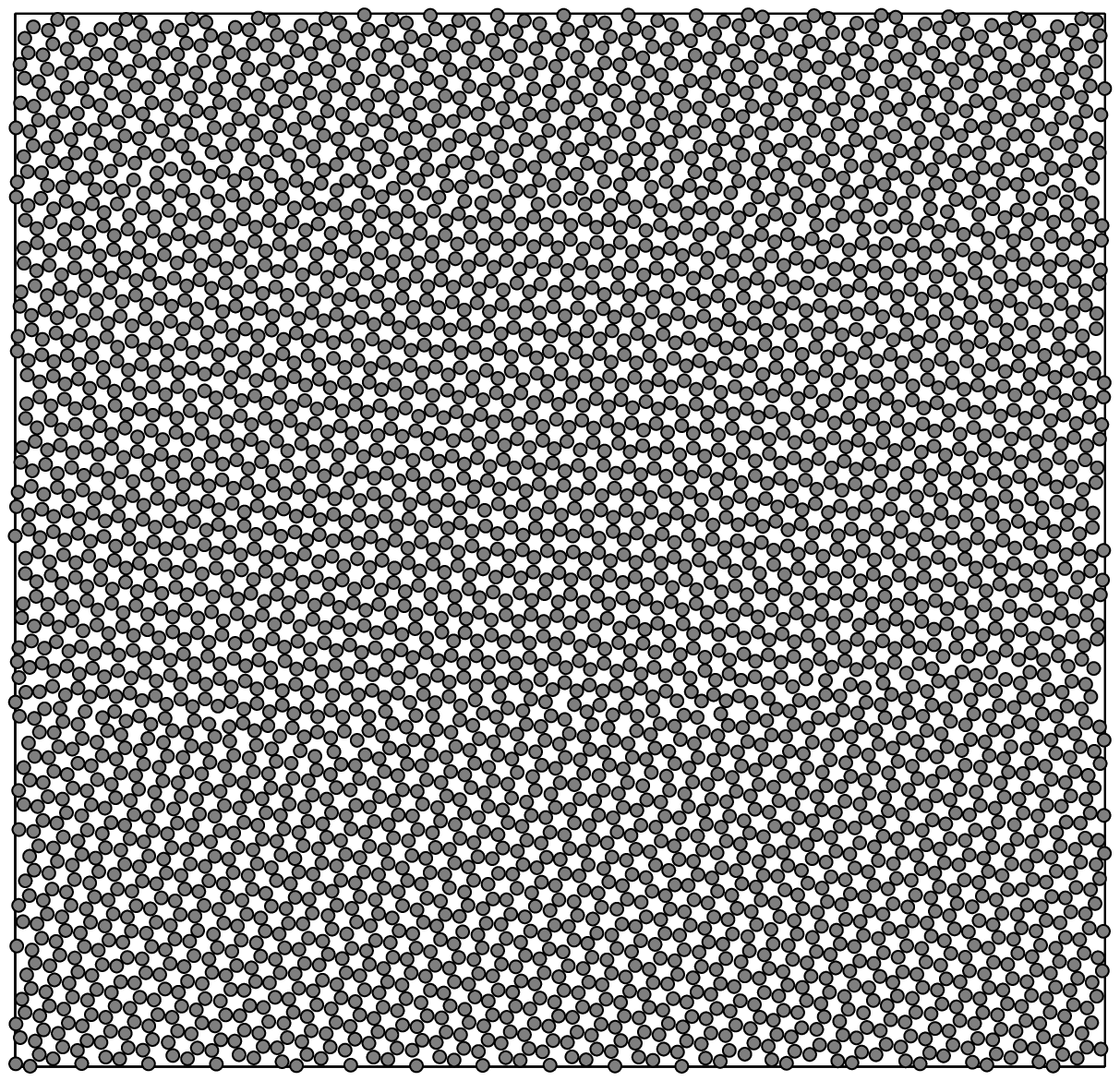}}     
        \hspace{0.1cm}
                   \subfigure[\footnotesize \label{fig:hugehcd}]{
        \includegraphics[height=7.5cm]{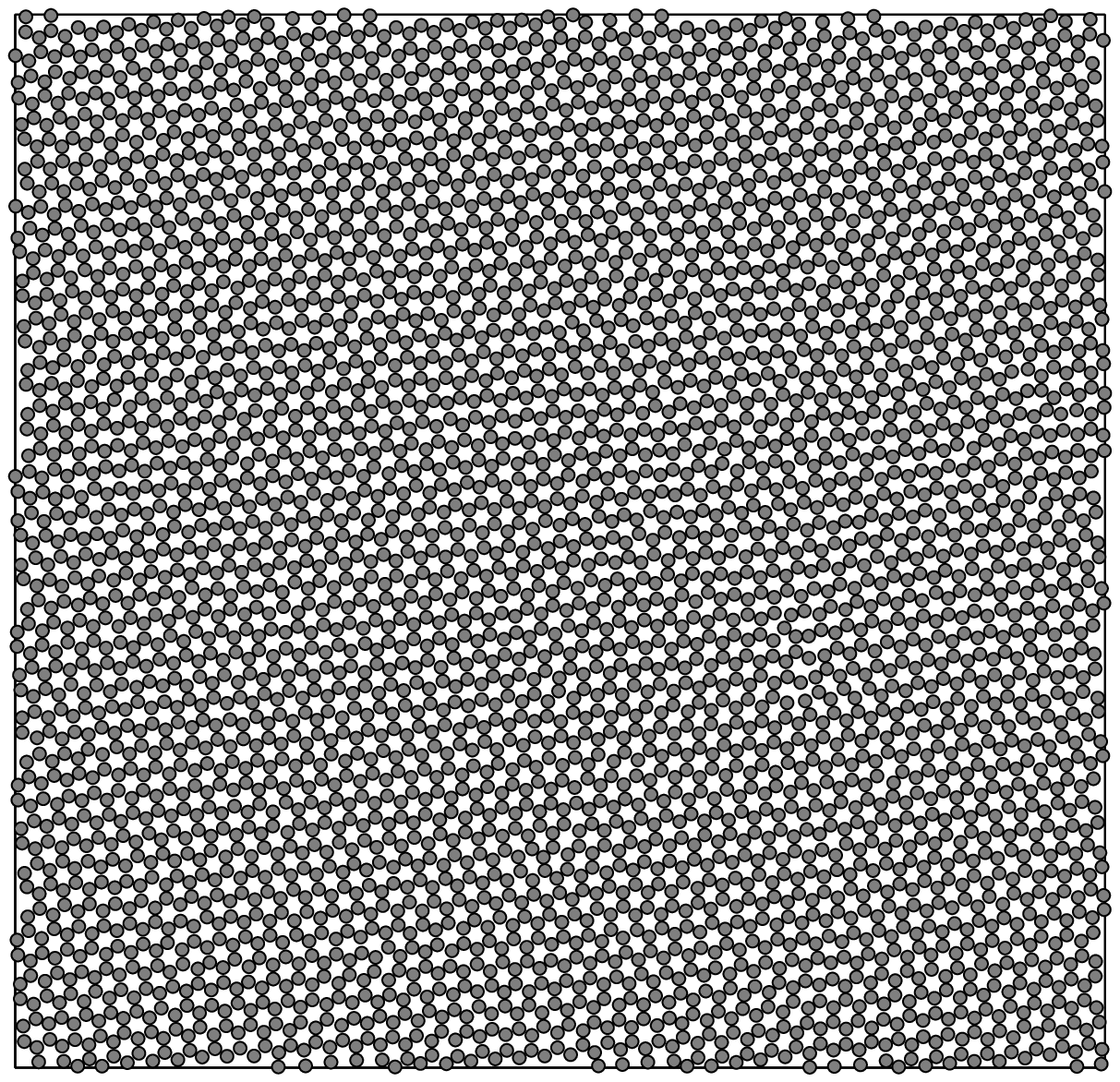}}   
\caption{\label{fig:hugelattices}\footnotesize Honeycomb lattices formed using the repulsive family of potentials discovered using trend optimization. (a) Two large regions of extremely well-formed honeycomb lattice meet to form a grain boundary.  (b) Large regions of the honeycomb lattice are formed with a few isolated defects.  These defects continue to move, even after the lattice is frozen, until they exit the boundary or are annihilated by collisions with other defects.  }
\end{center}
\end{figure}

Figure \ref{fig:hugelattices} shows two sample lattice configurations obtained for 4128 particles simulated using the potential labelled $V_{C}$ in Figure \ref{fig:potentials} that was generated with trend optimization on the Template Measure.  The parameter values for this potential are $a_0$=5.771, $a_1$=23.594, $a_2$=0.574, and $a_3$=1.816.  The final lattices exhibit large regions of extremely well-formed honeycomb lattice that we have not previously observed using potentials that contain local minima.  In Figure~\ref{fig:hugehc}, a prominent grain boundary lies between two large areas of well-formed honeycomb lattice.  As in Nature, this grain boundary forms when the cooling is not sufficiently slow.  In Figure~\ref{fig:hugehcd}, isolated defects are visible in the lattice; however, it should be noted that these defects are not yet ``frozen'' into the lattice, and given enough time will eventually leave the domain or disappear through mutual collision. 

The repulsive potentials found using trend optimization are remarkably effective at producing large regions of almost defect free honeycomb lattice despite their relatively simple shape.  The effectiveness of these potentials suggests that a far simpler basis of potential functions can be used in the parameterization of $V_\textsf{HC}(r)$.  In these potentials it is the exponential decay term that dominates over the Gaussian term.  The numerical simulations seem to suggest, and it would be interesting to pursue rigorously, that for a fixed density, the honeycomb lattice is a global minimizer (or ground state) of the exponential decay pairwise interaction potential.  This path is consistent with the numerical findings of Jagla \textit{et al} in~\cite{jagla1999a} for a ramp type potential. 
\section{Anisotropic Potentials\label{sec:AnisotropicPotentials}}

The methods presented thus far can be used to quickly generate isotropic potentials that produce high quality lattices. However, the potentials for self-assembling honeycomb lattices are not robust to variations in density -- the lattices only form if the initial density of the particles is very near the ideal density of the target lattice. If the initial density is much less than that of the target lattice, then a triangular lattice will be favored. \cite{rechtsman2006a} addressed this issue by searching for optimal potentials over a range of densities. However, the tolerance for variation in density is still very narrow and may indeed be a fundamental limitation of isotropic potentials. Allowing the potentials to have angular dependence leads to the robust formation of high-quality lattices from initial conditions with large variations in density. Moreover, admitting potentials with angular dependence allows for the construction of potentials that form more exotic lattices, such as the kagome lattice, which has not yet been accomplished with purely isotropic potentials \cite{rechtsman2006a}.   Certainly, the use of anisotropic potentials is well-motivated by abundant natural examples of anisotropic interaction potentials in Nature -- the water molecule serving as an ubiquitous prototype.

\begin{figure}[ht]
\begin{center}
 \includegraphics[width=5cm]{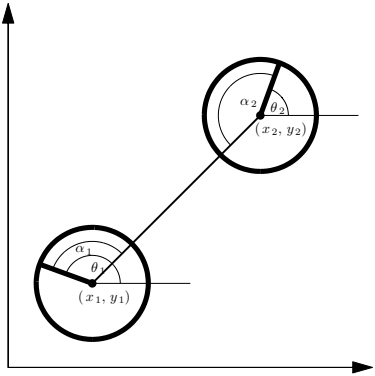}
  \caption{\footnotesize To implement an anisotropic potential, particles are now modeled as two-dimensional disks whose configuration is described by a location for the center of the disk, $(x,y)$, as well as a heading angle $\theta$.  The interaction potential between two particles is a function of the distance between the particles, as well as the angle, $\alpha$, that lies between the heading angle and the bearing toward the second particle.}
  \label{fig:hockeypuck}
\end{center}
\end{figure}

In order to introduce potentials with angular dependence, we must first extend the configuration space of the particle system to include an angular coordinate. We no longer consider the particles as point particles, but rather as two-dimensional sliding disks, each with radius $R$ and uniform mass density.   The configuration manifold of each particle is now $\mathbb{R}^2\times S^1$.  The configuration of the $i^\text{th}$ particle is described by the coordinate chart $(x_{i},y_{i},\theta_{i})$  as indicated in Figure \ref{fig:hockeypuck}, and we write the vector of coordinates describing the configuration of all $N$ particles as $\mathbf{x}=\big[(x_1,y_1,\theta_1),\cdots,(x_{N},y_{N},\theta_{N})\big]$.  
The total interaction potential over all particles now has the form
\begin{align} 
V(\mathbf{x}) = \sum_{i<j}^N V_\text{pair}(x_i-x_j,y_{i}-y_{j},\theta_i,\theta_j)
\end{align} 
where $V_\text{pair}:\mathbb{R}\times\mathbb{R}\times S^{1}\times S^{1} \rightarrow \mathbb{R}$ is the pairwise interaction potential between particles that depends not only on the relative displacement between particles, but also the angular displacement of each particle relative to the angular bearing of the other particle.  By construction, we make $V_\text{pair}$ symmetric with respect to particle interchange.  Specifically,
\begin{align}
V_\text{pair}(x_i-x_j,y_{i}-y_{j},\theta_i,\theta_j) :=  \psi(x_{i}-x_{j},y_{i}-y_{j},\theta_{i}) +  \psi(x_{j}-x_{i},y_{j}-y_{i},\theta_{j})\,.
\end{align} 
It remains to choose the functional dependence of $\psi(\cdot,\cdot,\cdot)$ to reflect the desired symmetry in the target lattice.  Consider the interaction potential 
\begin{align} 
\label{eq:psi}
\psi(\Delta x,\Delta y,\theta_{k}) := \frac{1}{2r^{12}} - \frac{1}{r^6}\left[1-\nu\sin^{2}\left(\frac{n\alpha_{k}}{2}\right)\right]
\end{align} 
where
\begin{align}
r := \sqrt{(\Delta x)^{2}+(\Delta y)^{2}} 
\end{align}
is the radial distance between the particles, and
\begin{align}
\alpha_{k} :=  \theta_{k}-\arctan(\Delta y,\Delta x)
\end{align}
is the angle between the heading of particle $k$ and the bearing toward the second particle (see Figure \ref{fig:hockeypuck}.)  This potential is simply a Lennard-Jones potential with added amplitude modulation in the azimuthal direction on the attractive term.  The periodicity of the trigonometric functions ensures that the potential has $n$-fold radial symmetry, and that each particle has preferred directions along which it feels the attractive pull of neighboring particles.  The free parameter $\nu$ is chosen to adjust the shape of the potential.  When $\nu$ is zero, the potential collapses to the isotropic Lennard-Jones potential.  For values of $\nu$ between zero and unity, the potential has $n$ potential wells symmetrically distributed in the azimuthal direction.  Taking values of $\nu$ greater than unity raises the repulsive regions between the potential wells.  Hence, $\nu$ is a parameter that determines how strongly the anisotropic potential prefers the binding site directions.  In simulations, we have observed that taking larger values of $\nu$ produces lattices with less defects since particles that do not align with the preferred binding directions fall into these more repulsive regions between the wells and create a configuration with much higher energy.   These defects are quickly removed by vibrations in the lattice.  In practice, a compromise must be met since the larger regions of repulsion created by higher values of $\nu$ increase the time taken for self-assembly to occur -- particles only bind with one another if they approach each other along an ever more narrower binding direction.   In the simulations that follow, we have used a value of $\nu=1.5$.

\begin{figure}[ht]
\begin{center}
\subfigure[\footnotesize \label{fig:angpothc}]{
  \includegraphics[width=4.5cm]{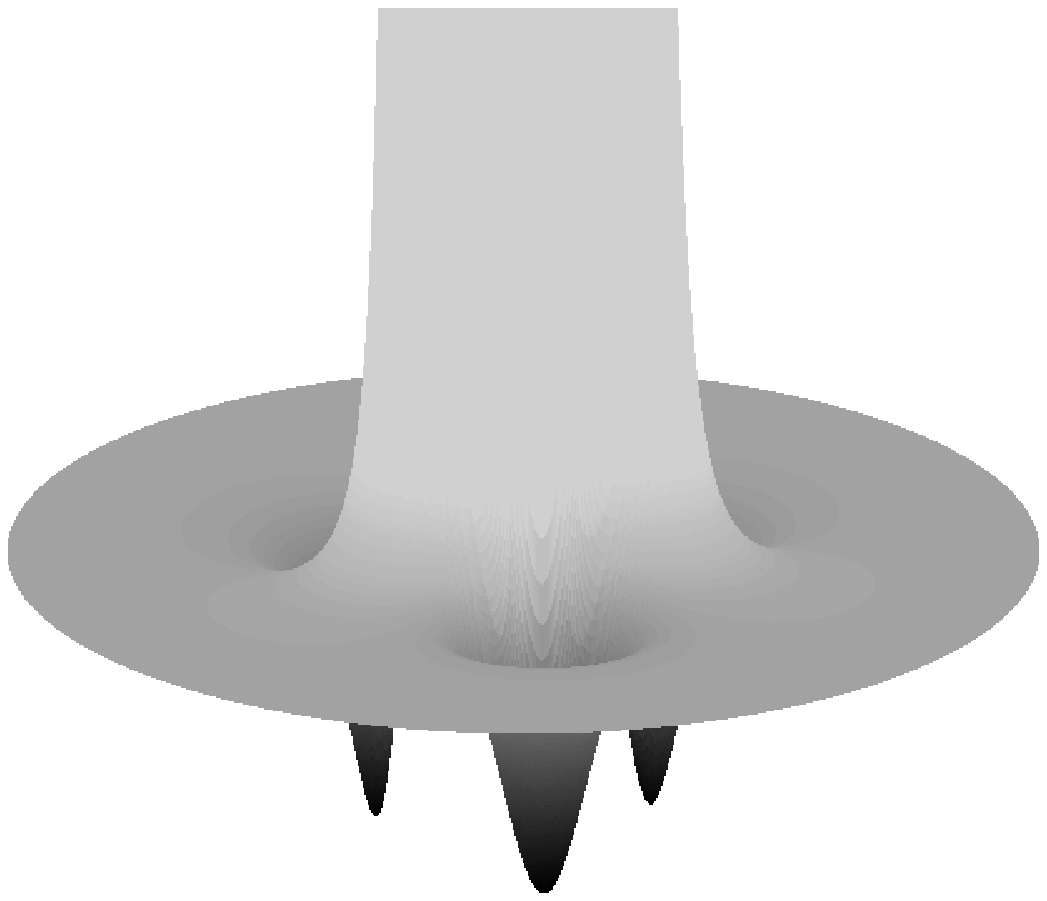}
  }
  \hspace{2.0cm}
    \subfigure[\footnotesize   \label{fig:angpotsq}]{
  \includegraphics[width=4.5cm]{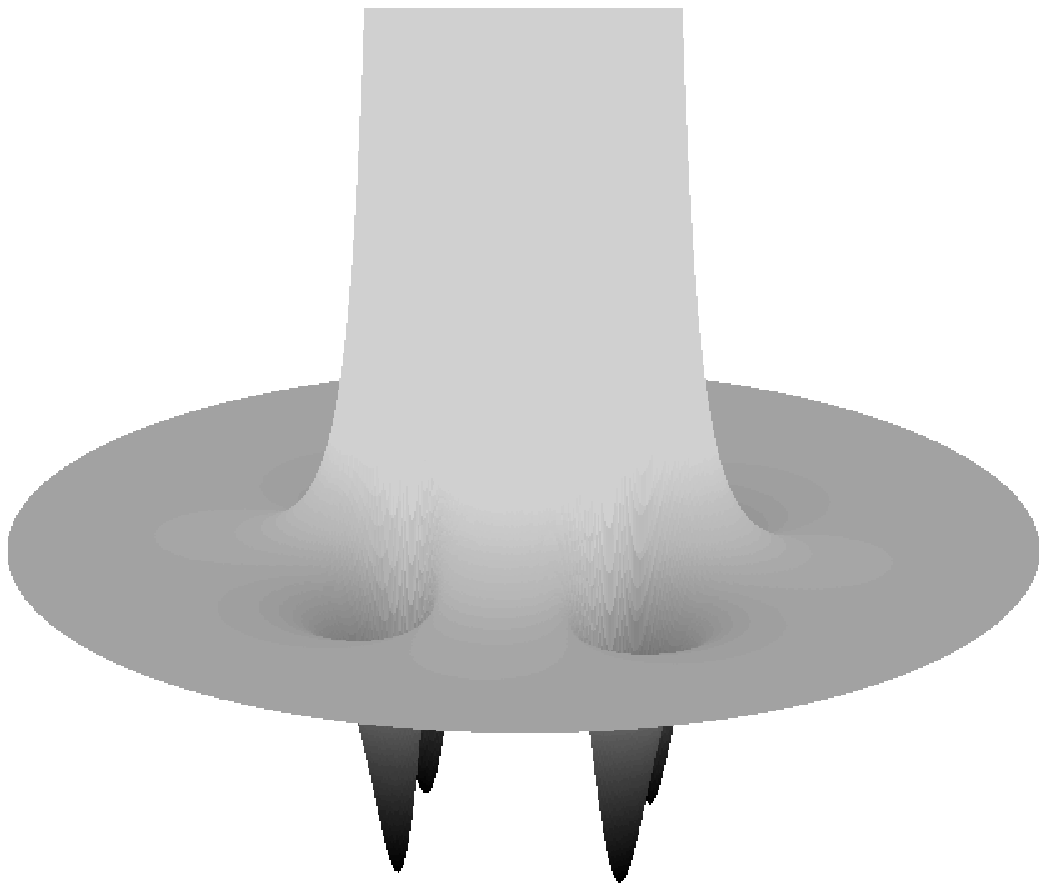}
  }\\
    \subfigure[\footnotesize   \label{fig:angpothclattice}]{
  \includegraphics[width=6.5cm]{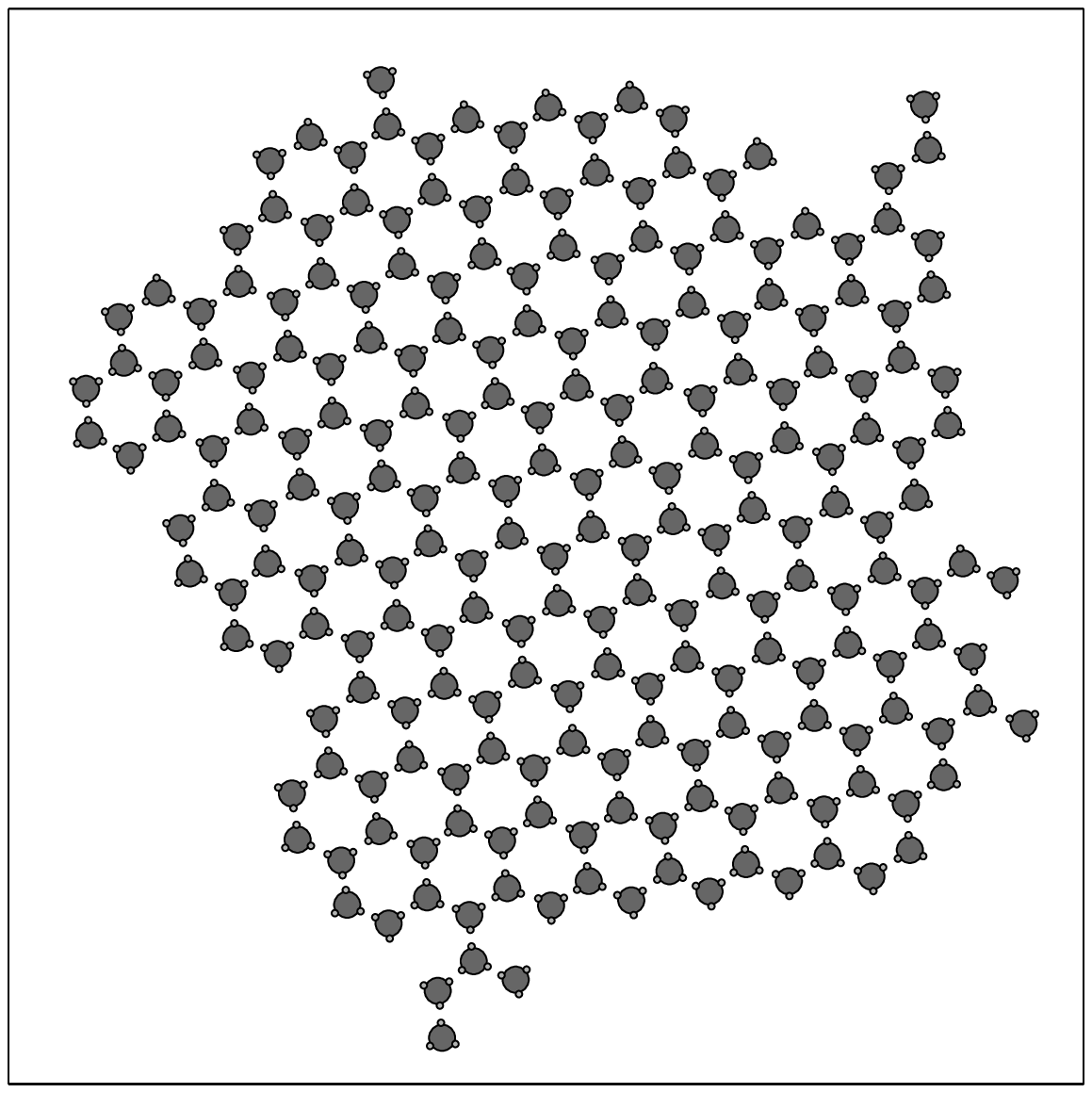}
  }
  \hspace{0.1cm}
  \subfigure[\footnotesize  \label{fig:angpotsqlattice}]{
  \includegraphics[width=6.5cm]{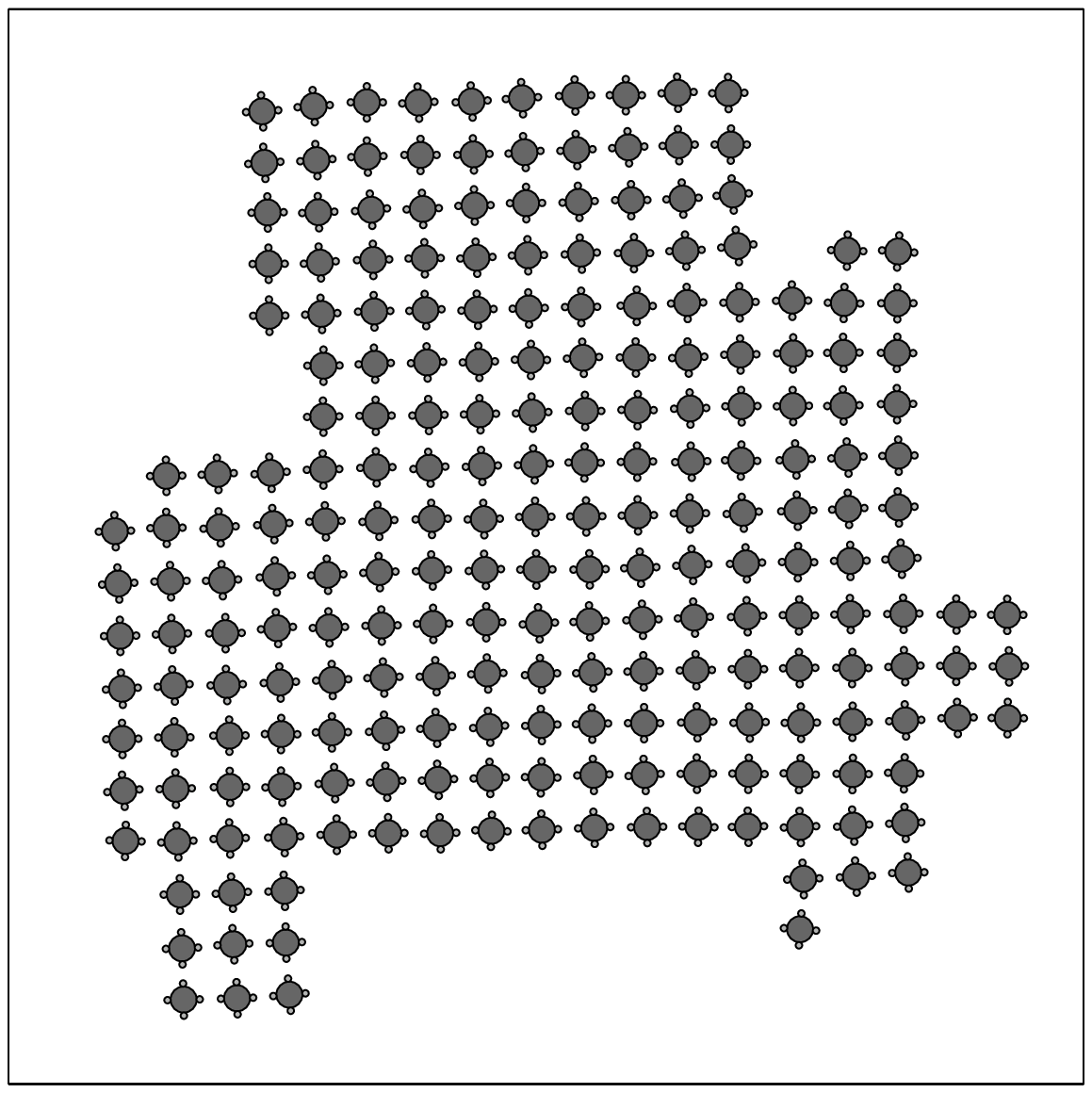}
  }
  \caption{ \label{fig:anghc} \footnotesize The potential with three-fold symmetry shown in (a) yields the honeycomb lattice shown in (c).  Similarly, the potential with four-fold symmetry shown in (b) yields the square lattice shown in (d).}
\end{center}
\end{figure}

By changing the integer value of $n$, we can induce the formation of lattices with desired $n$-fold symmetry.  Surface plots of potentials with 3-fold and 4-fold symmetry ($n$=3 and $n$=4) as a function of the radial coordinate $r$ and the azimuthal angle $\alpha$, are provided in Figures \ref{fig:angpothc} and \ref{fig:angpotsq}.   A honeycomb lattice and a square lattice produced with these potentials using particles initialized at low density are shown in Figures \ref{fig:angpothclattice} and \ref{fig:angpotsqlattice}.  

For the creation of more exotic lattices, we alter the azimuthal modulation of the potential even further.  Each particle in the kagome lattice, for example, has binding sites at angles of $0^{\circ}$, $60^{\circ}$,  $180^{\circ}$, and $240^{\circ}$~\cite{mekata2003a}.  To ensure preferred binding along these directions, we must modify the azimuthal dependence of the Lennard-Jones potential accordingly.  Before doing so, we first express the potential function $\psi$ of line \eqref{eq:psi} in the following equivalent way:
\begin{align} 
\label{eq:psi2}
\psi(\Delta x,\Delta y,\theta_{k}) = \frac{1}{2r^{12}} - \frac{1}{r^6}\left[1-\nu \,S(\alpha_{k})\right]
\end{align} 
where
\begin{align} 
S(\alpha_{k}):=\sin^{2}\left(\frac{n\alpha_{k}}{2}\right)\,.
\end{align} 
As written here, $\psi$ produces a potential with $n$-fold symmetry.  In order to produce a potential that favors the kagome lattice, we must simply introduce a new definition for the functional form of $S(\cdot)$.

\begin{figure}[ht]
\begin{center}
\subfigure[\footnotesize   \label{fig:CS}]{
  \includegraphics[width=4.4cm]{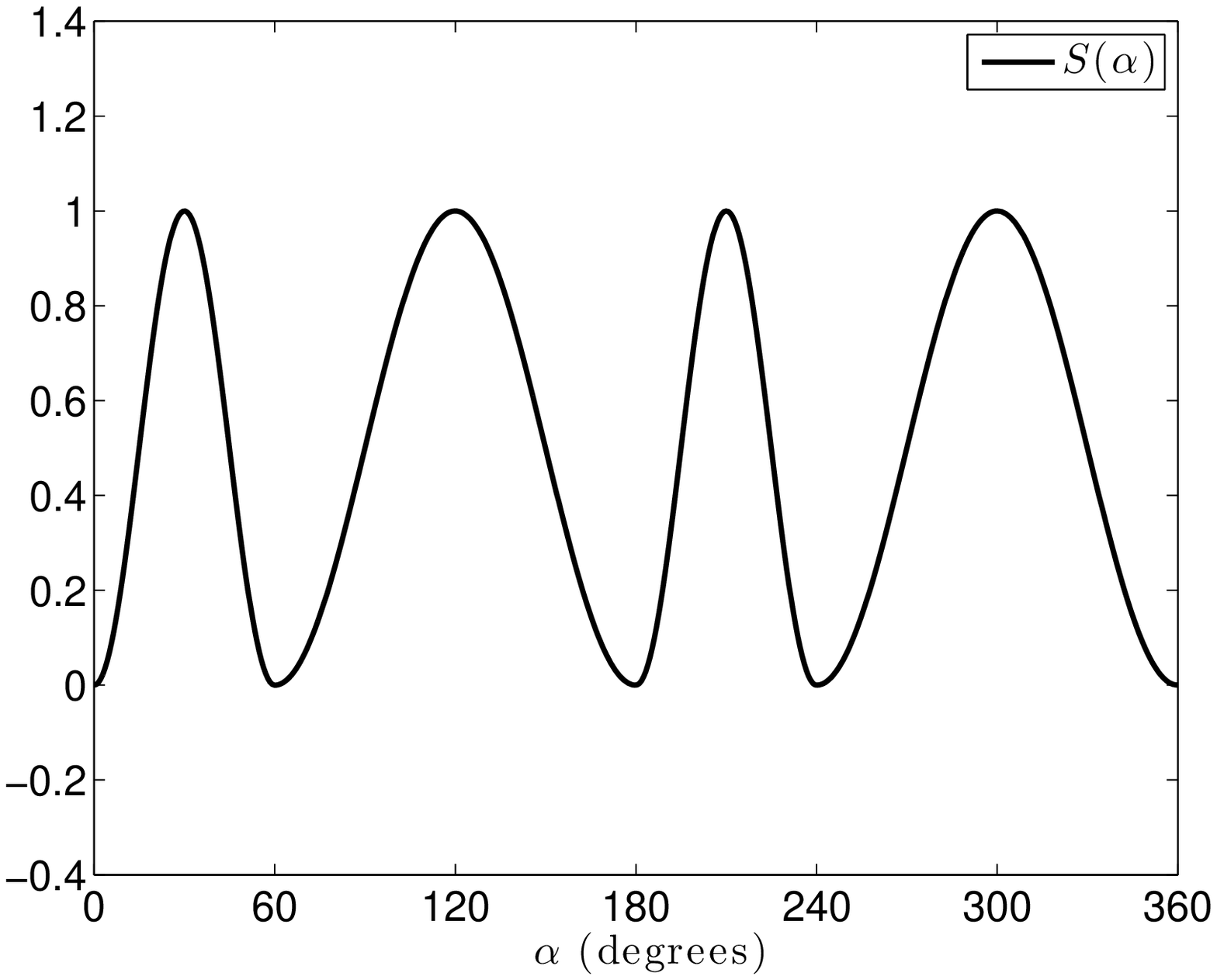}
  }
  \hspace{0.5cm}
  \subfigure[\footnotesize   \label{fig:angpotkag}]{
  \includegraphics[width=4.4cm]{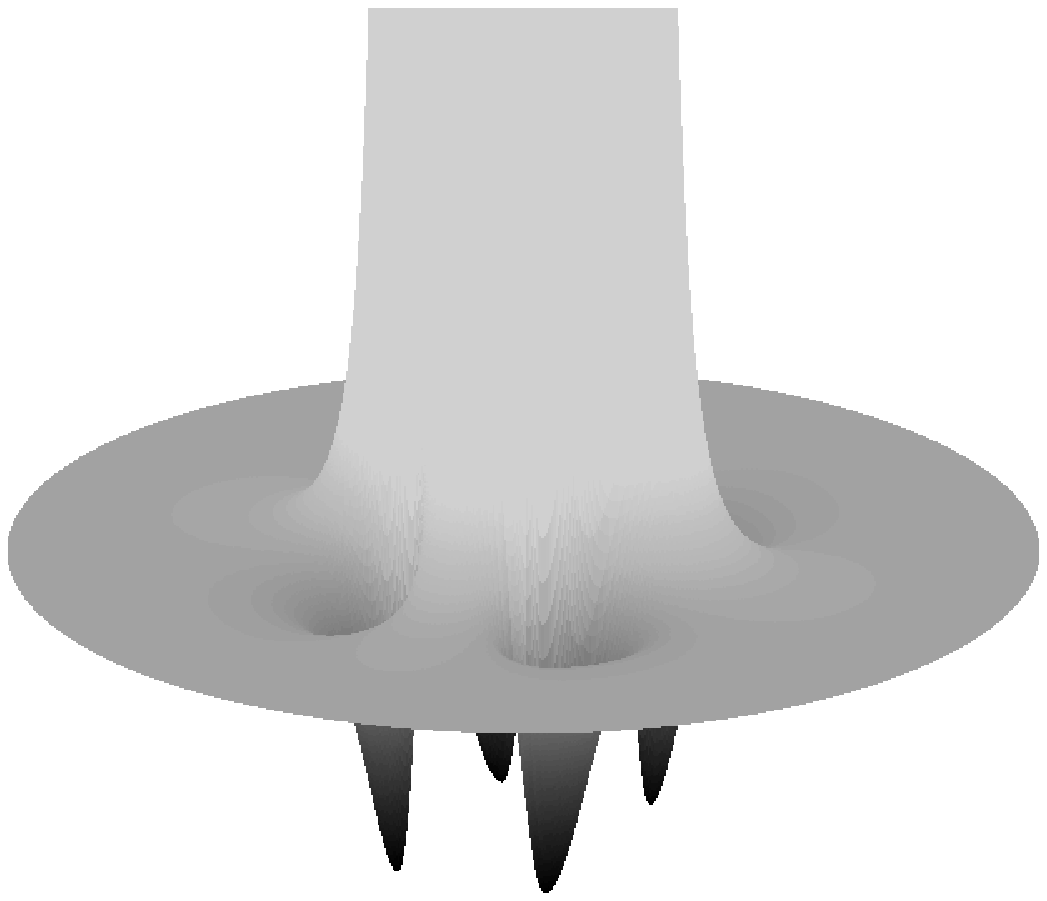}
  }\\
  \subfigure[\footnotesize   \label{fig:angpotkaglattice}]{
  \includegraphics[width=6.5cm]{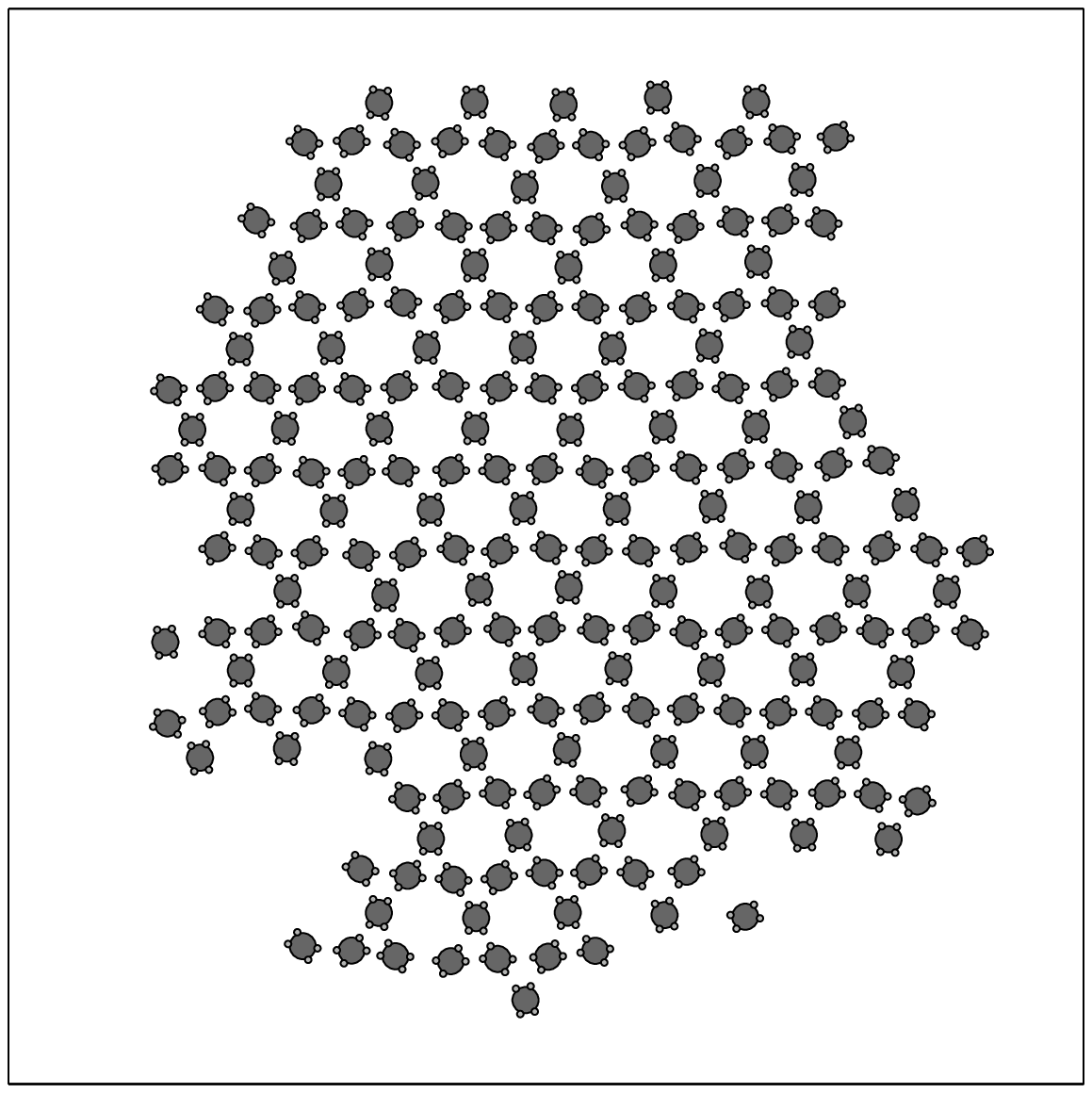}
  }
  \caption{ \label{fig:angkag}\footnotesize The function in (a) is used to azimuthally modulate the amplitude of a Lennard-Jones potential to encourage binding directions that favor the kagome lattice.  The resulting potential surface is shown in (b), while a kagome lattice formed with this potential is shown in (c).}
\end{center}
\end{figure}

Let $B:=\left[b_{1},\cdots,b_{n}\right]$ denote the ordered list of $n$ desired binding directions measured in radians satisfying 
\begin{align} 
0 = b_{1} < \cdots <b_{n}< 2\pi\,.
\end{align}
Note that without loss of generality, we may prescribe that the first binding direction lies along the ray corresponding to zero radians.  For the honeycomb lattice, $B_\textsf{HC}:=\left[0,\frac{2\pi}{3},\frac{4\pi}{3}\right]$, while for the kagome lattice we have $B_\textsf{kagome}:=\left[0,\frac{\pi}{3},\pi,\frac{4\pi}{3}\right]$.  Then, for $\alpha \in [0,2\pi)$, we use the elements in the list $B$ to define $S(\alpha)$ piecewise as follows:
\begin{align} 
S(\alpha) :=  \begin{cases} 
\sin^{2}\left(\frac{\alpha-b_{i}}{b_{i+1}-b_{i}}\pi\right)&\text{if } b_{i} \leq \alpha < b_{i+1}\,, \hspace{0.5cm}\text{\small for $i=1,\cdots,n-1\,$,}\\ 
\sin^{2}\left(\frac{\alpha-b_{n}}{2\pi-b_{n}}\pi\right)&\text{if }  b_{n} \leq \alpha < 2\pi\,.\end{cases}
\end{align} 
When $S(\alpha)$ defined in this way is substituted into the expression for the anisotropic potential function $\psi$ in line \eqref{eq:psi2}, it provides the necessary azimuthal modulation of the Lennard-Jones potential to produce potential wells along the binding directions specified in the list $S$.  A plot of $S(\alpha)$ using $B_\textsf{kagome}$ is provided in Figure \ref{fig:CS}.  Notice that $S(\alpha)$ has local minima precisely at the binding site angles of $0^{\circ}$, $60^{\circ}$,  $180^{\circ}$, and $240^{\circ}$ consistent with the kagome lattice (recall from line \eqref{eq:psi2} that minima in $S(\alpha)$ lead to minima in the interaction potential).  Consequently, the angular dependence in the resulting interaction potential favors the bond structure peculiar to the kagome lattice.  The potential and a lattice resulting from this potential are shown in Figures \ref{fig:angpotkag} and \ref{fig:angpotkaglattice} respectively.  

\section{Conclusions and Future Work\label{sec:Conclusion and Future Directions}}

We have presented and compared methods for generating potentials that lead to the self-assembly of specified target lattices. In particular, we have addressed the problem of designing pairwise interaction potentials that induce the formation of a honeycomb lattice when a planar system of particles is cooled.  

We have demonstrated that reasonably high quality lattices can be produced using a heuristic computation-free geometric method.  The geometric method provides principles that utilize purely geometric information to design potentials that by construction favor and stabilize the target lattice.   

A trend optimization algorithm has been introduced that quickly and robustly finds optimal shapes of the interaction potential that lead to the self-assembly of lattices of high quality.  The success of the trend method lies in its ability to quickly locate minima in a noisy and expensive objective function.  Moreover, the potentials discovered using the trend optimization procedure robustly form high quality lattices with respect to variations in the initial conditions of the particles.  We have seen that the trend optimization method has discovered a family of potentials characterized by very simple exponential decay profiles that routinely lead to the formation of the honeycomb lattice.

Our trend optimization algorithm robustly and routinely finds optimal values of the objective, although it must be noted that as currently implemented, the algorithm does not provide rigorous guarantees of convergence.  Convergence can be guaranteed, however, by incorporating a \textit{polling} step as required in the Surrogate Management Framework.  It would be interesting to investigate  the effects of polling on the optimal results and the efficiency of the method.  One expects that the addition of a polling step to guarantee convergence will represent only a marginal increase in the overall computational cost of the algorithm.  

The geometric method has also been extended to the design of anisotropic potentials.  Azimuthal dependence of the interaction potential allows for the formation of the kagome lattice which has not previously been performed using isotropic potentials.   Incorporating anisotropy into the potentials allows for the formation of lattices over a wide range of particle densities.

An auxiliary contribution of this paper is the development of two metrics for objective analysis of lattice quality.  The development of these metrics was necessary for comparison of the lattices produced by the proposed potentials. 

We anticipate that the methods presented here will naturally extend to three dimensions without impediment.  Rechtsman \textit{et al} have recently investigated the design of potentials for self-assembly of three-dimensional lattice structures~\cite{rechtsman2007a}.  Most notably, they have demonstrated the formation of the diamond and wurtzite lattices.  As a matter of course, we intend to apply the methods presented here to the design of potentials in the three-dimensional self-assembly problem and expect the trend optimization method to provide a significant computational savings in the design of potentials.  Of particular interest is to determine if trend optimization recovers the three-dimensional result of~\cite{rechtsman2007a}, or if a new, simpler, and perhaps more robust family of solutions is discovered.

A natural extension of the methods presented is to investigate the use of multi-specie and multi-body potentials for the self-assembly of quasicrystals and Penrose tilings.  Generating lattices with prescribed geometry and structure is highly motivated by the desire to produce photonic crystals with specified optical properties.  Also, we intend to explore the use of the trend optimization method for designing potentials that lead to the the formation of hexahedral meshes over complicated three-dimensional volumes.  Producing hexahedral meshes is a notoriously difficult problem that is currently a major stumbling block for continuum mechanics computations that use a finite element method.

More broadly, we expect to use trend optimization to understand more deeply the fundamental limitations and extraordinary possibilities of self-assembly.  It will be interesting to investigate, for example, the role of optimal potential design in natural systems.  In this regard, a study of self-assembly is a study of life itself.  Are there mechanisms at the level of local interactions that allow for the differentiation of cells that self-assemble to form bone, for instance, from those that form brain tissue?  How much control authority over the superstructure formed via self-assembly is provided by control over the local interactions?  Are there natural systems in which a small amount of flexibility in the properties of the local interactions allows for large and beneficial changes in structure at the macroscopic level?  We believe that insights to these questions may be afforded by finding optimal potentials through the method of trend optimization.

\end{document}